\documentclass[final,1p,times]{elsarticle}
\usepackage[T1]{fontenc}    
\usepackage{amsfonts}
\usepackage{amssymb}
\usepackage{amsmath}
\usepackage{amsthm}
\usepackage{amscd}
\usepackage{color}
\usepackage{graphicx}
\usepackage{epsfig}
\usepackage{subfigure}
\usepackage{fancyhdr}
\usepackage{dcolumn}
\usepackage{time}
\usepackage{shadow}
\usepackage{hyperref}
\hypersetup{
    unicode=false,          
    pdftoolbar=true,        
    pdfmenubar=true,        
    pdffitwindow=false,     
    pdfstartview={FitH},    
    pdftitle={Paper},       
    pdfauthor={John Dawson},     
    pdfsubject={Phys Rev article},   
    pdfcreator={John Dawson},   
    pdfproducer={John Dawson}, 
    pdfkeywords={keyword1} {key2} {key3}, 
    pdfnewwindow=true,      
    colorlinks=true,        
    linkcolor=red,          
    citecolor=blue,         
    filecolor=green,        
    urlcolor=cyan           
}
\definecolor{dark}{gray}{.5}
\definecolor{light}{gray}{.75}
\definecolor{darkmagenta}{rgb}{.5,0,.5}
\makeatletter
\def\cleardoublepage{\clearpage\if@twoside \ifodd\c@page\else
    \hbox{}
    \thispagestyle{plain}
    \newpage
    \if@twocolumn\hbox{}\newpage\fi\fi\fi}
\makeatother \clearpage{\pagestyle{plain}\cleardoublepage}
\newcolumntype{d}[1]{D{.}{.}{#1}}
\newcommand{\etal}{\textit{et~al.}}        
\newcommand{\vs}{vs.}                      
\newcommand{\QED}%
   {$\mathcal{Q\kern-.1em \lower.6ex\hbox{$\mathcal{E}$}\kern-.1667em D}$}
\newcommand{\QCD}%
   {$\mathcal{Q\kern-.1em \lower.6ex\hbox{$\mathcal{C}$}\kern-.1667em D}$}

\newcommand{\SUSYext}%
   {$\mathcal{S \kern-0.08em \lower 0.5ex \hbox{$\mathcal{U}$}
   \kern-0.05em S \kern-0.2em \lower 0.5ex
   \hbox{$\mathcal{Y}$}}\kern-0.05em{}_{\text{ext}}$}
\newcommand{\Schrodinger}{Schr{\"o}dinger}     
\newcommand{\kB}{k_{\text{B}}}                 
\newcommand{\rD}{\text{D}}

\newcommand{\boldeta}{\boldsymbol{\eta}}

\newcommand{\bnu}{\boldsymbol{\nu}}

\newcommand{\bphi}{\boldsymbol{\phi}}

\newcommand{\bnabla}{\boldsymbol{\nabla}}

\newcommand{\sphi}{\phi^{\star}}

\newcommand{\sbphi}{\boldsymbol{\phi}^{\star}} 
\newcommand{\sPhi}{\Phi^{\star}}

\newcommand{\ssigma}{\sigma^{\star}}

\newcommand{\lambdaCH}{\lambda^{\text{CH}}}
\newcommand{\bD}{\mathbf{D}}

\newcommand{\bG}{\mathbf{G}}

\newcommand{\bj}{\mathbf{j}}
\newcommand{\bk}{\mathbf{k}}

\newcommand{\bv}{\mathbf{v}}

\newcommand{\bx}{\mathbf{x}}

\newcommand{\sj}{j^{\star}}
\newcommand{\sbj}{\mathbf{j}^{\star}}
\newcommand{\sJ}{J^{\star}}

\newcommand{\sK}{K^{\star}}
\newcommand{\sD}{D^{\star}}
\newcommand{\spD}{D^{\phantom\star}}

\newcommand{\sX}{X^{\star}}
\newcommand{\sw}{w^{\star}}
\newcommand{\sW}{W^{\star}}
\newcommand{\calA}{\mathcal{A}}

\newcommand{\calC}{\mathcal{C}}
\newcommand{\calD}{\mathcal{D}}

\newcommand{\calJ}{\mathcal{J}}

\newcommand{\calN}{\mathcal{N}}

\newcommand{\Imag}[1]{\ensuremath{\mathcal{I}m \{ \, #1 \, \} }}  
\newcommand{\Quad}[1]{\quad\text{#1}\quad}         
\newcommand{\Qquad}[1]{\qquad\text{#1}\qquad}      
\newcommand{\Set}[1]{\bigl ( \, #1 \, \bigr )}     
\newcommand{\Setc}[1]{ \{ \, #1 \, \}}             
\newcommand{\rd}{\mathrm{d}}
\newcommand{\Partial}[4]
   {\Bigl ( \frac{\partial #1 }{\partial #2 } \Bigr )_{\! #3, #4 }}
\newcommand{\RAnabla}{\overrightarrow{\nabla}}
\newcommand{\LAnabla}{\overleftarrow{\nabla}}
\newcommand{\RAbnabla}{\overrightarrow{\boldsymbol{\nabla}}}
\newcommand{\LAbnabla}{\overleftarrow{\boldsymbol{\nabla}}}

\newcommand{\Det}[1]{\det [ \, #1 \, ]}
\newcommand{\DetB}[1]{\det \Bigl \lbrack \, #1 \, \Bigr \rbrack }

\newcommand{\ExpB}[1]{\exp \Bigl \{ \, #1 \, \Bigr \} }

\newcommand{\Ln}[1]{\ln [ \, #1 \, ]}

\newcommand{\LnB}[1]{\ln \Bigl \lbrack \, #1 \, \Bigr \rbrack }
\newcommand{\Tr}[1]{\mathrm{Tr} [ \, #1 \, ]}

\newcommand{\TrB}[1]{\mathrm{Tr} \Bigl \lbrack \, #1 \, \Bigr \rbrack }

\newcommand{\bra}[1]%
   {\ensuremath{\langle \, #1 \, |}}
\newcommand{\Bra}[1]%
   {\ensuremath{\langle \, #1 \, |}}
\newcommand{\bigbra}[1]%
   {\ensuremath{\Bigl \langle \, #1 \, \Bigr |}}
\newcommand{\ket}[1]%
   {\ensuremath{| \, #1 \, \rangle}}
\newcommand{\Ket}[1]%
   {\ensuremath{| \, #1 \, \rangle}}
\newcommand{\bigket}[1]%
   {\ensuremath{\Bigl | \, #1 \, \Bigr \rangle}}
\newcommand{\braket}[2]%
   {\ensuremath{\langle \, #1 \, | \, #2 \, \rangle}}
\newcommand{\Braket}[2]%
   {\ensuremath{\langle \, #1 \, | \, #2 \, \rangle}}
\newcommand{\matrixelement}[3]%
   {\ensuremath{\langle \, #1 \, | \, #2 \, | \, #3 \, \rangle}}
\newcommand{\MEangle}[3]%
   {\ensuremath{\langle \, #1 \, | \, #2 \, | \, #3 \, \rangle}}
\newcommand{\MEangleB}[3]%
   {\ensuremath{\Bigl \langle \, #1 \, \Big | \, #2 \, \Big | \, #3 \, \Bigr \rangle}}
\newcommand{\MatEl}[3]%
   {\ensuremath{\langle \, #1 \, | \, #2 \, | \, #3 \, \rangle}}
\newcommand{\MEparen}[3]%
   {\ensuremath{( \, #1 \, | \, #2 \, | \, #3 \, )}}
\newcommand{\pbra}[1]%
   {\ensuremath{( \, #1 \, |}}
\newcommand{\pket}[1]%
   {\ensuremath{| \, #1 \, )}}    
\newcommand{\Brap}[1]%
   {\ensuremath{( \, #1 \, |}}
\newcommand{\Ketp}[1]%
   {\ensuremath{| \, #1 \, )}}    
\newcommand{\pbraket}[2]%
   {\ensuremath{( \, #1 \, | \, #2 \, )}}
\newcommand{\braV}[1]%
   {\ensuremath{\langle \, #1 \, \Vert}}
\newcommand{\ketV}[1]%
   {\ensuremath{\Vert \, #1 \, \rangle}}
\newcommand{\Comm}[2]%
   {\ensuremath{[ \, #1, #2 \, ]}}
\newcommand{\AntiComm}[2]%
   {\ensuremath{\{ \, #1, #2 \, \}}}
\newcommand{\Pbracket}[2]%
   {\ensuremath{\{ \, #1, #2 \, \} }}
\newcommand{\PBracket}[2]%
   { \ensuremath{ \{ \, #1, #2 \, \}_{\lower1.0ex\hbox{\scriptsize \text{PB}}} } }
\newcommand{\wedgeComm}[2]
   {\ensuremath{[ \, #1, #2 \, ]_{\lower1.0ex\hbox{\scriptsize $\wedge$}} }}
\newcommand{\Expect}[1]
   {\ensuremath{\langle \, #1 \,  \rangle}}
\newcommand{\expect}[1]%
   {\ensuremath{\langle \, #1 \,  \rangle}}
\newcommand{\Expectbig}[1]%
   {\ensuremath{\Bigl \langle \, #1 \, \Bigr \rangle}}
\newcommand{\expectbig}[1]%
   {\ensuremath{\Bigl \langle \, #1 \, \Bigr \rangle}}
\newcommand{\Expectb}[1]%
   {\ensuremath{\bigl \langle \, #1 \, \bigr \rangle}}
\newcommand{\ExpectB}[1]%
   {\ensuremath{\Bigl \langle \, #1 \, \Bigr \rangle}}
\newcommand{\expectc}[2]%
   {\ensuremath{\langle \, \{ \, #1 , #2 \, \} \, \rangle}}
\newcommand{\expectq}[2]%
   {\ensuremath{\langle \, [ \, #1 , #2 \, ] \, \rangle}}
\newcommand{\expectT}[1]%
   {\ensuremath{\langle \, \mathcal{T} \{ \, #1 \, \} \, \rangle}}
\newcommand{\expectaT}[1]%
   {\ensuremath{\langle \, \mathcal{T}^{\ast} \{ \, #1 \, \} \, \rangle}}
\newcommand{\ExpectT}[1]%
   {\ensuremath{\langle \, \mathcal{T} \{ \, #1 \, \} \, \rangle}}
\newcommand{\ExpectaT}[1]%
   {\ensuremath{\langle \, \mathcal{T}^{\ast} \{ \, #1 \, \} \, \rangle}}
\newcommand{\ExpectTB}[1]%
   {\ensuremath{\Bigl \langle \, \mathcal{T}  \Bigl \{ \, #1 \, \Bigr \} \, \Bigr \rangle}}
\newcommand{\expectTabig}[1]%
   {\ensuremath{\biggl \langle \, \mathcal{T}^{\ast} \biggl \{ \, #1 \,%
\biggr \} \, \biggr \rangle}}
\newcommand{\Tproduct}[1]%
   {\ensuremath{\mathcal{T} \{ \, #1 \, \} } }
\newcommand{\aTproduct}[1]%
   {\ensuremath{\mathcal{T}^{\ast} \{ \, #1 \, \} } }
\newcommand{\Nproduct}[1]%
   {\ensuremath{\mathcal{N} \{ \, #1 \, \} } }
\newcommand{\ctpTproduct}[1]%
   {\ensuremath{\mathcal{T}_{\mathcal{C}} \{ \, #1 \, \} }}
\newcommand{\KTproduct}[1]%
   {\ensuremath{\mathcal{T}_{\mathcal{K}} \{ \, #1 \, \} }}
\newcommand{\tauordered}[1]%
   {\ensuremath{\mathcal{T}_{\tau} \{ \, #1 \, \} }}
\newcommand{\Norder}[1]%
   {\ensuremath{: #1 : \, }}
\newcommand{\ExpectTproduct}[1]%
   {\ensuremath{\langle \, \mathcal{T} \{ \, #1 \, \} \, \rangle}}
\newcommand{\ExpectTaproduct}[1]%
   {\ensuremath{\langle \, \mathcal{T}^{\ast} \{ \, #1 \, \} \, \rangle}}
\newcommand{\ExpectTCproduct}[1]%
   {\ensuremath{\langle \, \mathcal{T}_{\mathcal{C}} \{ \, #1 \, \} \, \rangle}}
\newcommand{\expectComm}[2]%
   {\ensuremath{\langle \, [ \, #1 , #2 \, ] \, \rangle}}
\newcommand{\expectPbracket}[2]%
   {\ensuremath{\langle \, \{ \, #1 , #2 \, \} \, \rangle}}
\newcommand{\threej}[6]%
{\begin{pmatrix} #1 & #2 & #3 \\ #4 & #5 & #6 \end{pmatrix}}
\newcommand{\sixj}[6]%
{\begin{Bmatrix} #1 & #2 & #3 \\ #4 & #5 & #6 \end{Bmatrix}}
\newcommand{\ninej}[9]%
{\begin{Bmatrix} #1 & #2 & #3 \\ #4 & #5 & #6 \\%
 #7 & #8 & #9 \end{Bmatrix}}
\newcommand{\reducedme}[3]%
{\langle \, #1 \, \Vert \, #2 \, \Vert \, #3 \, \rangle }
\journal{Annals of Physics}
\begin{document}
%
%
\begin{frontmatter}
\title{Auxiliary Field Loop Expansion of the Effective Action for a class of Stochastic Partial Differential Equations }
\author[add1,add2]{Fred Cooper} 
\ead{cooper@santafe.edu}
\address[add1]{The Santa Fe Institute, 1399 Hyde Park Road, Santa Fe, NM 87501, USA}
\address[add2]{Theoretical Division and Center for Nonlinear Studies,
   Los Alamos National Laboratory, Los Alamos, NM, 87545, USA}
%
\author[add3]{John F. Dawson}
\ead{john.dawson@unh.edu}
\address[add3]{Department of Physics,
   University of New Hampshire,
   Durham, NH 03824, USA}
%
%
\begin{abstract}

We present an alternative to the perturbative (in coupling constant)  diagrammatic approach for studying stochastic dynamics of a class of reaction diffusion systems.  Our approach is based on an auxiliary field loop expansion for the path integral representation for the generating functional of the noise induced  correlation functions of the fields describing these systems. The systems we consider include Langevin systems describable by the set of  self interacting classical fields $\phi_i(x,t)$ in the presence of external noise $\eta_i (x,t)$, namely $(\partial_t - \nu \nabla^2 ) \phi - F[\phi] = \eta$, as well as chemical reaction annihilation processes obtained by applying the many-body approach of 
Doi-Peliti to the Master Equation formulation of these problems. 
 We consider two different effective actions, one  based on the Onsager-Machlup (OM) approach, and the other  due to Jannsen-deGenneris based on the Martin-Siggia-Rose (MSR) response function approach.  For the simple models we consider, we determine an analytic expression for  the Energy landscape (effective potential) in both formalisms and show how to obtain the more physical effective potential of the Onsager-Machlup approach from the MSR effective potential in leading order in the auxiliary field loop expansion. For the KPZ equation we find that our approximation, which is non-perturbative and obeys broken symmetry Ward identities, does not lead to the appearance of a fluctuation induced symmetry breakdown. This contradicts the results of earlier studies.
\end{abstract}
\end{frontmatter}
%
%
%
\section{\label{s:Intro}Introduction}
The effective action $\Gamma[\phi]$  for stochastic partial differential equations extends the role played by the  action for field theories of  classical dynamical systems.  It is defined in terms of  functional Legendre transformation of the generating functional for the connected correlation functions (see for example the book by Rivers \cite{r:Rivers:1990yf}). 
The effective action accounts for the physics of the entire system, composed of the various degrees of freedom, represented by the fields, as well as the effect on them by stochastic agents in the form of noise.  The first derivative of $\Gamma[\phi]$  with respect to the field gives the equation for the evolution of the field, averaged over different realizations of the noise chosen from a given probability distribution.  Higher derivatives of the effective action determine the one particle-irreducible (1-PI) vertices, from which all the noise induced  correlation functions of the field can be reconstructed. The 1-PI vertices play a crucial role in identifying the eventual renormalization of the parameters found in the theory without noise.  The effective action in quantum field theory is reviewed in great detail in the 
books by Rivers \cite{r:Rivers:1990yf} and Itzykson and Zuber \cite{ref:ItzyksonZuber}, and  the extension of this approach to studying reaction diffusion equations was pioneered by  Hochberg and collaborators  [give references].   For studying phase structure, one considers the energy landscape probed by homogeneous fields using the value of the effective action for homogeneous  fields divided by the space time volume.  The utility of the auxiliary field loop expansion is that in lowest order it leads to a self consistent Hartree-like approximation, which preserves underlying symmetries that gives a qualitative analytic picture of the Energy Landscape.
As a recent example, we have used it to give a simple qualitative picture of the phase structure of the
Bose-Hubbard model \cite{PhysRevA.88.023607}. 
When the dynamics leads to quartic (and higher) self interactions in the Lagrangian, auxiliary fields can convert the topology of the coupled field equations to be trilinear.
This simplification leads to a dramatic  topological simplification of the  Schwinger Dyson equations for the correlation functions. This also simplifies dramatically the expansion of the generating functional of the two particle irreducible graphs and allows one to study dynamics in approximations which in lowest order are related to the approach of Kraichnan \cite{r:Kraichnan:1959yq,r:Kraichnan:1961hl} in his study of plasma turbulence.  Related methods have recently been used by Doherty in the study of the Kardar-Parisi-Zhang (KPZ ) equation\cite{r:Doherty:1994qo}. 

Most text books on quantum field theory (see for example Ref.~\cite{ref:ItzyksonZuber}) discuss the effective action and the effective potential in  the semi-classical approximation, which keeps only the gaussian fluctuations  around the classical motion.  Some more recent textbooks \cite{r:Calzetta:2008pb} describe approximations based on the generating functional for the two particle irreducible (2-PI) graphs, which includes the Hartree-Fock approximation.  However approximations to this 2-PI generating functional are notorious for violating Ward-Takahashi identities \cite{r:Hees:2002hc}.  This has been a great stumbling block for using the 2-PI approach when there are broken symmetries.  

In the recent literature, most discussions of the effective action for stochastic partial differential equations are based on a loop expansion in terms of the strength of the noise correlation function.  This approach is related to  the semi-classical approximation to the effective action of quantum field theory. Another approach has been to use the self-consistent Hartree approximation.  For the Kardar-Parisi-Zhang (KPZ) equation \cite{PhysRevLett.56.889}, recent discussions on the loop expansion is found in Refs.~\cite{r:PhysRevE.60.6343} and for the Hartree approximation \cite{r:Amaral:2007ty}.  In population biology, a recent discussion is found in Ref.~\cite{r:Dodd:2009qa}, and for pair annihilation and Gribov processes a recent discussion is found in Ref.~\cite{r:Hochberg:2004bv}. In the semi-classical approximation it is tacitly assumed that the noise is a small perturbation on the classical dynamics and that therefore perturbation theory in the strength of the noise is valid.  This separation is often not present in many dynamical systems acted upon by noise, whether the noixe is internal or external. This shortcoming of the semi-classical approximation was seen even at weak couplings in the theory of dilute Bose gases where the fluctuations are thermal in nature. In that situation the semiclassical approximation did not give  a true picture of the phase structure \cite{r:Andersen:2004uq}.  For the phase structure of dilute Bose gases, we were able to show that an approximation directly related to the one presented here \cite{PhysRevLett.105.240402,PhysRevA.85.023631} was able to predict the correct phase diagram.   The reason for the success of LOAF was that the auxiliary field loop expansion preserves the broken symmetry Ward identities.

One example we study in this paper is the phase structure of the massless KPZ equation using the auxiliary field loop expansion.  As mentioned above,
efforts based on the usual loop expansion \cite{r:PhysRevE.60.6343} as well as the Hartree approximation \cite{r:Amaral:2007ty} used methods which gave wrong results when applied to the phase structure of dilute Bose gases.  Unlike these previous approaches, we find no evidence for dynamical symmetry breaking.  Recently Bork and Ogarkov \cite{r:Bork:2014yg} have done a renormalization group improvement of the one loop result of Hochberg, \etal\ \cite{r:PhysRevE.60.6343} and also come to the conclusion that there is no symmetry breaking in the KPZ problem.

In this paper we present a new approach for determining the energy landscape for stochastic field theory evolutions , based on the work in Ref.~\cite{r:Bender:1977bh}, that is non-perturbative in both coupling constant strength and the strength of the noise correlation functions. 
This approach introduces auxiliary fields, which render the Lagrangian for the stochastic field theory quadratic in the original fields.  
In lowest order one gets a ``self consistent'' Gaussian approximation, but \emph{not} a Gaussian truncation, which preserves symmetries.   
One also can determine in a systematic way the corrections to the lowest order approximation. Each term in this loop expansion also satisfies the Ward identities. 
This method has been successfully used to determine approximately, in leading order, the phase structure of dilute Bose \cite{PhysRevLett.105.240402,PhysRevA.85.023631} and Fermi gases \cite{PhysRevLett.71.3202}, as well as the Bose-Hubbard model \cite{PhysRevA.88.023607}.
The effective potential here plays a role similar to the usual potential of classical mechanics as it maps the energy landscape of the system being studied.

Another topic we want to address is the relationship between the two path integral formulations for the generating functional of the correlation functions. 
In the literature there are two formulations for obtaining the effective action and effective potential.  The first formulation is based on the ideas of Onsager and Machlup (OM) \cite{r:Onsager:1953nr,r:Graham:1973ek,r:Zinn-Justin:1986lr}, and is in terms of the original fields present in the stochastic differential equation.   
A second formulation is due to Janssen \cite{r:Janssen:1976xy} and de~Dominicis (JD)\cite{r:Dominicis:1976dq} and introduces a field conjugate to the original field, which is a Lagrange multiplier field, to obtain a Lagrangian from which the equation of motion can be obtained.  This last method is based on the response theory formalism of Martin-Siggia-Rose (MSR) \cite{r:Martin:1973lr,r:peliti,r:Janssen:1976xy,r:Jouvet:1979fx}.  
The JD action also naturally occurs for the case of annihilation-diffusion chemical reactions when we start from a master equation.
Although formally one can recover the OM action from the JD action by integrating over the conjugate field, how one obtains the \emph{approximate} effective action for the OM formulation from the \emph{approximate} JD effective action has not been discussed (to our knowledge) in the literature. 
Here we will explicitly show that in the LOAF approximation, once we determine the value of the Lagrange multiplier field in terms of the original field,  and use this in the JD effective action, then we find $\Gamma_{OM} [ \phi] = \Gamma_{MSR} [\phi, \sphi[\phi] ]$.  
Here $\Gamma$ is the generating functional of one particle irreducible graphs.  Thus if we determine the LOAF approximation to the effective potential in the MSR formulation, we can reconstruct the \emph{physical} effective potential of the OM formalism, and get the same answer as if we calculated the potential directly using the OM formalism.  Thus nothing is given up by using the MSR formalism, and one gets the bonus of obtaining the response functions as well as the correlation functions. 

The use of auxiliary fields in many-body theory and quantum field theory has a long history starting with Hubbard and Stratonovich \cite{r:Hubbard:1959kx,r:Stratonovich:1958vn}.  The auxiliary fields render the classical action quadratic in the original fields in the path integral formulation of the theory.  
Because of this one can reproduce many self-consistent gaussian fluctuation approximations by evaluating the path integral over the auxiliary fields by steepest descent.  In quantum field theory and many-body field theory applications, the leading order approximation (LOAF) determines the large-N behavior of the $O(N)$ model \cite{PhysRevD.10.2491}, BCS theory of superconductivity \cite{PhysRevLett.71.3202}, and the LOAF theory for Bose-Einstein condensation \cite{PhysRevLett.105.240402,PhysRevA.85.023631}.  
This path integral approach allows a complete reorganization of the Feynman graphs of the theory in terms of the self-consistently obtained propagators for the original fields, and the leading order propagator for the auxiliary field \cite{r:Bender:1977bh}. 

Another topic we would like to elucidate in this paper is how to obtain the renormalization group flow of the coupling constants directly from the Effective potential.
When the original theory starts from a quartic self interaction of coupling  strength $\lambda$, the auxiliary fields that are utilized are quadratic in the original fields and the leading order self interaction gets replaced by a trilinear local interaction involving two of the original fields and the auxiliary field. 
The scattering in lowest order in the auxiliary field loop expansion proceeds by the intermediary of the composite field propagator. This propagator in lowest order is the geometric sum of an infinite number of bubble graph diagrams.  The second derivative of the effective potential with respect to the auxiliary field gives the value of the inverse of the composite field propagator.  Due to the constraint equation satisfied by the auxiliary field, one finds by studying the renormalized Green functions that this is a renormalization invariant.  This invariant, in leading order, is the inverse of the renormalized coupling constant \cite{PhysRevD.70.105008}. By choosing to define the  coupling constant at a particular value of the field, we will explicitly show how to derive the renormalization group equation for the renormalized coupling constant.  We will then compare our results for the renormalization group flow to those found using perturbation theory.

Although we concentrate on the static properties of the fields in this paper, the formalism presented here is perfectly well adapted to determining the time evolution of the field, averaged over noise configurations, as well as the noise induced correlation functions of the fields.  This approach has been used in the past to study the dynamics of phase transitions when there is chiral symmetry breaking \cite{r:CCMS01}, or phase separation in Bose-Einstein condensates \cite{r:Chien:2013bx}.
We also show how to use Auxiliary fields to produce approximations such as the bare vertex approximation (BVA) which is  related to the 2-PI-1/N approximation.

Many approaches to studying stochastic partial differential equations have been based on applying the dynamical renormalization group (RG) to the perturbative coupling constant expansion of the various noise induced one-particle irreducible graphs.  In applying LOAF one has to realize that in lowest order there is only the equivalent of  ``mass'' and dimensionful ``coupling constant'' renormalization.  Wave function and noise strength renormalization only occurs starting at next to leading order.  Several stochastic problems because of their directedness and simplicity do not have any wave function and noise strength renormalization.  This occurs in the chemical reaction diffusion annihilation problem $A + A \rightarrow 0$ and the related Cole-Hopf transformed KPZ equation.  In those cases one can determine from the effective potential in LOAF approximation an equation for the momentum dependent effective dimensionless reaction rate which compares well with the result of summing the geometric series of an infinite number of Feynman graphs that govern the elastic scattering process.  For more complicated problems where it is known that there are other renormalization effects, such as wave function and noise strength renormalization, one can either obtain these by going to next order in the loop expansion, or supplement the results obtained from the LOAF effective potential by the perturbative coupling constant results obtained for the running of these quantities.  These latter results are well known in the literature and have been used in discussions of the RG flow for the \emph{dimensionless} coupling constants.  See for example, the discussion of the KPZ effective action renormalization by Bork and Ogarkov \cite{r:Bork:2014yg} as well as the work of Zanella and Calzetta \cite{PhysRevE.66.036134}.  Here we will not repeat these perturbative calculations. 

The paper is organized as follows. 
In section~\ref{s:PIFormalism} we review the path integral approach to stochastic partial differential equations of the reaction-diffusion type when there is external noise whose statistics are known.  We give the generating functional for the connected noise-induced field correlation functions for both the OM formulation and the JD path integral representation of the MSR theory. 
In section~\ref{s:KPZ} we formulate the auxiliary field loop expansion method for the KPZ equation, and determine the effective action in the LOAF approximation for the KPZ equation.  We obtain the effective action in leading order in the LOAF expansion in both formulations and derive the OM effective potential from the JD-MSR effective potential.  
In section~\ref{s:DRrenorm}, we determine the renormalized effective potential in arbitrary spatial dimension $d$.  
In section~\ref{s:VeffSB} we show that in LOAF there is no fluctuation induced symmetry breakdown, in contrast to earlier work based on the Loop expansion as well as a calculation using the Hartree approximation. We also determine the running of the coupling constant in the LOAF approximation. 
In section~\ref{s:DR} we discuss the annihilation reaction diffusion process process $A+A \rightarrow 0$.  From the Doi-Peliti formalism one obtains a path integral of the MSR type which can be related to a Langevin process with imaginary multiplicative noise.  We then derive the effective action and effective potential in the LOAF approximation.  We compare our result for the running of the coupling constant with the known exact result of summing the infinite series of Feynman graphs.  
In section~\ref{s:ColeHopf} we obtain the effective potential for the Cole-Hopf transformed KPZ equation and relate that problem to the annihilation one.  We compare our result for the running of the dimensionless coupling constant with the exact result. 
In section~\ref{GL.s:GinsburyLandau} we discuss the Ginzburg-Landau model of spin relaxation and  determine the renormalized effective potential in the LOAF approximation. We compare our results for the running of the coupling constant with perturbative renormalization group results. 
In section~\ref{SchwingerDyson} we show how to obtain the Schwinger-Dyson equations that allow for approximation related to the 2-PI-1/N approximation.
In section~\ref{s:Conclude} we summarize our results.  In the appendices we include some Gaussian identities and functional relations used in this paper. 

%
%
\section{\label{s:PIFormalism}Path integral formalism for stochastic differential equations}

In this section we briefly review the path integral formulation for the field correlation functions induced by external noise that has gaussian correlations.  This formulation was originally developed by Onsager and Machlup \cite{r:Onsager:1953nr} and later by Graham \cite{r:Graham:1973ek} and  Zinn-Justin \cite{r:Zinn-Justin:1986lr}.  More recently it has been reviewed and elaborated on by Hochberg, Molina-Paris, Perez-Mercader and Visser \cite{r:PhysRevE.60.6343}.  The path integral version of the MSR formalism was developed by Jansen \cite{r:Janssen:1976xy} and de~Dominicus \cite{r:Dominicis:1976dq}, and is sometimes referred to as the Jansen-de~Dominicus action in the literature.  

A generic system of coupled stochastic equations for $\bphi = \Setc{\phi_1,\phi_2,\dotsc,\phi_n}$ species with external noise $\boldeta = \Setc{\eta_1,\eta_2,\dotsc,\eta_N}$ can be written in the form (here $\phi(x)$ is shorthand for $\phi(\vec x, t) $  ,

\begin{equation}\label{G.e:1}
   \bD_x \bphi(x)
   -
   F[ \, \bphi \, ]
   =
   \boldeta(x) \>,
   \qquad
   \bD_x
   =
   \partial_t -  \bnu \, \nabla^2 \>,
\end{equation}
In what follows  we will use  $x = (\vec{x} ,t)$ to represent coordinates \emph{and} time.
Here $\boldeta(x)$ is taken from the external gaussian distribution
\begin{equation}\label{G.e:2}
   P[ \, \boldeta \, ]
   =
   \calN \, 
   \ExpB{
      -
      \frac{1}{2} \iint \rd x \, \rd x' \,
      \boldeta(x) \cdot \bG_{\eta}^{-1}(x,x') \cdot \boldeta(x')
       } \>.
\end{equation}
with $\calN$ chosen so that  the path integration of $P[ \, \boldeta \, ]$  over the noise functions $ \eta_i(x)$  yields unity. That is 
\begin{equation}
   \int \calD \boldeta \, P[ \, \boldeta \, ] 
   \equiv
   \prod_i \int \calD  \eta_i
   = 
   1 \>,
\end{equation}
The path integral $\int \calD f(x)$ is defined in \ref{append-A} as the continuum limit of a product of ordinary integrals on a space-time lattice. 
Then
\begin{subequations}\label{G.e:3}
\begin{align}
   \Expect{ \eta_i(x) }
   &=
   \int \calD \boldeta \,\eta_i(x) P[ \, \boldeta \, ]
   =
   0 \>,
   \label{e:Expects-eta-a} \\
   \Expect{ \eta_i(x) \, \eta_j(x') }
   &=
   \int \calD \boldeta \,\eta_i(x) \, \eta_j(x') P[ \, \boldeta \, ]
   =
   G_{\eta\;i,j}(x,x') \>.
\end{align}
\end{subequations}
For the case of white noise, $G_{\eta\;i,j}(x,x') = \calA \, \delta_{i,j} \, \delta(x-x')$.
A generating functional for the fields $\bphi(x)$ is defined by
\begin{equation}\label{G.e:5}
   Z[\,\bj\,]
   =
   e^{W[\,\bj\,]}
   =
   \ExpectB{ \ExpB{ \int \rd x \> \bj(x) \cdot \bphi(x)} }
   =
   \int \calD \boldeta \, P[\,\boldeta\,] \, 
   \ExpB{ \int \rd x \> \bj(x) \cdot \bphi[\,\boldeta \,](x) } \>,
\end{equation}
where $\bphi[\, \boldeta \,](x)$ is a functional solution of Eq.~\eqref{G.e:1} (see, for example, Ref.~\cite{r:Bender:1983lq}). 
The fact that $P[\eta] $ is normalized to unity leads to the result that $Z[\, 0 \,] = 1$.
In what follows we well choose all our normalization constants to ensure this condition. 
%
%
\subsection{\label{ss:PI-OM}Onsager-Machlup form}
 
Changing variables in the path integral \eqref{G.e:5} from $\boldeta$ to $\bphi$ gives
\begin{equation}\label{G.e:6}
   Z[\,\bj\,]
   =
   e^{ W[\, \bj \, ] }
   =
   \calN^\prime \! \int \calD \bphi \, \calJ[ \, \bphi \, ] \,
   e^{ - S[\,\bphi;\bj \,] } \>,
\end{equation}
where now
\begin{align}
    S[\,\bphi;\bj \,]
    &=
    \frac{1}{2} \,
    \iint \rd x \, \rd x' \>
    \Bigl \{ \,
       \bigl [ \, \bD_x \bphi(x) - F[ \, \bphi \, ] \, \bigr ] 
       \cdot
       \bG_{\eta}^{-1}(x,x') \cdot
       \bigl [ \, \bD_{x'} \bphi(x') - F[ \, \bphi \, ] \, \bigr ] 
    \Bigr \}
    \label{G.e:7} \\
    & \hspace{5em}
    -
    \int \rd x \>
    \bj(x) \cdot \bphi(x) \>,
    \notag
\end{align}
and where $\calJ[ \, \bphi \, ]$ is the Jacobian determinant
\begin{equation}\label{G.e:8}
   \calJ[ \, \bphi \, ]
   \equiv
   \Big | \, \frac{\delta \boldeta}{\delta \bphi} \, \Big |
   =
   \DetB{ \bD_x - \frac{\delta F[ \, \bphi \, ]}{\delta \bphi} } \>.
\end{equation}
Here we choose $\calN'$ so that $Z[\, 0 \,]= 1$.  The equation for the noise averaged field $\Expect{\bphi(x)}$ is obtained from
\begin{equation}
   \Expect{\bphi(x)}
   =
   \int \calD \boldeta \, P[\, \boldeta \, ] \, \bphi[\,\boldeta\,](x)
   =
   \frac { \delta W[\, \bj \,]} {\delta \bj(x)} \Big |_{\bj=0} \>.
\end{equation}
Higher correlations functions are obtained by successive functional differentiation of this equation.  Since the normalization factors are independent of the sources
$\bj$, they do not enter into the determination of the correlation functions.  Since we have normalized $P[\, \boldeta \,]$ to unity, this implies that the normalization constants are chosen so that $Z[\, 0 \,] = 1$.
 
We call \eqref{G.e:7} the OM action, which is the form used in Ref.~\cite{r:Onsager:1953nr}.
The Jacobian determinant $\calJ[ \, \bphi \, ]$ can be replaced by a path integral over fermonic fields or ignored if we use an appropriate choice of the lattice version of the forward time derivative, which is the Ito regularization \cite{r:Cardy:1999fk,r:Kamenev:2011fk}.  We will assume Ito regularization in what follows. 

%
%
\subsection{\label{ss:PI-MSR}Martin-Siggia-Rose form}

In this section, we describe the Janssen-de~Dominicis  (JD) \cite{r:Janssen:1976xy,r:Dominicis:1976dq} path integral representation of the MSR formalism \cite{r:Martin:1973lr} for the classical generating functional.  We start with the functional identity discussed in detail in \ref{append-B}, 
\begin{align} \label{MRS.e:1} 
   1 
   &=
   \int \calD \boldeta \,
   \delta 
   \bigl [ \, \bD_x \bphi - F[\,\bphi\,] - \boldeta \, \bigr ]
\end{align}
Changing path integration variables from $\boldeta$ to $\bphi$ we then obtain 
   \begin{align}
   1 
   &=
   \int \calD \bphi \, \calJ[\, \bphi \, ] \,
   \delta 
   \bigl [ \, \bD_x \bphi - F[\,\bphi\,] - \boldeta \, \bigr ]
   \label{MRS.e:1.1} \\  
   &=
   \calN_1 \!
   \iint \calD \bphi \, \calD \sbphi \, \calJ[\, \bphi \, ] \,
   \ExpB{ \! - \!
      \int \! \rd x \, 
      \bigl \{ \,
         \sbphi(x) \cdot
         \bigl [ \, \bD_x \bphi(x) - F[\,\bphi(x)\,] + \boldeta(x) \, \bigr ] \!
         } \>,
   \notag 
\end{align}
where $\calJ[\, \bphi \, ]$ is the Jacobian defined in Eq.~\eqref{G.e:8} and the integration over the conjugate field $\sbphi$ runs along the imaginary axis and provides a generalization of the Fourier transform representation of the $\delta$ function \cite{r:Moshe:2003uq} as discussed in \ref{append-B}. Inserting this identity into Eq.~\eqref{G.e:5} for the generating functional, we find
\begin{align}
   Z[\,\bj\,]
   &=
   \int \calD \boldeta \, P[\,\boldeta\,] \, 
   \ExpB{ \int \rd x \> \bj(x) \cdot \bphi[\,\boldeta\,](x) }
   \label{MSR.e:2} \\
   &=
   \calN_1 \!
   \iiint 
   \calD \bphi \, 
   \calD \sbphi \, 
   \calD \boldeta \, \calJ[\, \bphi \, ] \, P[\,\boldeta\,]
   \notag \\
   & \hspace{4em}
   \times
   \ExpB{ 
      \int \rd x \, 
      \bigl \{ \,
         - \sbphi(x) \cdot
         \bigl [ \, \bD_x \bphi(x) - F[\,\bphi(x)\,] - \boldeta(x) \, \bigr ] 
         +
         \bj(x) \cdot \bphi(x) \,
      \bigr \} } \>.
   \notag
\end{align}
The integration over $\boldeta$ can now be done.  We find
\begin{align}
   &\int \calD \boldeta \,
   \ExpB{
      - 
      \frac{1}{2} \iint \rd x \, \rd x' \,
      \boldeta(x) \cdot \bG_{\eta}^{-1}(x,x') \cdot \boldeta(x')
      +
      \int \rd x \> \sbphi(x) \cdot \boldeta(x)
         }
   \label{MSR.e:3} \\
   & \hspace{4em}
   =
   \calN_2
   \ExpB{ 
      \frac{1}{2} 
      \iint \rd x \, \rd x' \>
      \sbphi(x) \cdot
      \bG_{\eta}(x,x') \cdot
      \sbphi(x') 
         } \>.
   \notag
\end{align}
Adding currents for the $\sbphi(x)$ field, Eq.~\eqref{MSR.e:2} becomes
\begin{equation}\label{MSR.e:4}
   Z[ \, \bj,\sbj \, ]
   =
   e^{ W[ \, \bj,\sbj \, ] } 
   =
   \calN_3\!
   \iint \calD \sbphi \, \calD \bphi \, 
   e^{ - S[ \, \bphi,\sbphi;\bj,\sbj \, ] } \>,
\end{equation}
where
\begin{align}
   &S[ \, \bphi,\sbphi;\bj,\sbj \, ]
   =
   -
   \frac{1}{2} 
   \iint \rd x \, \rd x' \>
   \sbphi(x) \cdot
   \bG_{\eta}(x,x') \cdot
   \sbphi(x') 
   \label{MSR.e:5} \\
   & \quad
   +
   \int \rd x \,
   \bigl \{ \,
      \sbphi(x) \cdot
      \bigl [ \, \bD_x \bphi(x) - F[ \, \bphi(x) \, ] \, \bigr ]
      -
      \sbj(x) \cdot \bphi(x)
      -
      \bj(x) \cdot \sbphi(x) \,
   \bigr \} \>.
   \notag
\end{align}
The final overall normalization constant is chosen so that $Z[\, 0, 0 \,] = 1$.
We call \eqref{MSR.e:5} the MSR or JD action, which is the form used for the KPZ equation in Ref.~\cite{r:Bork:2014yg}.
Here we have again assumed Ito regularization.  The normalization factor $\calN_3$ is independent of the sources and does not enter the calculation of the connected correlation functions.  In what follows we will use the generic $\calN$ for describing a normalization factor, since it does not enter into the dynamical equations. 

We notice that the MSR action is quadratic in the field $\sbphi$.  If we now set $\sbj = 0$ , and perform the Gaussian path integral over $\calD \sbphi$, we then recover the OM action.   What we will show below in specific examples is how to recover the approximate effective action which generates  the 1-PI graphs of the OM formalism from the effective action coming from the JD formalism. 
%
%
\subsection{\label{Auxfields}Auxiliary fields}

The key to simplifying the discussion of both the phase structure as well as the time evolution of the correlation functions is the introduction of auxiliary fields. 
The choice of auxiliary fields is not unique and one hope that a clever choice will represent a variable such as a composite field which is a bound state that
is important for understanding the phase structure and dynamics of the problem.  Auxiliary fields have been used for a long time to facilitate a calculation of the approximate phase structure of quantum field theories, and quantum many-body theories.  Once one has a path integral representation for the generating functional of the correlation function, such as given in Eqs.~\eqref{G.e:6} and \eqref{MSR.e:4}, a similar approach can be taken for stochastic PDE's that have an action of either the MO or MSR form.   For the purpose of obtaining an analytically calculable map of the phase structure as described by the effective potential, one introduces collective fields into the aforementioned action so that it is rendered quadratic in the original fields.  Once that is done, the Gaussian path integrals over the original fields can be performed (see \ref{append-A})  and the remaining path integrals over the auxiliary fields can be done by stationary phase, resulting in an expansion in terms of loops of composite field propagators around the leading order self-contently determined ``mean-field'' propagators (see BCG).   At leading order in the auxiliary field loop expansion (LOAF), one has found good qualitative agreement with experiment for the phase diagram of dilute Bose and Fermi gases as well as for the Bose-Hubbard model.  Although the effective potential for stochastic partial differential equations has been discussed in great detail by Hochberg and collaborators in the semi-classical loop expansion, and by Amaral and Roditi for the KPZ equation in a self-consistent Gaussian approximations, neither of these approximations gave a reliable picture of  Bose-Einstein condensation in the theory of dilute Bose gases, a LOAF approximation similar to the one we are presenting here gave a quite reasonable picture of the BEC phase diagram.  In what follows we will calculate the LOAF effective potential using both the OM and the JD actions for a variety of simple systems.  In particular we will show the absence of spontaneous symmetry breakdown for the KPZ effective potential.  We will explicitly show how to obtain the effective potential of the OM form from the effective potential of the JD form by determining the ``conjugate momentum,'' or constraint field, from the effective potential of the JD form.

We will also show how to obtain the renormalization group (RG) equations for the coupling constant flows directly from the effective potential, and compare this result with the perturbative approach for determining RG equations.  Another aspect of auxiliary fields is that when the original action has only trilinear quartic interactions, the introduction of auxiliary fields that are bilinear in the original fields then reduces all interactions to trilinear or bilinear.  In that case the exact equations for the noise averages field equations and correlation functions simplify in structure.  This allows one to make approximations to the dynamics similar to the direct-interaction approximation in plasma turbulence by Kraichnan \cite{r:Kraichnan:1959yq,r:Kraichnan:1989eu},
as well as in dynamical simulations of the KPZ equation by Doherty, \etal\ \cite{r:Doherty:1994qo},
and to what has been sometimes called the 2-PI-1/N expansion in quantum field theory.
The Schwinger-Dyson equations that govern this type of approximation are discussed in section~\ref{SchwingerDyson}.

%
\section{\label{s:KPZ}The KPZ equation}

For the KPZ equation describing ballistic surface growth,  $\bphi(x) \!\rightarrow\! \phi(\bx,t) \ge 0$, is a single field representing the height of a surface at a point $\bx$ on the surface at time $t$, and $F[ \phi ] = f_0 + \lambda \, | \bnabla \phi(x) |^2 / 2$.  

%
%
\subsection{\label{KPZ.ss:PI-OM}Onsager-Machlup form}

Assuming white noise, scaling the current by the amplitude of the noise $\calA$, and assuming Ito regularization, the OM version of the action, Eqs.~\eqref{G.e:6} and \eqref{G.e:7}, becomes
\begin{equation}\label{KPZ.OM.e:1}
   Z[\,j\,]
   =
   e^{ W[\, j \, ] / \calA }
   =
   \calN\! \int \calD \phi \, 
   e^{ - S[\,\phi;j \,] / \calA } \>,
\end{equation}
where now
\begin{equation}\label{KPZ.OM.e:2}
    S[\,\phi;j \,]
    =
    \int \rd x \>
    \bigl \{ \,
       | \, D_x \phi(x) - F[ \, \phi \, ] \, |^2 / 2
       -
       j(x) \, \phi(x) \,
    \bigr \} \>.
\end{equation}
We introduce the auxiliary fields $\sigma(x)$ and $\chi(x)$ into the path integral \eqref{KPZ.OM.e:1} using the identity
\begin{equation}\label{KPZ.OM.e:3}
   1
   =
   \int \calD \sigma \,
   \delta
   \bigl [ \, \sigma  -  \lambda F[ \, \phi \, ] \, \bigr ]
   =
   \calN \!
   \iint \calD \sigma \, \calD \chi \,
   \ExpB{
      -
      \int \rd x \,
      \frac{ \chi(x) }{ \lambda^2  \calA } \,
      \Bigl [ \, 
     \sigma(x)  -  \lambda  F[ \, \phi(x) \, ] \, 
      \Bigr ]
        } \>,
\end{equation}
where again integration over the $\chi$ field is along the imaginary axis.  
We have rescaled the conjugate momentum variable so as to make the trilinear coupling between the $\sigma(x)$ and $\phi(x)$ fields independent of $\lambda$, as done in Ref.~\cite{r:Coleman:1974ve}.  This scaling just changes the normalization of the path integral which does not enter into evaluation of the connected correlation functions. 
The generating functional \eqref{KPZ.OM.e:1} then becomes:
\begin{equation}\label{KPZ.OM.e:4}
   Z[\,j,s,r\,]
   =
   e^{ W[\,j,s,r\,] /\calA }
   =
   \calN\! \iiint \calD \phi \, \calD \sigma \calD \chi \, 
   e^{ - S[\,\phi,\sigma,\chi;j,s,r \,] /\calA } \>,
\end{equation}
with
\begin{align}
   S[\,\phi,\sigma,\chi;j,s,r \,]
   &=
   \int \! \rd x \,
   \Bigl \{ \,
      \frac{1}{2} \, [ \,  D_x \, \phi(x) - \sigma(x) / \lambda \, ]^2
      + 
      \frac{ \chi(x) }{ \lambda^2 } \,
      \bigl [ \, \sigma(x)  - \lambda  F[ \, \phi(x) \, ] \, \bigr ]
      \label{KPZ.OM.e:5} \\
      & \qquad\qquad
      -
      j(x) \, \phi(x)
      -
      s(x) \, \sigma(x)
      -
      r(x) \, \chi(x) \,
   \Bigr \} \>,
   \notag   
\end{align}
where we have added currents $s(x)$ and $r(x)$ for the auxiliary fields.  
By parts integration, we can write \eqref{KPZ.OM.e:5} in the form,
\begin{align}
   &S[\,\phi,\sigma,\chi;j,s,r \,]
   =
   \frac{1}{2} 
   \iint \! \rd x \, \rd x' \, 
   \phi(x) \, G^{-1}[\chi](x,x') \, \phi(x')
   \label{KPZ.OM.e:6} \\
   & 
   +
   \int \! \rd x \,
   \Bigl \{ \,
      \frac{ \sigma^2(x) + 2 \, \chi(x) \, \sigma(x)}{2 \lambda^2}
      -
      [\, j(x) - \sD_x \sigma(x) / \lambda \,] \, \phi(x)
      -
      s(x) \, \sigma(x)
      -
      [\,  r(x) + f_0 / \lambda \,] \, \chi(x) \,
   \Bigr \} \>,
   \notag
\end{align}
with
\begin{equation}\label{KPZ.OM.e:7}
   G^{-1}[\chi](x,x')
   =
   \delta(x-x') \,
   \bigl \{ \,
      \sD_x \spD_x
      -
      \LAbnabla \chi(x) \cdot \RAbnabla
   \bigr \} \>,
\end{equation}
and $\sD_x = - \partial_t - \nu \nabla^2$.  
Here we use the convention:
\begin{equation}
   A \LAnabla  B = (\nabla A) B \>,
   \Qquad{and}  
   A \RAnabla B = A (\nabla B) \>.
\end{equation}
Note that once we rewrite the action in terms of the auxiliary fields, the coefficient of $\sigma^2(x)$ is $1/ ( 2\lambda^2)$.  So the renormalized coupling constant can be determined from the second derivative of the effective action with respect to $\sigma(x)$, evaluated at zero momentum \cite{r:Coleman:1974ve}.
 
It is convenient at this point to set
\begin{equation}\label{KPZ.OM.e:9}
   X(x)
   =
   \begin{pmatrix}
      \sigma(x) \\ \chi(x)
   \end{pmatrix} \>,
   \Qquad{and}
   I
   =
   \begin{pmatrix}
      1 & 1 \\ 1 & 0
   \end{pmatrix} \>,
\end{equation}
and put
\begin{gather}
   K(x)
   =
   K_0(x) + K_1(x)
   =
   \begin{pmatrix}
      s(x) \\ r(x)
   \end{pmatrix}
   +
   \begin{pmatrix}
      0 \\ f_0 / \lambda
   \end{pmatrix} \>,
   \label{KPZ.OM.e:8} \\
   J[X](x)
   =
   J_0(x)
   +
   J_1(x)
   =
   j(x) - \sD_x \, \sigma(x) / \lambda \>,
   \notag \\
   \sigma^2(x) + 2 \, \chi(x) \, \sigma(x) 
   = 
   X^{T}(x) \, I \, X(x) \>,
   \notag
\end{gather}
so that Eq.~\eqref{KPZ.OM.e:6} becomes
\begin{align}
   S[\,\phi,X;J,K \,]
   &=
   \frac{1}{2} 
   \iint \rd x \, \rd x' \, 
   \phi(x) \, G^{-1}[X](x,x') \, \phi(x')
   \label{KPZ.OM.e:10} \\
   & \hspace{2em}
   +
   \int \rd x \,
   \Bigl \{ \,
      \frac{ X^{T}(x) \, I \, X(x) }{ 2 \lambda^2 }
      -
      J[X](x) \, \phi(x)
      -
      K^T(x) \, X(x) \,
   \Bigr \} \>.
   \notag
\end{align}
The action is now quadratic in the field $\phi$ and can be integrated over $\phi$.  The generating functional \eqref{KPZ.OM.e:4} then becomes
\begin{equation}\label{KPZ.OM.e:11}
   Z[\,J,K\,]
   =
   e^{ W[\,J,K\,] / \calA }
   =
   \calN\! \int \calD X \, 
   e^{ - S_{\text{eff}}[\,X;J,K \,] / (\epsilon \calA) } \>,
\end{equation}
where now
\begin{align}
   &S_{\text{eff}}[\,X;J,K \,]
   =
   -
   \frac{1}{2} 
   \iint \rd x \, \rd x' \, 
   J[X](x) \, G[X](x,x') \, J[X](x')
   \label{KPZ.OM.e:12} \\
   & \quad
   +
   \int \rd x \,
   \Bigl \{ \,
      \frac{ X^{T}(x) \, I \, X(x) }{ 2 \lambda^2 }
      -
      K^T(x) \, X(x)
      +
      \frac{\calA}{2} \,
      \Tr{ \Ln{ G^{-1}[X](x,x) } } \, 
   \Bigr \} \>.
   \notag
\end{align}
Here $\epsilon$, which is eventually set to unity, is used to count the order of the LOAF expansion.  
The remaining integral is done by expanding the effective action \eqref{KPZ.OM.e:12} about the saddle point $X_0(x)$, defined by the equations,
\begin{equation}\label{KPZ.OM.e:13}
   \frac{\delta S_{\text{eff}}[\,X;J,K \,] }{ \delta X_i(x) } \Big |_{X_0} = 0 \>,
\end{equation}
which yields the saddle point equations,
\begin{subequations}\label{KPZ.OM.e:14}
\begin{align}
   &\frac{ \sigma_0(x) + \chi_0(x) }{ \lambda^2 }
   =
   -
   \spD_x \phi_0[X_0](x) / \lambda
   +
   s(x) \>,
   \label{KPZ.OM.e:14-a} \\
   &\frac{ \sigma_0(x) - f_0 }{ \lambda^2 }
   =
   \frac{1}{2} \,
   \Bigl \{ \,
      | \, \bnabla  \phi_0[X_0](x) \, |^2
      +
      \calA \,
      \bigl [ \,
         \bnabla \cdot \bnabla' \, G[X_0](x,x') \, 
      \bigr ]_{x=x'} \,
   \Bigr \}
   +
   r(x) \>,
   \label{KPZ.OM.e:14-b}
\end{align}
\end{subequations}
where we have defined
\begin{equation}\label{KPZ.OM.e:15}
   \phi_0[X_0](x)
   =
   \int \rd x' \, G[X_0](x,x') \, J[X_0](x') \>.
\end{equation}
Expanding the effective action about the saddle point, 
\begin{align}
   S_{\text{eff}}[\, X;J,K \,] 
   &=
   S_{\text{eff}}[\, X_0;J,K \,]
   \label{KPZ.OM.e:16} \\
   & \quad
   +
   \frac{1}{2} 
   \iint \rd x \, \rd x' \,
   D^{-1}_{ij}[\,J,K \,](x,x') \,
   ( X_i(x) - X_{0\,i}(x) ) \, ( X_j(x') - X_{0\,j}(x') )
   +
   \dotsb \>,
   \notag
\end{align}
where
\begin{equation}\label{KPZ.OM.e:17} 
   D^{-1}_{ij}[\,J,K\,](x,x')
   =
   \frac{ \delta^2 S_{\text{eff}}[\, X;J,K \,] }
        { \delta X_i(x) \, \delta X_j(x') } \Big |_{X_0}
   =
   I_{ij} \, \delta(x-x')
   +
   \Sigma_{ij}[J,K](x,x') \>,
\end{equation}
and carrying out the remaining path integral gives
\begin{align}
   &W[\, X_0,J,K \,]
   =
   -
   S_{\text{eff}}[\, X_0;J,K \,]
   -
   \frac{\epsilon \calA}{2} \,
   \int \rd x \,
   \Tr{ \Ln{ D^{-1}[\,J,K\,](x,x) } }
   +
   \dotsb
   \notag \\
   & \hspace{3em} 
   =
   \frac{1}{2} 
   \iint \rd x \, \rd x' \,
   J[\,X_0\,](x) \, G[\,X_0\,](x,x') \, J[\,X_0\,](x')
   \label{KPZ.OM.e:18} \\
   & \hspace{4em}
   -
   \int \rd x \, 
   \Bigl \{ \,
      \frac{ X_0^T(x) \, I \, X_0(x) }{ 2 \lambda^2 }
      -
      K^T(x) \, X_0(x)
      +
      \frac{\calA}{2} \, \Tr{ \Ln{ G^{-1}[\,X_0\,](x,x) } }
      \notag \\
      & \hspace{5em}
      +
      \frac{\epsilon \calA}{2} \,
      \Tr{ \Ln{ D^{-1}[\,J,K\,](x,x) } } \,      
   \Bigr \}
   +
   \dotsb \>.
   \notag
\end{align}
Note that this is an expansion in powers of $\epsilon$, not $\calA$ as in the usual loop expansion.  The advantages of this auxiliary field expansion is discussed in detail in Refs.~\cite{r:Bender:1977bh,r:Cooper:2011ix}.
To first order in $\epsilon$, the fields are then given by the expansion:
\begin{subequations}\label{KPZ.OM.e:19}
\begin{align}
   \phi(x)
   &=
   \frac{\delta W[\, J,K \,] }{ \delta J(x) }
   =
   \phi_0[X_0](x)
   +
   \epsilon \, \phi_{1}[X_0](x)
   +
   \dotsb
   \label{KPZ.OM.e:19a} \\
   X_i(x)
   &=
   \frac{\delta W[\, J,K \,] }{ \delta K_{i}(x) }
   =
   X_{0\,i}(x)
   +
   \epsilon \, X_{1\,i}(x)
   \dotsb \>,
   \label{KPZ.OM.e:19b}
\end{align}
\end{subequations}
where $\phi_{0}[X_0](x)$ is given by Eq.~\eqref{KPZ.OM.e:15}, and
\begin{subequations}\label{KPZ.OM.e:20}
\begin{align}
   &\phi_{1}[X_0](x)
   =
   \frac{\calA}{2} \!\!
   \iint \! \rd x_1 \, \rd x_2 \, 
   \TrB{ D[\,J,K\,](x_1,x_2) \, 
         \frac{\delta D^{-1}[\,J,K\,](x_2,x_1)}{\delta J(x)} } \,,
   \label{KPZ.OM.e:20a} \\
   &X_{1\,i}(x)
   =
   \frac{\calA}{2} \!\!
   \iint \! \rd x_1 \, \rd x_2 \, 
   \TrB{ D[\,J,K\,](x_1,x_2) \, 
         \frac{\delta D^{-1}[\,J,K\,](x_2,x_1)}{\delta K_{i}(x)} } \>.
   \label{KPZ.OM.e:20b}
\end{align}
\end{subequations}
The effective action $\Gamma[ \, \Phi,X  \, ]$ is defined by the functional Legendre transformation (see for example Ref.~\cite{r:Rivers:1990yf}),
\begin{align}
   \Gamma[ \, \phi,X  \, ]
   &=
   \int \rd x \, 
   \{ \, 
      J_0(x) \, \phi(x)
      +
      K^T_0(x) \, X(x) \,
   \}
   -
   W[ \, J,K \, ]
   \label{KPZ.OM.e:21} \\
   &
   =
   \int \rd x \, 
   \{ \, 
      [ \, J(x) - J_1(x) \, ] \, \phi(x)
      +
      [ \, K^T(x) - K_{1}^T(x) \, ] \, X(x) \,
   \}
   -
   W[ \, J,K \, ] \>,
   \notag   
\end{align}
where $J_1(x)$ and $K_{1}(x)$ are defined in Eq.~\eqref{KPZ.OM.e:8}.
From \eqref{KPZ.OM.e:19}, to first order in $\epsilon$ we can replace $\phi_{0}[X_0](x)$ by $\phi(x)$ and $X_{0}(x)$ by $X(x)$ in the expression \eqref{KPZ.OM.e:18} for $W[\,J,K\,]$, which gives an effective action
\begin{align}
   &\Gamma[ \, \phi,X  \, ]
   =
   \frac{1}{2} 
   \iint \rd x \, \rd x' \,
   \phi(x) \, G^{-1}[\,X\,](x,x') \, \phi(x')
   \label{KPZ.OM.e:22} \\
   & \quad
   +
   \int \rd x \, 
   \Bigl \{ \,
      \frac{ X^T(x) \, I \, X(x) }{ 2 \lambda^2 }
      + 
      \sigma(x) \, D \, \phi(x) / \lambda
      -
      f_0 \, \chi(x) / \lambda
      +
      \frac{\calA}{2} \, 
      \Tr{ \Ln{ G^{-1}[\, X \,](x,x) } }  \,
   \Bigr \}
   +
   \text{O}[ \, \epsilon \, ] \>.
   \notag
\end{align}
The effective potential is the natural generalization of the classical potential when there are stochastic or quantum fluctuations present.  It is defined usually for static fields by evaluating the effective action for these fields and then dividing by the space-time volume.   For some fields, such as the vector potential in electrodynamics, and here the KPZ field $\phi$, only the curl or gradient of the field contributes to the energy, and one must look at the potential as a function of the derivatives of the fields entering the Lagrangian. That is the case for the KPZ equation. 
To find the effective potential for the KPZ equation, we take as an ansatz for the fields:
\begin{equation}\label{KPZ.OM.e:23}
   \phi(x)
   =
   - \bx \cdot \bv \>,
   \Qquad{and}
   X(x)
   =
   \begin{pmatrix}
      \sigma(x) \\
      \chi(x)
   \end{pmatrix}
   =
   \begin{pmatrix}
      \sigma \\
      \chi
   \end{pmatrix} \>,
\end{equation}
where $( \, \bv, \sigma, \chi \, )$ are all constants, independent of space and time.  The leading order effective potential from Eq.~\eqref{KPZ.OM.e:22} is then given by
\begin{equation}\label{KPZ.OM.e:23.1}
   V_{\text{eff}}[ \, v,\sigma,\chi \, ]
   =
   \frac{ \Gamma[\,  v,\sigma,\chi \,] }{ \Omega }
   =
   \frac{ \sigma^2 }{2 \lambda^2}
   +
   \frac{ \chi }{ \lambda } \, 
   \Bigl [ \, \frac{\sigma}{ \lambda } - f_0 - \lambda \frac{v^2}{2} \, \Bigr ]
   +
   \frac{\calA}{2} \, \Tr{ \Ln{ G^{-1}[\,\chi\,](x,x) } } \>,
\end{equation}
where $\Omega$ is the space-time volume.  Expanding the Green function $G^{-1}[\,\chi\,](x,x')$ in a Fourier-Laplace series,
\begin{equation}\label{KPZ.OM.e:23.2}
   G^{-1}[\,\chi\,](x,x')
   =
   \int \! \frac{\rd^d k}{(2\pi)^d} \int \! \frac{\rd z}{2\pi i} \,
   \tilde{G}^{-1}[\,\chi\,](\bk,\omega) \,
   e^{ i \, \bk \cdot (\bx - \bx') + z (t - t') } \>,
\end{equation}
where
\begin{equation}\label{KPZ.OM.e:24}
   \tilde{G}^{-1}[\,\chi\,](\bk,z)
   =
   [ \, \nu k^2 - z \, ] \,
   [ \, \nu k^2 + z \, ]
   -
   \lambda \, \chi \, k^2
   =
   \omega_k^2[\chi] - z^2 \>,
\end{equation}
and where we have put $\omega_k^2[\chi] = \nu^2 \, k^2 \, ( \, k^2 - \chi / \nu^2 \, )$.
So we find
\begin{equation}\label{KPZ.OM.e:26}
   \Tr{ \Ln{ G^{-1}[\,\chi\,](x,x) } }
   =
   \int \! \frac{\rd^d k}{(2\pi)^d} \int \! \frac{\rd z}{2\pi i} \,
   \Ln{ \omega_k^2[\chi] - z^2 }
   =
   \int \! \frac{\rd^d k}{(2\pi)^d} \, 
   \bigl \{ \, | \, \omega_k[\chi] \, | + C_{\infty} \, \bigr \} \>,
\end{equation}
where $C_{\infty}$ is an infinite constant which is absorbed into the overall effective potential normalization.  Inserting this result into Eq.~\eqref{KPZ.OM.e:23} gives
\begin{equation}\label{KPZ.OM.e:27}
   V_{\text{eff}}[ \, v,\sigma,\chi \, ]
   =
   \frac{ \sigma^2 }{2 \lambda^2}
   +
   \frac{ \chi }{ \lambda } \, 
   \Bigl [ \, \frac{ \sigma }{ \lambda } - f_0 - \lambda \frac{v^2}{2} \, \Bigr ]
   +
   \frac{\nu \, \calA}{2} \,
   \int \! \frac{\rd^d k}{(2\pi)^d} \, 
   |\, k \,| \, \sqrt{ k^2 - \chi / \nu^2 } \>.
\end{equation}
The gap equations are now
\begin{subequations}\label{KPZ.OM.e:28}
\begin{align}
   \frac{ \partial V_{\text{eff}}[ \, v,\sigma,\chi \, ] }{ \partial \sigma }
   &=
   \frac{ \sigma + \chi }{ \lambda }
   = 
   0 \>,
   \label{KPZ.OM.e:28a} \\
   \frac{ \partial V_{\text{eff}}[ \, v,\sigma,\chi \, ] }{ \partial \chi }
   &=
   \frac{1}{\lambda}
   \Bigl [ \, \frac{\sigma}{\lambda} - f_0 - \lambda \frac{v^2}{2} \, \Bigr ]
   -
   \frac{\calA}{4 \, \nu } \,
   \int \! \frac{\rd^d k}{(2\pi)^d} \, 
   \frac{ |\, k \,| }{ \sqrt{ k^2 - \chi / \nu^2 } }
   =
   0 \>.
   \label{KPZ.OM.e:28b}
\end{align}
\end{subequations}
So from \eqref{KPZ.OM.e:28a} setting $\chi = -\sigma$, the effective potential Eq.~\eqref{KPZ.OM.e:27} reads
\begin{equation}\label{KPZ.OM.e:29}
   V_{\text{eff}}[ \, v,\sigma \, ]
   =
   -
   \frac{\sigma^2}{2 \lambda^2}
   +
   \frac{\sigma}{\lambda} \, 
   \Bigl [ \, f_0 + \lambda \frac{v^2}{2} \, \Bigr ]
   +
   \frac{\nu \, \calA}{2} \,
   \int \! \frac{\rd^d k}{(2\pi)^d} \, 
   |\, k \,| \, \sqrt{ k^2 + \sigma / \nu^2 } \>. 
\end{equation}
The gap equation \eqref{KPZ.OM.e:28b} becomes
\begin{equation}\label{KPZ.OM.e:30}
   \frac{\sigma}{\lambda^2}
   =
   \frac{1}{\lambda} \,
   \Bigl [ \, f_0 + \lambda \frac{v^2}{2} \, \Bigr ] 
   +
   \frac{\calA}{4 \, \nu } \,
   \int \! \frac{\rd^d k}{(2\pi)^d} \, 
   \frac{ |\, k \,| }{ \sqrt{ k^2 + \sigma / \nu^2 } } \>.
\end{equation}
The integrals in \eqref{KPZ.OM.e:29} and \eqref{KPZ.OM.e:30} diverge.  They are made finite by renormalization methods, which will be discussed in Section~\ref{s:DRrenorm}.

%
%
\subsection{\label{KPZ.ss:PI-MSR}Martin-Siggia-Rose form}

Again assuming white noise, scaling the star field $\sphi(x)$ and the star current $\sj(x)$ by the amplitude of the noise $\calA$, the MSR version of the action, Eqs.~\eqref{MSR.e:4} and \eqref{MSR.e:5}, becomes
\begin{equation}\label{KPZ.MSR.e:1}
   Z[ \, j,\sj \, ]
   =
   e^{ W[ \, j,\sj \, ] / \calA } 
   =
   \calN\!
   \iint \calD \sphi \, \calD \phi \, 
   e^{ - S[ \, \phi,\sphi;j,\sj \, ] / \calA } \>,
\end{equation}
where
\begin{align}
   S[ \, \phi,\sphi;j,\sj \, ]
   &
   =
   \int \rd x \>
   \bigl \{ \,
      -
      [ \, \sphi(x) \, ]^2 / 2
      +
      \sphi(x) \,
      \bigl [ \, D_x \phi(x) - F[ \, \phi(x) \, ] \, \bigr ] 
      \label{KPZ.MSR.e:2} \\
      & \qquad\qquad\qquad
      -
      \sj(x) \, \phi(x)
      -
      j(x) \, \sphi(x) \,
   \bigr \} \>.
   \notag
\end{align}
Again introducing auxiliary fields $\sigma(x)$ and $\chi(x)$ by means of Eq.~\eqref{KPZ.OM.e:3}, the generating functional \eqref{KPZ.MSR.e:1} becomes
\begin{equation}\label{KPZ.MSR.e:3}
   Z[ \, j,\sj \, ]
   =
   e^{ W[ \, j,\sj \, ] / \calA } 
   =
   \calN\!
   \iiiint 
   \calD \sphi \, \calD \phi \, \calD \sigma \, \calD \chi \>
   e^{ - S[ \, \phi,\sphi,\sigma,\chi;j,\sj,r \, ] / \calA } \>,
\end{equation}
where now
\begin{align}
   &S[\,\phi,\sphi,\sigma,\chi;j,\sj,r \,]
   =
   \int \!\rd x \,
   \Bigl \{ \,
      \frac{1}{2} \,
      \bigl \{ \,
         \sphi(x) \, \spD_x \, \phi(x) + \phi(x) \, \sD_x \, \sphi(x)
         -
         [ \, \sphi(x) \, ]^2
         \notag \\
         & \hspace{5em}
         -
         [ \bnabla \phi(x) ] \, \chi(x) \cdot [ \bnabla \phi(x) ] \,
      \bigr \}
      +
      \chi(x) \,  \sigma(x) / \lambda^2
      \label{KPZ.MSR.e:4}  \\
      & \quad
      -
      \sj(x) \, \phi(x)
      -
      [ \, j(x) + \sigma(x) / \lambda \, ] \, \sphi(x)
      -
      s(x) \, \sigma(x)
      -
      [ \, r(x) + f_0/\lambda \, ] \, \chi(x) \,
   \Bigr \} \>.
   \notag
\end{align}
Let us first note that by setting $\sj(x)$ and $r(x)$ to zero and integrating $\sphi$ along the imaginary axis, integrating over $\chi$ and then $\sigma$, reproduces the action \eqref{KPZ.OM.e:2} we used in the OM formalism.  So it must be possible to obtain the approximate effective action for the OM formalism from that of the 
JD formalism. We will explicitly show how to do this for the KPZ problem. 
Introducing a two-component notation with the definitions:
\begin{align*}
   \Phi(x)
   &=
   \begin{pmatrix}
      \phi(x) \\ \sphi(x)
   \end{pmatrix} ,
   &
   J[\,X\,](x)
   &=
   \begin{pmatrix}
      j(x) \\ \sj(x) + \sigma(x)/\lambda
   \end{pmatrix} ,
   \\
   X(x)
   &=
   \begin{pmatrix}
      \sigma(x) \\ \chi(x)
   \end{pmatrix} ,
   &
   K(x)
   &=
   \begin{pmatrix}
      s(x) \\ r(x) + f_0/\lambda
   \end{pmatrix} .
\end{align*}
Then the generating functional \eqref{KPZ.MSR.e:1} can be written as
\begin{equation}\label{KPZ.MSR.e:6}
   Z[\,J,K\,]
   =
   \calN \iint \calD \Phi \, \calD X \,
   e^{ - S[\,\Phi,X;J,K \,] /\calA } \>,
\end{equation}
and the action \eqref{KPZ.MSR.e:2} in this notation becomes
\begin{align}
   S[ \, \Phi,X;J,K \, ]
   &=
   \frac{1}{2} 
   \iint \rd x \, \rd x' \,
   \sPhi(x) \, G^{-1}[\,X\,](x,x') \, \Phi(x')
   \label{KPZ.MSR.e:7} \\
   & \qquad
   +
   \int \rd x \, 
   \Bigl \{ \,
      \frac{ \sX(x) \, X(x) }{ 2 \lambda^2 }
      -
      \sJ[\,X\,](x) \, \Phi(x)
      -
      \sK(x) \, X(x) \,
   \Bigr \} \>,
   \notag
\end{align}
where
\begin{equation}\label{KPZ.MSR.e:7.1}
   G^{-1}[\,X\,](x,x')
   =
   \delta(x-x') \, 
   \begin{pmatrix}
      \sD_x & -1 \\
      - \LAbnabla \, \chi(x) \cdot \RAbnabla \>, & 
      \spD_x \\
   \end{pmatrix} \>.
\end{equation}
Here $\sPhi(x) = \Set{ \sphi(x), \phi(x) }$ and $\sX(x) = \Set{ \chi(x),\sigma(x) }$ with corresponding definitions for $\sJ[\,X\,](x)$ and $\sK(x)$.
The action is now quadratic in the fields $\Phi$ and can be integrated out.
The generating functional \eqref{KPZ.MSR.e:4} then becomes
\begin{equation}\label{KPZ.MSR.e:9}
   Z[\,J,K\,]
   =
   e^{ W[\,J,K\,] / \calA }
   =
   \calN\! \int\! \calD X \, 
   e^{ - S_{\text{eff}}[\,X;J,K \,] / (\epsilon \calA) } \>,
\end{equation}
where now
\begin{align}
   S_{\text{eff}}[\,X;J,K \,]
   &=
   -
   \frac{1}{2} 
   \iint \rd x \, \rd x' \, 
   \sJ[X](x) \, G[X](x,x') \, J[X](x')
   \label{KPZ.MSR.e:10} \\
   & \quad
   +
   \int \rd x \,
   \Bigl \{ \,
      \frac{ \sX(x) \, X(x) }{ 2 \lambda^2 }
      -
      \sK(x) \, X(x)
      +
      \frac{\calA}{2} \,
      \Tr{ \Ln{ G^{-1}[X](x,x) } } \, 
   \Bigr \} \>,
   \notag
\end{align}
and we have again introduced $\epsilon$ to count orders of approximation.  
The remaining integral is done by expanding the effective action \eqref{KPZ.MSR.e:7} about the stationary point $X_0(x)$, defined by the equation,
\begin{equation}\label{KPZ.MSR.e:11}
   \frac{\delta S_{\text{eff}}[\,X;J,K \,] }{ \delta X_i(x) } \Big |_{X_0} = 0 \>.
\end{equation}
The saddle point equations in this case become
\begin{subequations}\label{KPZ.MSR.e:12}
\begin{align}
   \chi_0(x)
   &=
   \sphi_0[X_0](x)
   +
   s(x) \>,
   \label{KPZ.MSR.e:12-a} \\
   \sigma_0(x)
   &=
   f_0
   +
   \frac{\lambda}{2} \,
   \Bigl \{ \,
      | \, \bnabla  \phi_0[X_0](x) \, |^2
      +
      \calA \,
      \bigl [ \,
         \bnabla \cdot \bnabla' \, G_{11}[X_0](x,x') \, 
      \bigr ]_{x=x'} \,
   \Bigr \}
   +
   r(x) \>,
   \label{KPZ.MSR.e:12-b}
\end{align}
\end{subequations}
where
\begin{equation}\label{KPZ.MSR.e:13}
   \Phi_0[X_0](x)
   =
   \int \rd x' \, G[X_0](x,x') \, J[X_0](x') \>.
\end{equation}
Expanding the effective action about the saddle points and 
carrying out the remaining path integral gives
\begin{align}
   W[\, X_0,J,K \,]
   &=
   -
   S_{\text{eff}}[\, X_0;J,K \,]
   -
   \frac{\epsilon \calA}{2} \,
   \int \rd x \,
   \Tr{ \Ln{ D^{-1}[\,J,K\,](x,x) } }
   +
   \dotsb
   \notag \\
   &=
   \frac{1}{2} 
   \iint \rd x \, \rd x' \,
   J[\,X_0\,](x) \, G[\,X_0\,](x,x') \, J[\,X_0\,](x')
   \label{KPZ.MSR.e:14} \\
   & \hspace{1em}
   -
   \int \rd x \, 
   \bigl \{ \,
      \frac{ X_0^T(x) \, I \, X_0(x) }{ 2 \lambda^2 }
      -
      K^T(x) \, X_0(x)
      +
      \frac{\calA}{2} \, \Tr{ \Ln{ G^{-1}[\,X_0\,] } }
      \notag \\
      & \hspace{6em}
      +
      \frac{\epsilon \calA}{2} \,
      \Tr{ \Ln{ D^{-1}[\,J,K\,](x,x) } } \,      
   \bigr \} \>,
   \notag
\end{align}
where
\begin{equation} \label{KPZ.MSR.e:15}
   D^{-1}_{ij}[\,J,K\,](x,x')
   =
   \frac{ \delta^2 S_{\text{eff}}[\, X;J,K \,] }
        { \delta X_i(x) \, \delta X_j(x') } \Big |_{X_0}
   =
   \delta_{ij} \, \delta(x-x')
   +
   \Sigma_{ij}[J,K](x,x') \>.
\end{equation}
The effective action $\Gamma[ \, \Phi,X  \, ]$ is defined by the Legendre transformation 
\begin{align}
   \Gamma[ \, \Phi,X  \, ]
   &=
   \int \! \rd x  
   [ \, 
      \sJ_0(x) \, \Phi(x)
      +
      \sK_0(x) \, X(x) \,
   ]
   -
   W[ \, J,K \, ]
   \label{KPZ.MSR.e:16} \\
   &
   =
   \int \! \rd x \, 
   \{ \, 
      [ \, \sJ(x) - \sJ_1(x) \, ] \, \Phi(x)
      +
      [ \, \sK(x) - \sK_{1}(x) \, ] \, X(x) \,
   \}
   -
   W[ \, J,K \, ] \>.
   \notag
\end{align}
To first order in $\epsilon$ we can replace $\Phi_{0}[X_0](x)$ by $\Phi(x)$ and $X_{0}(x)$ by $X(x)$ in the expression \eqref{KPZ.MSR.e:11} for $W[\,J,K\,]$, which gives an effective action
\begin{align}
   &\Gamma[ \, \Phi,X  \, ]
   =
   \frac{1}{2} 
   \iint \rd x \, \rd x' \,
   \sPhi(x) \, G^{-1}[\,X\,](x,x') \, \Phi(x')
   \label{KPZ.MSR.e:17} \\
   &
   +
   \int \rd x \, 
   \Bigl \{ \,
      \frac{ \sX(x) \, X(x) }{ 2 \lambda^2 }
     - 
      \sigma(x) \, \sphi(x) / \lambda
      -
      f_0 \, \chi(x) / \lambda
      +
      \frac{\calA}{2} \, 
      \Tr{ \Ln{ G^{-1}[\, X \,](x,x) } }  \,
   \Bigr \}
   +
   \text{O}[ \, \epsilon \, ] \>,
   \notag
\end{align}
To find the effective potential, we take as an ansatz for the fields:
\begin{equation}\label{KPZ.MSR.e:18}
   \Phi(x)
   =
   \begin{pmatrix}
      \phi(x) \\ \sphi(x)
   \end{pmatrix}
   =
   \begin{pmatrix}
      - \bx \cdot \bv \\ \sphi
   \end{pmatrix} \>,
   \Qquad{and}
   X(x)
   =
   \begin{pmatrix}
      \sigma(x) \\
      \chi(x)
   \end{pmatrix}
   =
   \begin{pmatrix}
      \sigma \\
      \chi
   \end{pmatrix} \,,
\end{equation}
where $( \, \bv, \sphi, \sigma, \chi \, )$ are all constants.
The leading order effective potential from Eq.~\eqref{KPZ.MSR.e:12} is given by
\begin{equation}\label{KPZ.MSR.e:19}
   V_{\text{eff}}[ \, \Phi,X \, ]
   =
   \frac{\Gamma[ \, \Phi,X  \, ]}{\Omega}
   =
   -
   \frac{ [ \, \sphi \, ]^2 }{2}
   -
   \frac{v^2}{2} \, \chi
   + 
   \frac{\sigma \, \chi }{ \lambda^2 }
   -
   \frac{ f_0 \, \chi }{ \lambda }
   -
   \frac{ \sigma \, \sphi }{ \lambda }
   +
   \frac{\calA}{2} \, \Tr{ \Ln{ G^{-1}[\,\chi\,](x,x) } } \>,
\end{equation}
where $\Omega$ is the space-time volume.  Expanding the Green function in a Fourier-Laplace series, as in Eq.~\eqref{KPZ.OM.e:23.2}, we find
\begin{equation}\label{KPZ.MSR.e:21}
   \tilde{G}^{-1}[\,\chi\,](\bk,z)
   =
   \begin{pmatrix}
      \nu k^2 - z & -1 \\
      - \chi \, k^2 & \nu k^2 + z
   \end{pmatrix} \>.
\end{equation}
Then
\begin{equation}\label{KPZ.MSR.e:21.1}
   \Det{ \tilde{G}^{-1}[\,\chi\,](\bk,z) }
   =
   \omega^2_{k}[\,\chi\,] - z^2 \>,
\end{equation}
where
\begin{equation}\label{KPZ.MSR.e:22}
   \omega^2_{k}[\,\chi\,]
   =
   \nu^2 \, k^4 - \chi \, k^2
   =
   \nu^2 \, k^2 \, ( \, k^2 - \chi / \nu^2 \, ) \>.
\end{equation}
So then
\begin{align}
   \Tr{ \Ln{ G^{-1}[\,\chi\,](x,x) } }
   &=
   \int \! \frac{\rd^d k}{(2\pi)^d} \int \! \frac{\rd z}{2\pi i} \,
   \Ln{ \Det{ \tilde{G}^{-1}[\,\chi\,](\bk,z) } }
   \label{KPZ.MSR.e:22.1} \\
   &
   =
   \int \! \frac{\rd^d k}{(2\pi)^d} \int \! \frac{\rd z}{2\pi i} \,
   \Ln{ z^2 - \omega^2_{k}[\,\chi\,] }
   =
   \int \! \frac{\rd^d k}{(2\pi)^d} \,
   \bigl \{ \, | \, \omega_{k}[\,\chi\,] \, | + C_{\infty} \, \bigr \} \>,
   \notag   
\end{align}
where $C_{\infty}$ is an infinite constant which is absorbed into the overall effective potential normalization.  Inserting this result into Eq.~\eqref{KPZ.MSR.e:13} gives
\begin{equation}\label{KPZ.MSR.e:23}
   V_{\text{eff}}[ \, v,\sphi,\sigma,\chi \, ]
   =
   -
   \frac{ [ \, \sphi \, ]^2 }{2}
   -
   \frac{v^2}{2} \, \chi
   + 
   \frac{ \sigma \, \chi }{ \lambda^2 }
   -
   \frac{ f_0 \, \chi }{ \lambda }
   -
   \frac{ \sigma \, \sphi }{ \lambda }
   +
   \frac{\nu \, \calA}{2} \,
   \int \! \frac{\rd^d k}{(2\pi)^d} \, 
   |\, k \,| \, \sqrt{ k^2 - \frac{\chi }{ \nu^2 } } \>.
\end{equation}
The gap equations are now
\begin{subequations}\label{KPZ.MSR.e:24}
\begin{align}
   \frac{ \partial V_{\text{eff}}[ \, v,\sphi,\sigma,\chi \, ] }{ \partial \sigma }
   &=
   \frac{ \chi - \lambda \sphi }{ \lambda^2 }
   = 
   0 \>,
   \label{KPZ.MSR.e:24-a} \\
   \frac{ \partial V_{\text{eff}}[ \, v,\sphi,\sigma,\chi \, ] }{ \partial \chi }
   &=
   \frac{ \sigma }{ \lambda^2 }
   -
   \frac{ f_0 }{ \lambda }
   -
   \frac{v^2}{2}
   -
   \frac{\calA}{4 \, \nu } \,
   \int \! \frac{\rd^d k}{(2\pi)^d} \, 
   \frac{ |\, k \,| }{ \sqrt{ k^2 - \chi / \nu^2 } }
   =
   0 \>.
   \label{KPZ.MSR.e:24-b}  \\
   \frac{ \partial V_{\text{eff}}[ \, v,\sphi,\sigma,\chi \, ] }{ \partial \sphi }
   &=
   -
   \sphi
   -
   \sigma / \lambda
   =
   0 \>.
   \label{KPZ.MSR.e:24-c} 
\end{align}
\end{subequations}
From Eqs.~\eqref{KPZ.MSR.e:24-a} and \eqref{KPZ.MSR.e:24-c} we find that $\chi = \lambda \sphi = -\sigma$.  Eliminating $\sphi$ and $\chi$ from $V_{\text{eff}}$, the effective potential \eqref{KPZ.MSR.e:23} becomes,
\begin{equation}\label{KPZ.MSR.e:25}
   V_{\text{eff}}[ \, v,\sigma \, ]
   =
   - \frac{ \sigma^2 }{2 \lambda^2}
   +
   \frac{ \sigma }{ \lambda } \, 
   \Bigl [ \, f_0 + \lambda \, \frac{v^2}{2} \, \Bigr ]
   +
   \frac{\nu \, \calA}{2} \,
   \int \! \frac{\rd^d k}{(2\pi)^d} \, 
   |\, k \,| \, \sqrt{ k^2 + \sigma / \nu^2 } \>,
\end{equation}
with the gap equation,
\begin{equation}\label{KPZ.MSR.e:26}
   \frac{\sigma}{\lambda^2}
   =
   \frac{1}{\lambda} \,
   \Bigl [ \, f_0 + \lambda \, \frac{v^2}{2} \, \Bigr ]
   +
   \frac{\calA}{4 \, \nu } \,
   \int \! \frac{\rd^d k}{(2\pi)^d} \, 
   \frac{ |\, k \,| }{ \sqrt{ k^2 + \sigma / \nu^2 } } \>.
\end{equation}
Eqs.~\eqref{KPZ.MSR.e:25} and \eqref{KPZ.MSR.e:26} agree with the OM results in Eqs.~\eqref{KPZ.OM.e:29} and \eqref{KPZ.OM.e:30}.  So the two methods give the same effective potential in leading order in the auxiliary field expansions.

%
%
\section{\label{s:DRrenorm}Renormalization using dimensional regularization}

In this section, we renormalize the KPZ equation following the renormalization procedure of Coleman, Jackiw, and Politzer \cite{r:Coleman:1974ve}.
The second derivative of the effective potential, Eq.~\eqref{KPZ.MSR.e:25}, with respect to the auxiliary field $\sigma$ is given by
\begin{equation}\label{RE.e:1}
   - \frac{\partial^2 V_{\text{eff}}[ \, \sigma,d \, ] }{ \partial \sigma^2 }
   =
   \frac{1}{\lambda^2}
   +
   \frac{\calA}{8 \nu^3} \,
   \Bigl ( \frac{\sigma}{\nu^2} \Bigr )^{\! (d-2)/2} \,
   I[\, d \, ] \>,
\end{equation}
where
\begin{equation}\label{RE.e:2}
   I[\, d \, ]
   =
   \frac{\Omega^d}{(2\pi)^3} 
   \int_{0}^{\infty} \frac{ t^{(d-1)/2} \, \rd t }{ (t + 1)^{3/2} }
   =
   \frac{8}{ (4 \pi )^{(d+1)/2} } \,
   \frac{ \Gamma[ \, (d + 1)/2 \, ] \, \Gamma[ \, 1 - d/2 \, ] }
        { \Gamma[ \, d/2 \, ] } \>.
\end{equation}
Now the effective action $\Gamma$ is the generator of 2-PI vertices and $- \partial^2 V_{\text{eff}} / \partial \sigma^2$ is the inverse propagator of the composite $\sigma$-field.  So with our definitions, $- \partial^2 V_{\text{eff}}[ \, \mu^2, d \, ] / \partial \sigma^2$, evaluated at $\sigma/\nu^2 = \mu^2$, is the inverse of the square of the renormalized coupling constant,
\begin{equation}\label{RE.e:3}
   \frac{1}{ \lambda^2_r[ \, \mu^2, d \, ] }
   =
   \frac{1}{\lambda^2}
   +
   \frac{\calA}{8 \, \nu^3} \,
   \mu^{\! d-2} \,
   I[\, d \, ] \>.
\end{equation}
We can use \eqref{RE.e:3} to compare coupling constants at different scales,
\begin{equation}\label{RE.e:4}
   \frac{1}{ \lambda^2_r[ \, \mu^2, d \, ] }
   =
   \frac{1}{ \lambda^2_r[ \, \mu^2_0, d \, ] }
   +
   \frac{\calA}{8 \, \nu^3} \,
   \bigl [ \,
      \mu^{\! d-2} - \mu_0^{\! d-2} \,
   \bigr ] \,
   I[\, d \, ] \>.
\end{equation}
or equivalently define the $\beta$-function,
\begin{equation}\label{RE.e:5}
   \beta_{\lambda}[d]
   \equiv
   \mu \frac{ \rd \lambda_r[\mu,d] }{\rd \mu}
   =
   \frac{\calA \, \lambda_r^3[\mu]}{ 8 \nu^3 } \, J[d] \, \mu^{d-2} \>,
\end{equation}
where $J[d] = (d-2) I[d]$.  Here $J[d]$ has no singularities at $d=2$.  In particular in two dimensions, we obtain
\begin{equation}\label{RE.e:6}
    \beta_{\lambda}[2]
    =
    \frac{\calA \, \lambda_r^3[\mu]}{ 4 \pi \nu^3 } \>,
\end{equation}
in agreement with the one-loop answer of Hochberg, \etal\ \cite{r:Hochberg:2000jk}.  However unlike Hochberg, we are able here to calculate $\beta_{\lambda}[d]$ for all dimensions $d$. 

The full renormalization group equations are obtained by studying the flow of the dimensionless renormalized coupling constant, which is defined as 
\begin{equation}\label{RE.e:7}
   g_r[\mu^2] 
   = 
   \frac{ \calA_r[\, \mu \,] \, \lambda_r^2[\, \mu \,] }
        { 8 \, \nu^3[\, \mu \,] \, \mu^{d-2} }  \>.
\end{equation}
Here $\calA_r[\, \mu \,]$ and $\nu_r[\, \mu \,]$ also depend on a renormalization scale.  It is well known that the leading order auxiliary field loop expansion only has ``mass'' and coupling constant renormalization.  Thus if we want the full renormalization group approach we need to supplement the calculation presented here with a calculation of the renormalization of $\nu$ and $\calA$ at next order in the $\epsilon$ expansion.  
An example of this is given in Haymaker and Cooper \cite{r:Haymaker:1979fk}.  
Elsewhere in the literature, the running of these quantities has been done as a separate calculation.  
In the work of Bork and Ogarkov \cite{r:Bork:2014yg}, who study the effective potential in the loop expansion, these further renormalizations are evaluated in perturbation theory from Feynman graphs.  
In another perturbative approach to calculating the effective action by Zanella and Calzetta \cite{PhysRevE.66.036134}, a dynamical renormalization group approach to the perturbative effective action was used.  Since these calculation have already been done in the literature, we will not repeat them here.
Instead we will side-step this approach in Section~\ref{s:ColeHopf} by applying the auxiliary field loop expansion to the Cole-Hopf transformed effective action for the KPZ equation.  
In the Cole-Hopf form of the action, only the new coupling constant gets renormalized and one can calculate the full $\beta$-function.

Substituting \eqref{RE.e:4} into \eqref{RE.e:1}, the second derivative in terms of the renormalized coupling constant is now given by
\begin{equation}\label{RE.e:8}
   - \frac{\partial^2 V_{\text{eff}}[ \, \mu,\sigma,d \, ] }{ \partial \sigma^2 }
   =
   \frac{1}{ \lambda^2_r[ \, \mu^2, d \, ] }
   + 
   \frac{\calA}{8 \, \nu^3} \,
   \Bigl \{ \,
      \Bigl ( \frac{\sigma}{\nu^2} \Bigr )^{\! (d-2)/2}
      \!\!\!
      -
      \mu^{\! d-2} \,
   \Bigr \} \,
   I[\, d \, ] \>.
\end{equation}
Integrating this equation twice, and choosing the integration constants so that classical part of the effective potential is the renormalized classical potential for the KPZ equation, we obtain
\begin{equation}\label{RE.e:9}
   V_{\text{eff}}[ \, \mu,\sigma,d \, ]
   =
   -
   \frac{ \sigma^2 }{ 2 \lambda^2_r[ \, \mu^2, d \, ] }
   +
   \frac{ \sigma \, v^2 }{2}
   +
   \frac{f_r \, \sigma}{\lambda}
   -
   \frac{\nu \calA}{8} \,
   \Bigl [ \,
      \frac{4}{ d ( d + 2 ) }
      \Bigl ( \frac{ \sigma }{ \nu^2 } \Bigr )^{d/2+1} \!\!\!
      -
      \frac{ \mu^{d - 2} }{2} \,
      \Bigl ( \frac{\sigma }{ \nu^2 } \Bigr )^2 \,
   \Bigr ] \,
   I[\, d \, ] \>,
\end{equation}
with the gap equation,
\begin{equation}\label{RE.e:10}
   \frac{ \sigma }{\lambda^2_r[ \, \mu^2, d \, ] }
   =
   \frac{ v^2 }{2}
   +
   \frac{f_r}{\lambda}
   -
   \frac{\calA}{8 \nu} \,
   \Bigl [ \,
      \frac{2}{d}
      \Bigl ( \frac{ \sigma }{ \nu^2 } \Bigr )^{d/2} \!\!\!
      -
      \mu^{d - 2}\,
      \Bigl ( \frac{\sigma }{ \nu^2 } \Bigr ) \,
   \Bigr ] \,
   I[\, d \, ] \>.   
\end{equation}
For the massless KPZ equation, we would set $f_r = 0$.  The massless problem is what we will consider in what follows in order to revisit the problem of dynamical symmetry breaking discussed in Refs.~\cite{r:Hochberg:2000jk,r:Amaral:2007ty}.
This effective potential cannot depend on the renormalization point $\mu$.  
Using this fact by taking the derivative of the effective potential with respect to $\mu$ and setting it equal to zero, we could have determined the renormalization group equation for $\beta_{\lambda}$, Eq.~\eqref{RE.e:5}, this alternate way, as was done in Ref.~\cite{r:Hochberg:2000jk}.

In one and three dimension, 
\begin{equation}\label{RE.e:11}
   \frac{ \lambda^2_r[ \, \mu^2,d \, ] }{ \lambda^2 }
   =
   \begin{cases}
      \displaystyle
      \frac{1}{ 1 + \alpha / ( \pi \, \mu ) } \>,
      &
      \text{for $d = 1$,} 
      \\[10pt]
      \displaystyle
      \frac{1}{ 1 - \alpha \, \mu / \pi^2 } \>,
      &
      \text{for $d = 3$.} 
   \end{cases}
\end{equation}
where $\alpha = \calA \, \lambda^2 / ( 8 \nu^3 )$.  So for $d=1$, choosing $\mu = \infty$ sets $\lambda_r[ \, \infty,1 \, ] = \lambda$, whereas for $d=3$, choosing $\mu = 0$ also sets $\lambda_r[ \, 0,3 \, ] = \lambda$.  In two dimensions, we have to take the limit of Eq.~\eqref{RE.e:9} as $d \rightarrow 2$.  For $d=1$, we find for the effective potential and gap equations
\begin{subequations}\label{RE.e:12}
\begin{align}    V_{\text{eff}}[ \, v,\sigma \, ]
    &=
    - 
    \frac{\sigma^2}{2 \, \lambda^2}
    +
    \sigma \, \frac{v^2}{2}
    -
    \frac{\calA}{6 \pi \, \nu^2} \, \sigma^{3/2} \>,
    \label{RE.e:12-a} \\
    \frac{\sigma}{\lambda^2}
    &=
    \frac{v^2}{2}
    -
    \frac{\calA}{4 \pi \, \nu^2} \, \sigma^{1/2} \>.
    \label{RE.e:12-b}
\end{align}
\end{subequations}
For $d=2$, we get
\begin{subequations}\label{RE.e:13}
\begin{align}
    V_{\text{eff}}[ \, v,\mu,\sigma \, ]
    &=
    - 
    \frac{\sigma^2}{2 \, \lambda_r^2[\mu^2]}
    +
    \sigma \, \frac{v^2}{2}
    +
    \frac{\calA}{64 \pi \, \nu^3} \, \sigma^2 \,
    \Bigl \{ \,
       \LnB{ \frac{ \sigma }{ \nu^2 \mu^2 } }
       -
       \frac{3}{2} \,
    \Bigr \} \>,
    \label{RE.e:13-a} \\
    \frac{\sigma}{\lambda_r^2[\mu^2]}
    &=
    \frac{v^2}{2}
    +
    \frac{\calA}{32 \pi \, \nu^3} \, \sigma \,
    \Bigl \{ \,
       \LnB{ \frac{ \sigma }{ \nu^2 \mu^2 } }
       -
       1 \,
    \Bigr \} \>.
    \label{RE.e:13-b}
\end{align}
\end{subequations}
For $d=3$, we get
\begin{subequations}\label{RE.e:14}
\begin{align}
    V_{\text{eff}}[ \, v,\sigma \, ]
    &=
    - 
    \frac{\sigma^2}{2 \, \lambda^2}
    +
    \sigma \, \frac{v^2}{2}
    +
    \frac{\calA}{30 \pi^2 \, \nu^4} \, \sigma^{5/2} \>.
    \label{RE.e:14-a} \\
    \frac{\sigma}{\lambda^2}
    &=
    \frac{v^2}{2}
    +
    \frac{\calA}{12 \pi^2 \, \nu^4} \, \sigma^{3/2} \>.
    \label{RE.e:14-b}
\end{align}
\end{subequations}
Eqs.~\eqref{RE.e:12-b}, \eqref{RE.e:13-b}, and \eqref{RE.e:14-b} are easily inverted to find $\sigma$ as a function of $v$, which are substituted into the effective potentials to find $V_{\text{eff}}[v]$.  
As a check, we renormalized the KPZ equation using standard momentum cut-off methods, which agreed with the results we found in this section.   

%
%
\section{\label{s:VeffSB}Effective potential and conclusions about symmetry breaking}

One of the major reasons for using the LOAF approximation to study the KPZ equation is that it was known to cure some of the defects of both the one-loop approximation as well as the Hartree approximation when they were used to study symmetry breaking in the problem of dilute Bose gases.  Since we are considering the massless KPZ equation we want to maintain the masslessness of the theory after renormalization.  This is similar to making sure we are obtaining a massless particle (a Goldstone boson) which is present in the BEC theory when we include fluctuations.  In earlier treatments of the effective potential at a one loop level, perurbative in powers of $\calA$, or in a self-consistent gaussian approximation, related to the Hartree approximation, it was found that as a function of the coupling constant there could be a phase transition into the phase where $\Expect{v} \neq 0$.   Since it is well known that gaussian approximations violate Ward identities and that a one-loop calculation in the BEC problem did not allow one to explore the BEC phase transition, we thought it important to see whether an approximation that was non-perturbative in $\calA$ and which preserved Ward identities would lead to a different conclusion.  The reason the LOAF approximation is non-perturbative in $\calA$ is due to the non-linear nature of the gap equation which brings in all powers of $\calA$.  Interestingly, when we reexpand our result in powers of $\calA$, we recover the one-loop result which predicts the phase transition.  It is the non-analytic nature of the potential as a function of $\calA$ which shows that one cannot trust the one-loop result even at small $\calA$.

The condition for $V_{\text{eff}}[ \, v,\sigma,d \, ]$ to be a minimum is that
\begin{equation}\label{RE.e:15}
   \frac{ \rd V_{\text{eff}}[ \, v,\sigma,d \, ] }{\rd v}
   =
   \frac{\partial V_{\text{eff}}[ \, v,\sigma,d \, ] }{ \partial v } 
   \Big |_{\sigma}
   +
   \frac{ \partial \sigma[v]}{\partial v } \,
   \frac{\partial V_{\text{eff}}[ \, v,\sigma,d \, ] }{ \partial \sigma } 
   \Big |_{v}
   =
   \frac{\partial V_{\text{eff}}[ \, v,\sigma,d \, ] }{ \partial v } 
   \Big |_{\sigma}
   =
   \sigma \, v \>,  
\end{equation}
since the second term vanishes by the gap equation.  Therefore in the LOAF approximation, the minimum condition is when
\begin{equation}\label{RE.e:16}
   \frac{ \rd V_{\text{eff}}[ \, v,\sigma,d \, ] }{\rd \sigma}
   =
   \sigma \, v 
   =
   0 \>.
\end{equation}
Thus in order for $v \neq 0$, one requires that $\sigma=0$. 
But from the \emph{massless} gap equation \eqref{RE.e:10}, if $\sigma = 0$, then $v=0$. Therefore there can be no broken symmetry solution ($v \neq 0$) in the LOAF approximation.  However, for the \emph{massive} case when $f_r \ne 0$, non-zero solutions for $v$ are possible.  Our explicit calculation of $V_{\text{eff}}$ will illustrate this theorem.  Our result does not preclude symmetry breakdown in higher order in the auxiliary field loop expansion.

%
%
\subsection{\label{s:Veffd1}Effective potential, one dimension}

\begin{figure}[t]
   \centering
   \subfigure[$d=1$]
   { \label{f:Veff-KPZ-d1}
     \includegraphics[width=0.45\columnwidth]{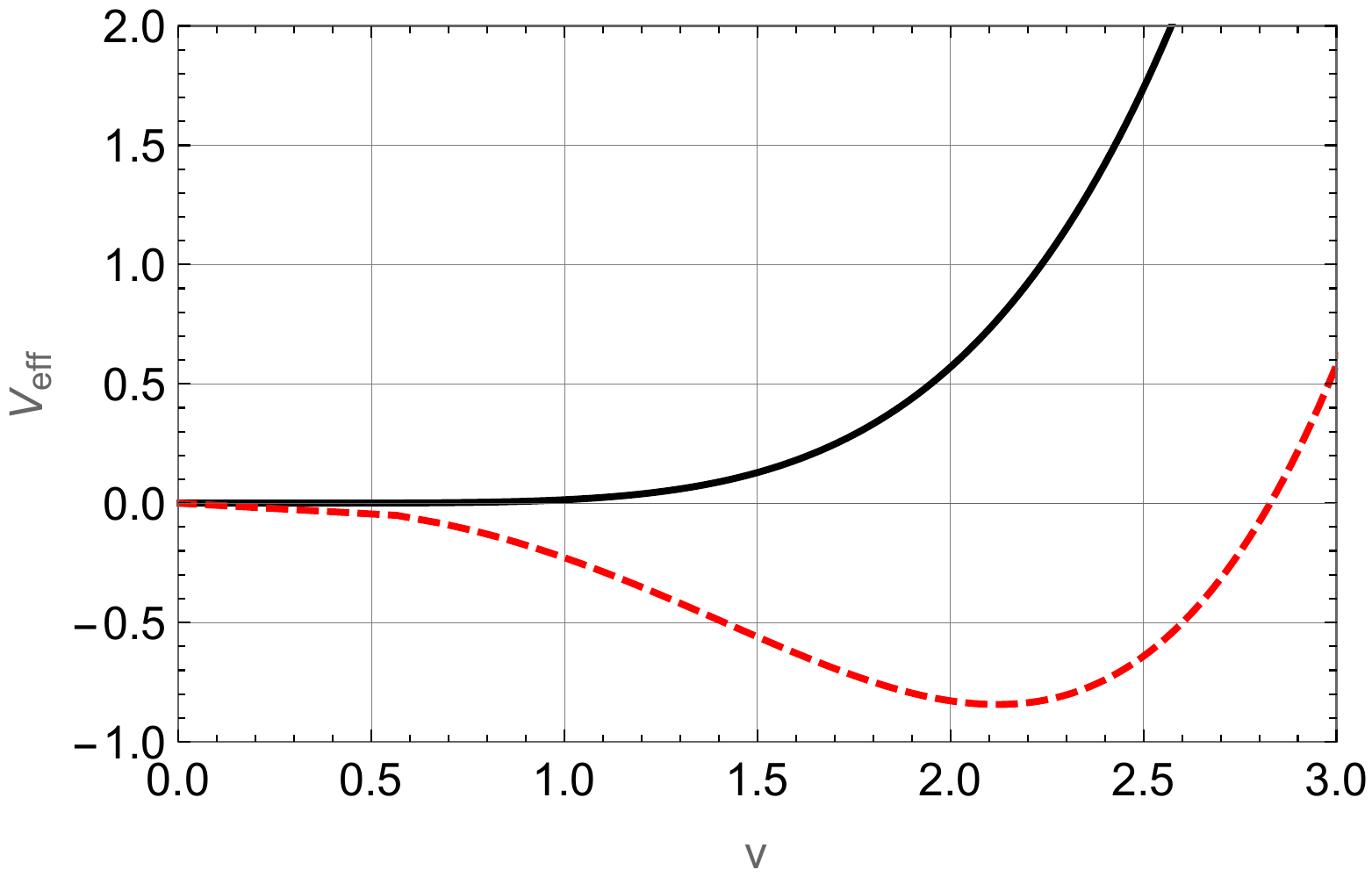} }
   \subfigure[$d=2$]
   { \label{f:Veff-KPZ-d2}
     \includegraphics[width=0.45\columnwidth]{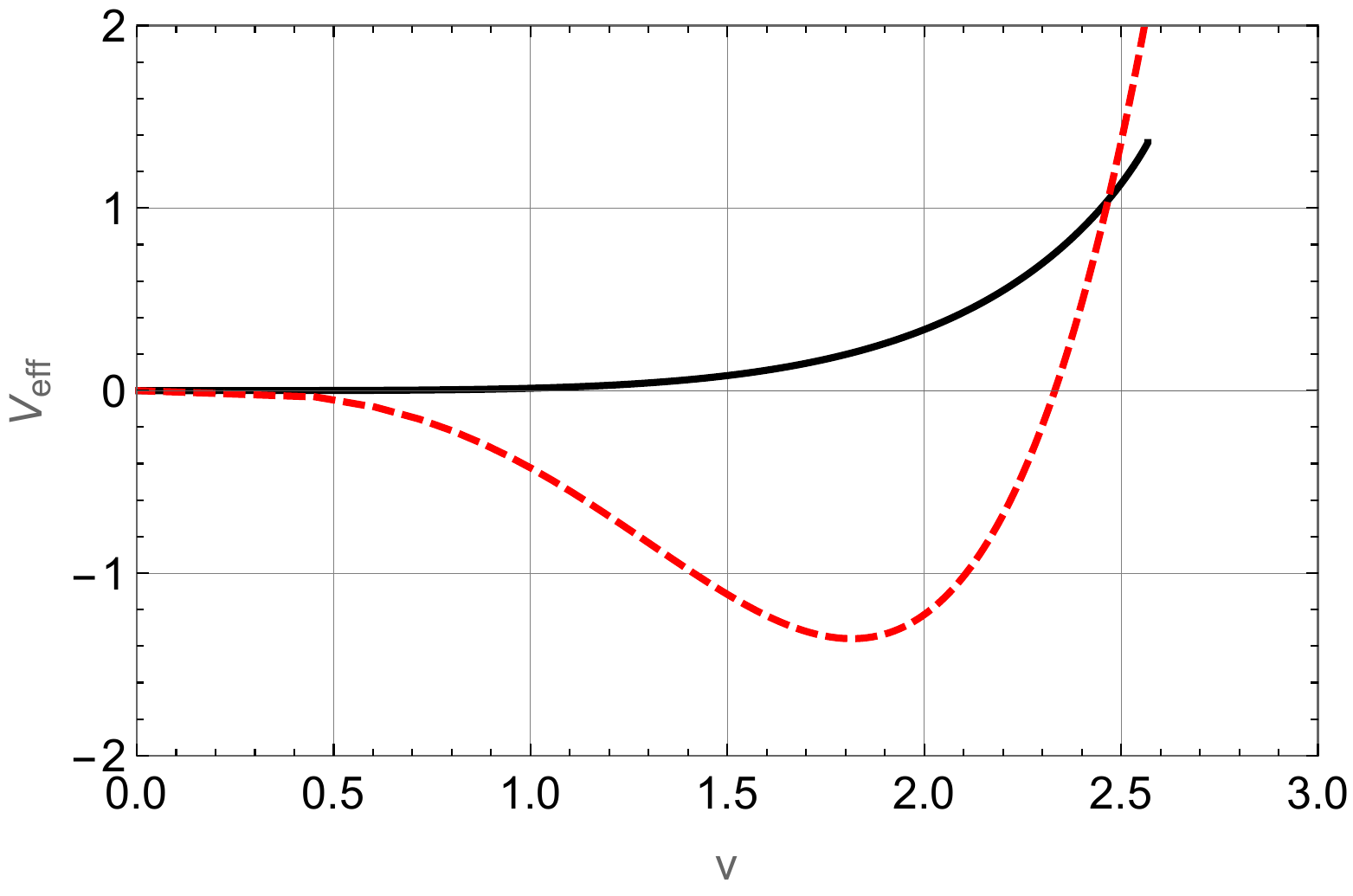} }
   \subfigure[$d=3$]
   { \label{f:Veff-KPZ-d3}
     \includegraphics[width=0.45\columnwidth]{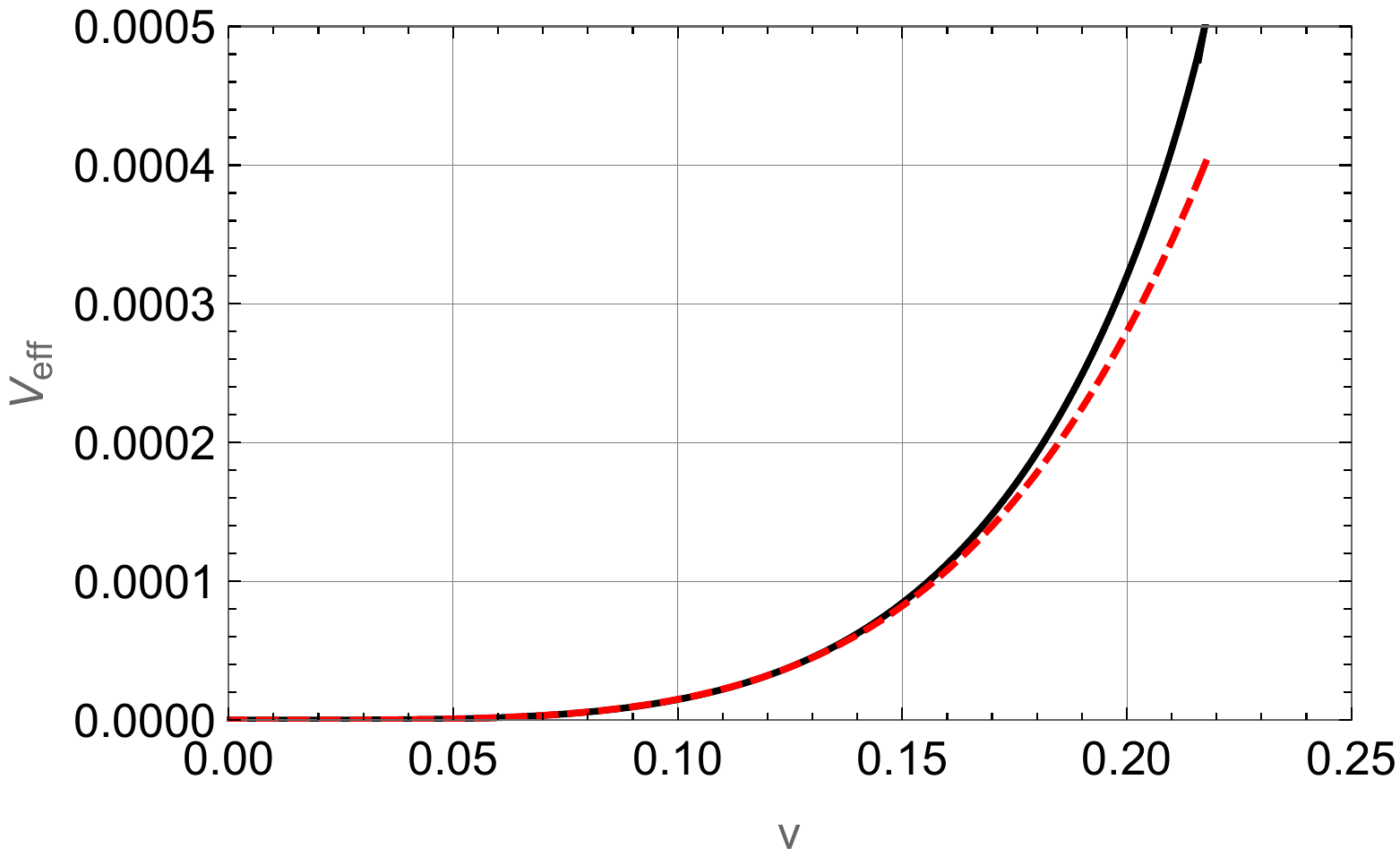} }
   \caption{\label{f:Veff-KPZ}Plots of $V_{\text{eff}}$ \vs\ $v$ for
   (a) $d=1$, where $\calA = 6\pi$, (b) $d=2$, where $\calA = 64\pi$
   with $\mu = 1$, and (c) $d=3$, where $\calA = 30\pi^2$, 
   for the case when $\lambda = \nu = 1$.  Solid black curves are the
   LOAF results whereas the dashed (red online) curves are the loop
   expansion results.} 
\end{figure}

To determine the effective potential for $d=1$ we have to solve the gap equation \eqref{RE.e:12-b} for $\sigma$ as a function of $v$.  This can be done explicitly in one dimension,
\begin{equation}\label{RE.e:17}
   \sigma^{1/2}
   =
   \sqrt{ \lambda^2 \, v^2 / 2 + b^2 / 4} - b/2 \>,
   \qquad
   b
   =
   \frac{\calA \lambda^2}{4 \pi \nu^2 } \>.
\end{equation}
This is substituted into \eqref{RE.e:12-a} to obtain $V_{\text{eff}}[v]$, which is shown in Fig.~\ref{f:Veff-KPZ-d1}.  If we re-expand this result in a series of $\calA$, we obtain the result of the loop expansion in Ref.~\cite{r:Hochberg:2000jk}, namely
\begin{equation}\label{RE.e:18}
   V_{\text{eff}}^{\text{one-loop}}[v]
   =
   \lambda^2 \, \frac{v^4}{8}
   -
   \frac{\calA}{6 \pi \nu^2} \, \Bigl ( \frac{ \lambda^2 v^2 }{2} \Bigr )^{3/2} \>.
\end{equation}
Eq.~\eqref{RE.e:18} has a double well construction for all $\lambda$, but $V_{\text{eff}}$ for the LOAF approximation, as a result of the theorem displayed in Eq.~\eqref{RE.e:16}, does not for any $\lambda$.
For the one-loop approximation, the minimum of the potential occurs at
\begin{equation}
v= \pm \frac{\lambda \calA}{2 \sqrt{2} \pi \nu^2}
\end{equation} 

%
%
%
%

%
%
\subsection{\label{s:Veffd2}Effective potential, two dimensions}

For $d=2$, the effective potential is given in Eq.~\eqref{RE.e:13-a}.  Solving \eqref{RE.e:13-b} for $v$ as a function of $\sigma$, this can be written as
\begin{equation}\label{RE.e:19}
   V_{\text{eff}}[\, \sigma \,]
   =
   \frac{\sigma^2}{2 \, \lambda_r^2[\mu]}
   -
   \frac{\calA \, \sigma^2}{32 \pi \, \nu^3} \,
   \Bigl \{ \,
      \LnB{ \frac{\sigma}{\mu^2 \nu^2} }
      -
      \frac{1}{2} \,
   \Bigr \} \>.
\end{equation}
When we expand $V_{\text{eff}}[ v ]$ is a power series in $\calA$, we obtain the one-loop result,
\begin{equation}\label{RE.e:20}
   V_{\text{eff}}^{\text{one-loop}}[v]
   =
   \frac{\lambda^2 v^4}{8}
   +
   \frac{ \calA \lambda^4 v^4 }{ 256 \pi \, \nu^3 } \,
   \Bigl \{ \,
      \LnB{ \frac{\lambda^2 v^2}{2 \mu^2 \nu^2} }
      -
      \frac{3}{2} \,
   \Bigr \} \>.
\end{equation}
We notice that the leading term in the correction to the classical answer has opposite signs for the LOAF approximation and the first term in a re-expansion in $\calA$.  Thus even small corrections of order $\calA^2$ as evidenced here can change the character of the answer.  The loop expansion for the effective potential leads to a double well structure for all $\lambda$ \cite{r:Hochberg:2000jk}, whereas the Hartree approximation leads to a double well structure for values of $\lambda$ greater than a critical value \cite{r:Amaral:2007ty}.
On the other hand, the LOAF approximation does not lead to double well structure as shown in Fig.~\ref{f:Veff-KPZ-d2}.

%
%
\subsection{\label{s:Veffd3}Effective potential, three dimensions}

For $d=3$, the effective potential is given in Eq.~\eqref{RE.e:14-a}.  Solving \eqref{RE.e:14-b} for $\sigma$ as a function of $v$ gives the results shown in Fig.~\ref{f:Veff-KPZ-d3}, which is compared with the one-loop result, obtained by reexpanding the effective potential as a series of $\calA$,
\begin{equation}\label{RE.e:21}
   V_{\text{eff}}^{\text{one-loop}}[v]
   =
   \frac{\lambda^2 v^4}{8}
   +
   \frac{ \calA \lambda^5 v^5 }{ 120 \sqrt{2} \, \pi^2 \, \nu^4 } \>.
\end{equation}
In $d=3$, the Hartree approximation leads to a phase transition as function of $\lambda$.  Neither the loop expansion nor the LOAF approximation displays this property.  This is seen in Fig.~\ref{f:Veff-KPZ-d3}.  

%
%
\section{\label{s:DR}The chemical reaction annihilation process}

As a second example of our LOAF approximation methods, we derive an effective potential for the  reaction-diffusion chemical annihilation process, $A + A \rightarrow 0$.  The Langevin equation for this process is closely related to the KPZ equation (see Section~\ref{s:ColeHopf} below). 

The many-body formulation of this annihilation and diffusion process is discussed thoroughly in the literature \cite{r:Vollmayr-Lee:1994nr,r:Tauber:2005rc,r:Cardy:1999fk,r:Tauber:2014bf}.  The standard procedure for obtaining a path integral for the generating functional is to start from a master equation for the process, develop a number algebra with annihilation and creation operators using a Hilbert space, define a conserving state vector $\Ket{\Psi(t)}$ and a \Schrodinger-like equation, pass over to a continuum description, and then write a path integral for the generating functional. 
The resulting generating functional is given by
\begin{equation}\label{RD.e:1}
   Z[\,j,\sj\,]
   =
   e^{W[\,j,\sj\,]} 
   = 
   \calN \!
   \iint \calD \phi \, \calD\sphi \,
   e^{ - S[\, \phi, \sphi;j,\sj \, ] } \>, 
\end{equation}
where the action $S[\, \phi, \sphi;j,\sj \, ]$ is given by
\begin{align}
   S[\, \phi, \sphi;j,\sj \, ]
   &=
   \int \!\rd x \, 
   \bigl\{ \,
      \sphi(x) \, \partial_t \, \phi(x)
      +  
      \nu \, 
      [\, \bnabla \sphi(x) \,] \cdot 
      [\, \bnabla \phi(x)  \,]
      \label{RD.e:2} \\
      & \qquad 
      -  
      \lambda \, [\, 1 - \phi^{\star 2}(x) \,] \, \phi^2(x)
      -
      \sj(x) \phi(x)
      -
      j(x) \, \sphi(x) \,
   \bigr \} \>.
   \notag
\end{align}
Here we have set $\int \!\rd x \equiv \int \rd^d x \int \rd t$.

%
%
\subsection{\label{ss:DR.Langevin}Langevin equation}

We first show that we can obtain a Langevin equation for this process by 
making a Doi shift \cite{r:Doi:1976fk} of the star-field, $\sphi(x) \rightarrow 1 + \sphi(x)$ in the path integral \eqref{RD.e:1}.  This leads to a new action of the form,
\begin{align}
   S_{\text{doi}}[\, \phi, \sphi;j,\sj \, ]
   &
   =
   \int \!\rd x \, 
   \bigl \{ \,
      \sphi(x) \, \partial_t \, \phi(x)
      +  
      \nu \, 
      [\, \bnabla \sphi(x) \,] \cdot 
      [\, \bnabla \phi(x)  \,]
      \label{RD.e:3} \\
      & \qquad\quad
      +
      2 \lambda \, \sphi(x) \, \phi^2(x)
      +
      \lambda \, [\sphi(x)]^2 \, \phi^2(x)
      -
      \sj(x) \, \phi(x)
      -
      j(x) \, \sphi(x) \,        
   \bigr \} \>.
   \notag
\end{align}
Here the kinetic term $\partial_t \, \phi(x)$ is integrated, evaluated at the end points and absorbed in the overall normalization (see Ref.~\cite{r:Cardy:1999fk}).
Using the identity,
\begin{equation}\label{RD.e:4}
   \ExpB{
      - \int \rd x \, \lambda \, [\sphi(x)]^2 \, \phi^2(x) }
   =
   \int \calD \eta \, P[\, \eta \,] \,
   \ExpB{
      i \, \sqrt{ 2 \lambda } \, \sphi(x) \, \phi(x) \, \eta(x) } \>,
\end{equation}
where
\begin{equation}\label{RD.e:5}
   P[\, \eta \,]
   =
   \calN \,
   \ExpB{
      - \int \rd x \, \eta^2(x) / 2 } \>,
\end{equation}
the path integral \eqref{RD.e:1} for zero currents becomes
\begin{equation}\label{RD.e:6}
   Z_0
   =
   \calN \!
   \iiint \calD \phi \, \calD\sphi \, \calD \eta \,
   P[\, \eta \,] \,
   e^{ - S_{\text{doi}}[\, \phi, \sphi, \eta \, ] } \>, 
\end{equation}
where the action is now given by
\begin{equation}\label{RD.e:7}
   S_{\text{doi}}[\, \phi, \sphi, \eta \, ]
   =
   \int \!\rd x \,
   \sphi(x) \,
   \bigl [ \,
      D_x \, \phi(x)
      +
      2 \, \lambda \, \phi^2(x)
      +
      i \, \sqrt{ 2 \lambda } \, \phi(x) \, \eta(x) \,
   \bigr ] \>,
\end{equation}
where $D_x = \partial_t - \nu \, \nabla^2$. 
So the path integral is to be evaluated only for values of $\phi(x)$ which satisfy the Langevin equation,
\begin{equation}\label{RD.e:9}
   D_x \, \phi(x)
    +
   2 \, \lambda \, \phi^2(x) =
   -
   i \, \sqrt{ 2 \lambda } \, \phi(x) \, \eta(x) \>.
\end{equation}
The noise source in Eq.~\eqref{RD.e:9} is multiplicative and purely imaginary, which is a surprise since the field $\phi(x)$ started out to be real.  However, as discussed in Ref.~\cite{r:Cardy:1999fk}, the imaginary nature of the field for the Doi-shifted path integral is required for the probability interpretation of the path integral.  
The imaginary component of the average field $\Expect{ \Imag{\phi(x)} } = 0$ vanishes (see Ref.~\cite{PhysRevE.74.057102}).

%
%
\subsection{\label{RD.ss:Seff-Veff}Effective Action and Effective Potential for annihilation}

The annihilation process is an ideal test-bed for the LOAF approximation, since one can show that the renormalized reaction rate can be exactly determined by summing an infinite geometric series of Feynman diagrams.  Here there is no wave function or noise strength renormalization, so that in our formalism one can directly calculate the approximate $\beta$ function for the running of the coupling constant by considering only the effective potential.  We will find that our approach, which does not rely on perturbative Feynman graphs, gives qualitatively good results for the $\beta$ function at all dimensions $d$.   

Another reason for studying the annihilation process is that the effective potential for this problem was obtained in a one loop approximation by Hochberg and Zorzano \cite{r:Hochberg:2004bv}, who were able to determine analytically the  effective potential for $d=2$ only.  In that dimension our results for the renormalization group equation (RGE) agrees with theirs.  However, because of the simpler way terms are grouped in our approach we are able to determine {\it analytically} the renormalized effective potential for all dimensions $d$.  

The starting point for our LOAF calculation is the path integral representation for the generating functional \eqref{RD.e:1} with the action given in Eq.~\eqref{RD.e:2}.
Auxiliary fields $\sigma(x)$ and $\ssigma(x)$ are introduced by means of a Hubbard-Stratonovich transformation \cite{r:Hubbard:1959kx,r:Stratonovich:1958vn}.  That is we add to the action \eqref{RD.e:2} an action of the form,
\begin{equation}\label{RD.e:10}
   S_{\text{HS}}[\, \phi,\sphi,\sigma,\ssigma \,]
   =
   -
   \int \!\rd x \,
   \bigl \{ \, 
      \bigl [\, \ssigma(x) - \lambda \, \phi^{\star 2}(x) \,\bigr ] \,
      \bigl [\, \sigma(x) - \lambda \, \phi^2(x) \,\bigr ] \,
   \bigr \} / \lambda \>,
\end{equation}
to obtain an action which becomes quadratic in $\phi$ and $\sphi$. 
Adding sources for the auxiliary fields and introducing a two-component notation,
\begin{subequations}\label{RD.e:11}
\begin{align}
   \Phi(x)
   &=
   \begin{pmatrix}
      \phi(x) \\ \sphi(x)
   \end{pmatrix} \>,
   \qquad
   &
   J(x)
   &=
   \begin{pmatrix}
      j(x) \\ \sj(x)
   \end{pmatrix} \>,
   \label{RD.e:11-a} \\
   X(x)
   &=
   \begin{pmatrix}
      \sigma(x) \\ \ssigma(x)
   \end{pmatrix} \>,
   \qquad
   &
   K(x)
   &=
   \begin{pmatrix}
      s(x) \\ s^{\star}(x)
   \end{pmatrix} \>,
   \label{RD.e:11-b}
\end{align}
\end{subequations}
The path integral and action, including auxiliary fields and currents, can then be written as
\begin{equation}\label{RD.e:12}
   Z[\, J, K \, ]
   =
   e^{W[\, J, K \, ]}
   =
   \iint \! \calD \Phi \, \calD X \,
   e^{ - S[\, \Phi, X; J, K \, ] } \>,
\end{equation}
where
\begin{align}
   S[\, \Phi, X; J, K \, ] 
   &=
   \frac{1}{2} 
   \iint \! \rd x \,\rd x' \,
   \sPhi(x) \, G^{-1}[\, X \, ](x,x') \, \Phi(x)
   \label{RD.e:13} \\
   & \hspace{2em}
   -
   \int \! \rd x \,
   \Bigl \{ \, 
      \frac{ \sX(x) \, X(x) }{ 2 \lambda }
      +
      \sJ(x) \, \Phi(x)
      +
      \sK(x) \, X(x) \,
   \Bigr \} \>,
   \notag
\end{align}
where
\begin{equation}\label{RD.e:14}
   G^{-1}[\, X \, ](x,x')
   =
   \delta(x-x') \,
   \begin{pmatrix}
      \spD_x & 2 \, \sigma(x) \\
      2 \, [\, \ssigma(x) - 1 \,] &  \sD_x
   \end{pmatrix} \>,
\end{equation}
and where $\spD_x = \partial_t - \nu \nabla^2$ and  $\sD_x = - \partial_t - \nu \nabla^2$.
Performing the integration over the fields $\phi, \sphi$, we obtain
\begin{equation}
   Z[\, J,K \,]
   =
   e^{ W[\, J,K \,] }
   =
   \iint \calD X \,
   e^{ - S_{\text{eff}}[\, X;J,K \,] }
\end{equation}
where
\begin{align}
   S_{\text{eff}}[\, X;J,K \,]
   &=
   -
   \frac{1}{2} \iint \! \rd x \, \rd x' \,
   \sJ(x) \, G[X](x,x') \, J(x)
   \label{RD.e:15} \\
   & \hspace{2em}
   -
   \int \! \rd x \,
   \Bigl \{ \,
      \frac{ \sX(x) \, X(x) }{ 2 \lambda }
      +
      K^{\star}(x) \, X(x)
      -
      \frac{1}{2} \Tr{ \Ln{ G^{-1}[X](x,x) } } \,
   \Bigr \} \>.
   \notag
\end{align}
Following the same procedure as in Sections~\ref{KPZ.ss:PI-OM} and \ref{KPZ.ss:PI-MSR} for the KPZ equation, we perform the integration over the auxiliary fields by steepest descent and obtain
\begin{align}
   W[\, J,K \,]
   &=
   \frac{1}{2} 
   \iint \! \rd x \,\rd x' \,
   \sJ(x) \, G[\, X_0 \, ](x,x') \, J(x)
   \label{RD.e:16} \\
   & \hspace{2em}
   +
   \int \,\rd x \,
   \Bigl \{ \, 
      \frac{ \sX_0(x) \, X_0(x) }{ 2 \lambda }
      +
      \sK(x) \, X_0(x)
      -
      \frac{1}{2} \Tr{ \Ln{ G^{-1}[\,X_0\,](x,x) } }\,
   \Bigr \}
   +
   \dotsb \>,
   \notag
\end{align}
where $X_0[\,J,K\,]$ is the saddle point, defined by
\begin{equation}\label{RD.e:17}
   \frac{ \delta S_{\text{eff}}[\, X;J,K \,] }
        { \delta X(x) } \Big |_{X=X_0}
   =
   0 \>.
\end{equation}
Legendre transforming \eqref{RD.e:16}, we obtain in leading order in the auxiliary field loop expansion, the effective action:
\begin{align}
   \Gamma[\, \Phi, X \,] \!
   &=
   \! \int \rd x \,
   \bigl \{
      \sJ(x) \, \Phi(x)
      +
      \sK(x) \, X(x) 
   \bigr \}
   -
   W[\, J,K \,]
   \label{RD.e:19} \\
   &=
   \frac{1}{2} \iint \rd x \, \rd x' \,
   \sPhi(x) \, G^{-1}[X](x,x') \, \Phi(x)
   \notag \\
   & \qquad
   -
   \int \,\rd x \,
   \Bigl \{ \, 
      \frac{ \sX(x) \, X(x) }{ 2 \lambda }
      -
      \frac{1}{2} \Tr{ \Ln{ G^{-1}[\,X\,](x,x) } }\,
   \Bigr \}
   +
   \dotsb \>.
   \notag
\end{align}
which is the generating functional of the 1-PI graphs.
The two-particle correlation functions are obtained from the inverse of the matrix of second derivatives of the effective action $\Gamma[\, \Phi, X \,]$ with respect to the fields.  

Restricting ourselves to constant fields $\Phi$ and $X$, the effective potential is given by
\begin{align}
   V_{\text{eff}}[\, \phi,\sphi,\sigma,\ssigma \,] 
   &= 
   \Gamma[\, \phi,\sphi,\sigma,\ssigma \,] / \Omega
   \label{RD.e:20} \\
   &
   =
   ( \, \ssigma - 1 \, ) \, \phi^2
   +
   \sigma \, \phi^{\star 2}
   -
   \frac{ \ssigma \, \sigma }{ \lambda }
   +
   \frac{1}{2} 
   \Tr{ \Ln{ G^{-1}[\sigma,\ssigma](x,x) } } \>.
   \notag
\end{align}
where $\Omega$ is the space time volume.  Expanding the Green function $G^{-1}[\sigma,\ssigma](x,x)$ in a Fourier-Laplace series, as in Eq.~\eqref{KPZ.OM.e:23.2}, we have
\begin{equation}\label{RD.e:21}
   \tilde{G}^{-1}[\, \sigma,\ssigma \,](\bk,z)
   =
   \begin{pmatrix}
      \nu k^2 + z & 2 \sigma \\
      2 ( \ssigma - 1 ) & \nu k^2 - z 
   \end{pmatrix} \>,
\end{equation}
so then
\begin{equation}\label{RD.e:22}
   \Det{ \tilde{G}^{-1}[\, \sigma,\ssigma \,](\bk,z) }
   =
   \omega_k^2[\sigma,\ssigma] - z^2 \>,
\end{equation}
where
\begin{equation}\label{RD.e:22.1}
   \omega_k^2[\sigma,\ssigma]
   =
   \nu^2 k^4 + 4 \, ( 1 - \ssigma ) \, \sigma \>,
\end{equation}
from which we find
\begin{align}
   \Tr{ \Ln{ G^{-1}[\sigma,\ssigma](x,x) } }
   &
   =
   \int \! \frac{\rd^d k}{(2\pi)^d} \int \! \frac{\rd z}{2\pi i} \,
   \Det{ \tilde{G}^{-1}[\, \sigma,\ssigma \,](\bk,z) }
   \label{RD.e:23} \\
   &
   =
   \int \!\! \frac{\rd^d k}{(2\pi)^d} \,
   \bigl \{ \, | \, \omega_{k}[\, \sigma,\ssigma \,] \, | + C_{\infty} \, \bigr \} \>,
   \notag
\end{align}
where again $C_{\infty}$ is absorbed into the overall effective potential normalization.  Inserting this result into \eqref{RD.e:20} gives
\begin{equation}\label{RD.e:23.1}
   V_{\text{eff}}[\, \phi,\sphi,\sigma,\ssigma \,] 
   =
   ( \, \ssigma - 1 \, ) \, \phi^2
   +
   \sigma \, \phi^{\star 2}
   -
   \frac{ \ssigma \, \sigma }{ \lambda }
   +
   \frac{\nu}{2} 
   \int \!\! \frac{\rd^d k}{(2\pi)^d} \,
   \sqrt{ k^4 + 4 \, \frac{( 1 - \ssigma ) \, \sigma }{ \nu^2 } } \>.
\end{equation}
The gap equations are obtained by minimizing the effective potential with respect to $\sigma$ and $\ssigma$,
\begin{subequations}\label{RD.e:24}
\begin{align}
   \frac{ \partial V_{\text{eff}} }
        { \partial \ssigma }
   &=
   \phi^2
   -
   \frac{\sigma}{\lambda}
   -
   \frac{\sigma}{\nu}
   \int \!\! \frac{\rd^d k}{(2\pi)^d} \,
   \frac{1}{ [ \, k^4 + 4 \, ( 1 - \ssigma ) \, \sigma / \nu^2 \,]^{1/2} }
   =
   0 \>,
   \label{RD.e:24-a} \\
   \frac{ \partial V_{\text{eff}} }
        { \partial \sigma }
   &=
   \phi^{\star 2}
   -
   \frac{\ssigma}{\lambda}
   +
   \frac{1 - \ssigma}{\nu}
   \int \!\! \frac{\rd^d k}{(2\pi)^d} \,
   \frac{1}{ [ \, k^4 + 4 \, ( 1 - \ssigma ) \, \sigma / \nu^2 \,]^{1/2} }
   =
   0 \>.   
   \label{RD.e:24-b}
\end{align}
\end{subequations}
Renormalization of these equations is carried out in the next section.

%
%
\subsection{\label{RD.ss:Renormalization}Renormalization and beta-function}

\begin{figure}[t!]
   \centering
   \includegraphics[width=0.75\columnwidth]{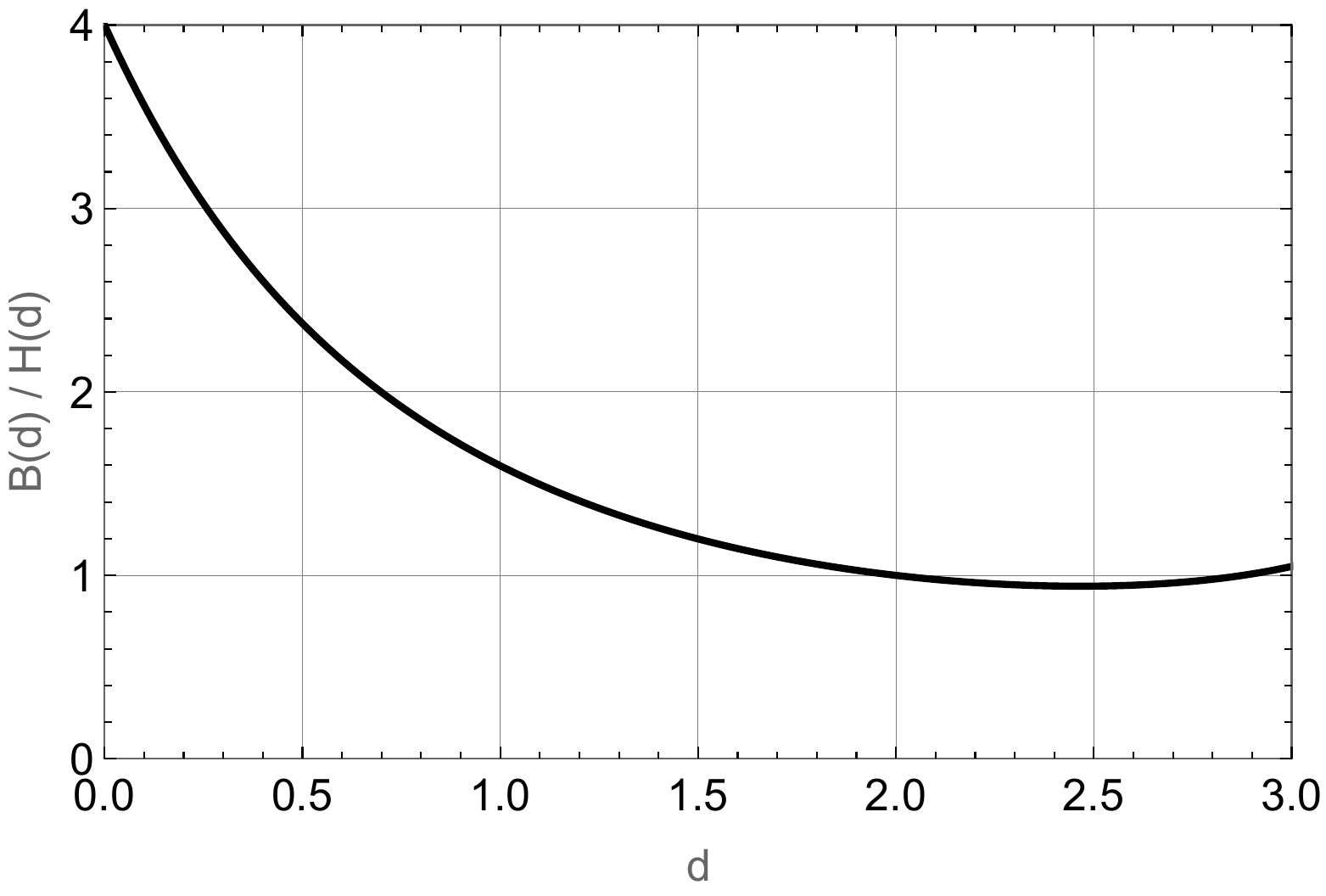}
   \caption{\label{RD.f:BHratio}The ratio $B[\, d \,] / H[\, d \,]$ as a function of $d$.}
\end{figure}

The second derivative of the effective potential is the negative of the inverse of the correlation function $D_{\sigma\ssigma}^{-1}(0,0)$ at zero momentum, and is the renormalized coupling constant,
\begin{equation}\label{RD.e:25}
   \frac{1}{\lambda_r[\, m^2,d \,]}
   \equiv
   D_{\sigma\ssigma}^{-1}(0,0)
   =
   -
   \frac{\partial^2 V_{\text{eff}}}{\partial \sigma \, \partial \ssigma}
   =
   \frac{1}{\lambda}
   +
   \frac{1}{\nu} \, \Sigma[\, m^2,d \,] \>,
\end{equation}
where $\Sigma[\, m^2,d \,] = \Sigma_{1}[\, m^2,d \,] - \Sigma_{2}[\, m^2,d \,]$ with
\begin{subequations}\label{RD.e:26}
\begin{align}
   \Sigma_{1}[\, m^2,d \,]
   &=
   \int \!\! \frac{\rd^d k}{(2\pi)^d} \,
   \frac{1}{ [ \, k^4 + m^4 \,]^{1/2} } \>,
   \label{RD.e:26-a} \\
   \Sigma_{2}[\, m^2,d \,]
   &=
   \frac{ m^4 }{ 2 }
   \int \!\! \frac{\rd^d k}{(2\pi)^d} \,
   \frac{1}{ [ \, k^4 + m^4 \,]^{3/2} } \>,
   \label{RD.e:26-b}   
\end{align}
\end{subequations}
and where we have set
\begin{equation}\label{RD.e:27}
   m^4
   \equiv
   4 \, ( 1 - \ssigma ) \, \sigma / \nu^2 \>.
\end{equation}
Using dimensional regularization, we obtain
\begin{subequations}\label{RD.e:28}
\begin{align}
   \Sigma_{1}[\, m^2,d \,]
   &=
   \frac{ m^{d-2} \, \Omega_d }{ 2 - d } \,
   \frac{ \Gamma[\, 3/2 - d/4 \,] \, \Gamma[\, d/4 \,] }
        { (2\pi)^{d} \, \sqrt{\pi} } \>,
   \label{RD.e:28-a} \\
   \Sigma_{2}[\, m^2,d \,]
   &=
   \frac{ m^{d-2} \, \Omega_d }{ 4 } \,
   \frac{ \Gamma[\, 3/2 - d/4 \,] \, \Gamma[\, d/4 \,] }
        { (2\pi)^{d} \, \sqrt{\pi} } \>,
   \label{RD.e:28-b}   
\end{align}
\end{subequations}
where the angular volume $\Omega_d$ in $d$-dimension is
\begin{equation}\label{RD.e:29}
   \Omega_d
   =
   \frac{ 2 \, \pi^{d/2} }{ \Gamma[\,d/2\,] } \>.
\end{equation}
So $\Sigma_1$ has ultraviolet divergences in dimensions two and higher.  
From Eqs.~\eqref{RD.e:28}, we find
\begin{equation}\label{RD.e:30}
   \Sigma[\, m^2,d \,]  
   = 
   \Sigma_1[\, m^2,d \,]
   - 
   \Sigma_2[\, m^2,d \,]
   =   
   \frac{ m^{d-2} }{2-d} \, H[\, d \,] \>,
\end{equation}
where
\begin{equation}\label{RD.e:31}
   H[\, d \,] 
   =
   \frac{ (d+2) \, \Gamma[ \, 3/2 - d/4 \, ] \, \Gamma[ \, d/4 \, ] }
        { (4\pi)^{d/2 + 1/2} \, \Gamma[\, d/2 \,] } \>.
\end{equation}
So for any mass $\mu$, we define the renormalized coupling constant $\lambda_r[\, \mu^2,d \, ]$ by the equation,
\begin{equation}\label{RD.e:32}
   \frac{1}{\lambda_r[\, \mu^2,d \, ] } 
   =   
   \frac{1}{\lambda}
   +
   \frac{1}{\nu} \, \Sigma[\, \mu^2,d \,]
   =   
   \frac{1}{\lambda}
   +
   \frac{ \mu^{d-2} }{2-d} \, \frac{ H[\, d \,] }{ \nu } \>.
\end{equation}
Differentiating Eq.~\eqref{RD.e:32} with respect to $\mu$ gives the renormalization group equation,
\begin{equation}\label{RD.e:33}
   \mu \, \frac { \rd \lambda_r[\, \mu^2,d \,]}{\rd \mu} 
   = 
   \frac{ \lambda_r^2[\, \mu^2,d \,] }{ \nu } \, H[\, d \,] \>.
\end{equation}
At two different mass scales, the renormalized coupling constants are related by the equation,
\begin{equation}\label{RD.e:35}
   \frac{1}{\lambda_r[\, \mu^2,d \, ] } 
   =
   \frac{1}{\lambda_r[\, \mu_0^2,d \, ] }
   +
   \frac{ H[\, d \,] }{ \nu \, ( 2 - d ) } \,
   \bigl [ \,
      \mu^{d-2} - \mu_0^{d-2} \,
   \bigr ] \>.
\end{equation}
Let us define a dimensionless renormalized reaction rate $g_r[\, \mu^2,d \,]$ by
\begin{equation}\label{RD.e:36}
   g_r[\, \mu^2,d \,]
   =
   \mu^{d-2} \, \lambda_r[\, \mu^2,d \, ] / \nu \>,
\end{equation}
in which case Eq.~\eqref{RD.e:32} becomes
\begin{equation}\label{RD.e:37}
   \frac{1}{ g_r[\, \mu^2,d \,] }
   =
   \frac{1}{g_0}
   +
   \frac{ H[\, d \,] }{ 2 - d } \>,
\end{equation}
where the bare reaction rate $g_0$ is defined by
\begin{equation}\label{RD.e:38}
   g_0 
   = 
   \frac{\lambda \, \mu^{d-2}}{\nu} \>.
\end{equation}
Writing \eqref{RD.e:37} in the form,
\begin{equation}\label{RE.e:39}
    g_r[\, \mu^2,d \,]
    =
    \frac{ g_0 }{ 1 + g_0 \, H[\, d \,] / ( d - 2 ) } \>,
\end{equation}
we see that for $d<2$ there is a stable infrared fixed point at
\begin{equation}\label{RD.e:40}
   g_0 = g^\star \equiv ( 2-d ) / H[\, d \,] \>.
\end{equation}
Differentiating \eqref{RD.e:37} with respect to $\mu$ gives the $\beta$ function for our LOAF approximation.  We find
\begin{equation}\label{RD.e:41}
   \beta_g[\, \mu^2,d \,]
   = 
   \mu \, \frac{ \partial g_r[\, \mu^2,d \,] }{ \partial \mu } 
   = 
   (d-2) \, g_r[\, \mu^2,d \,] + g_r^2[\, \mu^2,d \,] \, H[\, d \,] \>.
\end{equation}
The exact $\beta_g$ function can be calculated by summing all the perturbative one loop graphs (see for example Eq.~(28) in Ref.~\cite{r:Vollmayr-Lee:1994nr}).  This leads to the exact answer,
\begin{equation}\label{RD.e:42}
   \frac{1}{ g_r^{\text{exact}}[\, \mu^2,d \,] }
   =
   \frac{1}{g_0}
   +
   \frac{ B[\, d \,] }{ 2 - d } \>,
\end{equation}
where
\begin{equation}\label{RE.e:43}
   B[\, d \,] 
   = 
   \frac{ 4 \, \Gamma[\,2 - d/2 \,] }{ ( 8\pi )^{d/2} } \>,
\end{equation}
leading to the exact $\beta_g$ function,
\begin{equation}\label{RD.e:43}
   \beta_g^{\text{exact}}[\, \mu^2,d \,]
   = 
   \mu \, \frac{ \partial g_r[\, \mu^2,d \,] }{ \partial \mu } 
   = 
   (d-2) \, g_r[\, \mu^2,d \,] + g_r^2[\, \mu^2,d \,] \, B[\, d \,] \>.
\end{equation}
In Fig.~\ref{RD.f:BHratio} we plot the ratio $B[\, d \,] / H[\, d \,]$ from $d=1$ to $d=3$, showing that the LOAF approximation yields a reasonable answer for the running of the coupling constant compared to the exact result for $1.5 < d < 3$. 

%
%
%
%

%
%
\subsection{\label{RD.ss:Veff}The effective potential}

From Eqs.~\eqref{RD.e:25}, \eqref{RD.e:27}, and \eqref{RD.e:35}, the second derivative of the effective potential at the reference mass $\mu^2$ is given by
\begin{equation}\label{RD.e:44}
   -
   \frac{ \partial^2 V_{\text{eff}}[\, \mu,\sigma,\ssigma,d \,] }
        { \partial \sigma \, \partial \ssigma }
   =
   \frac{1}{\lambda_r[\, \mu^2,d \, ] }
   +
   \frac{ H[\, d \,] }{ \nu \, ( 2 - d ) } \,
   \Bigl \{ \,
      \Bigl [ \, \frac{ 4 \, ( 1 - \ssigma ) \, \sigma }{ \nu^2 } \, \Bigr ]^{(d-2)/4} 
      - 
      \mu^{d-2} \,
   \Bigr \} \>.
\end{equation}
To obtain the renormalized effective potential we integrate Eq.~\eqref{RD.e:42} with respect to $\sigma$  and $\ssigma$, being mindful of the constants of integration and find
\begin{align}
   V_{\text{eff}}[\, \mu,d \,]
   &=
   ( \ssigma - 1 ) \, \phi^2
   -
   \sigma \, \phi^{\star 2} 
   -
   \frac{\ssigma \sigma}{ \lambda_r[\, \mu^2,d \, ] }
   \notag \\
   & \hspace{2em}
   +
   \frac{ ( \ssigma - 1 ) \, \sigma \, H[\,d\,] }{ \nu ( d - 2 ) } \,
   \Bigl \{ \,
      \mu^{d-2} - \frac{16}{(d+2)^2} \, m^{d-2} \,
   \Bigr \} \>,
   \label{RD.e:45}
\end{align}
where $m$ is given in Eq.~\eqref{RD.e:27}.

To obtain the effective potential in terms of only $\phi$ and $\sphi$ we need to evaluate the effective potential at the solution of the gap equations.
\begin{equation}\label{RD.e:46}
   \frac{\partial V_{\text{eff}} }{\partial \sigma} 
   = 
   0 \>,
   \Qquad{and}
   \frac{\partial V_{\text{eff}} }{\partial \ssigma} 
   = 
   0 \>.
\end{equation}
This leads the gap equations:
\begin{subequations}\label{RD.e:gaps}
\begin{align}  
   \frac{\ssigma}{\lambda_r} 
   &= 
   \phi^{\star \, 2} 
   +
   \frac{ ( \ssigma - 1 ) \, H[\, d \, ] }{ (d-2) \, \nu } \,
   \bigl \{ \,
      2^{ d/2 + 1 } \,  (\, d + 2 \, ) \,
      \bigl [\,
         \sigma ( \, 1 - \ssigma \, ) / \nu^2 \,
      \bigr ]^{(d-2)/4} 
      -
      \mu^{d-2} \,
   \bigr \} \>,
   \label{RD.e:gap1} \\
   \frac{\sigma}{\lambda_r} 
   &= 
   \phi^{2} 
   +
   \frac{ \sigma \, H[\, d \, ] }{ (d-2) \, \nu } \,
   \bigl \{ \,
      2^{ d/2 + 1 } \,  (\, d + 2 \, ) \,
      \bigl [\,
         \sigma ( \, 1 - \ssigma \, ) / \nu^2 \,
      \bigr ]^{(d-2)/4} 
      -
      \mu^{d-2} \,
   \bigr \} \>,
   \label{RD.e:gap2}
\end{align}
\end{subequations}
From Eqs.~\eqref{RD.e:gaps} one can solve numerically for $\sigma$ and $\ssigma$ as a function of $\phi$ and $\sphi$.
%
%
\subsubsection{d=2}

For the critical dimension $d=2$, we can take the limit of Eq.~\eqref{RD.e:45} as $d \rightarrow 2$ to obtain
\begin{align}
   V_{\text{eff}}[\, \mu,\sigma,\ssigma,d \,]
   &=
   ( \ssigma - 1 ) \, \phi^2
   +
   \sigma \, \phi^{\star 2} 
   -
   \frac{\ssigma \sigma}{ \lambda_r[\, \mu^2,d \, ] }
   \label{RD.e:47} \\
   & \hspace{2em}
   +
   \frac{ ( \ssigma - 1 ) \, \sigma \, H[\,d\,] }{ \nu } \,
   \Bigl \{ \,
      \frac{1}{4} \, \LnB{ \frac{m^4}{\mu^4} } - \frac{1}{2} \,
   \Bigr \} \>,
   \notag
\end{align}
and from  from Eq.~\eqref{RD.e:42} in the limit that $d \rightarrow 2$, 
\begin{equation}\label{RD.e:48}
   -
   \frac{ \partial^2 V_{\text{eff}}[\, \mu,\sigma,\ssigma,d \,] }
        { \partial \sigma \, \partial \ssigma }
   =
   \frac{1}{\lambda_r[\, \mu^2,d \, ] }
   -
   \frac{1}{ 8 \pi \, \nu } \, \LnB{ \frac{m^4}{\mu^4} } \>.  
\end{equation}
In two dimensions, since $H[2]$ = $1 / (2 \pi)$, we obtain the same RG equation for $\lambda$ as found in Ref.~\cite{r:Hochberg:2004bv}, namely
\begin{equation}\label{RD.e:49}
   \mu \, \frac { \rd \lambda_r}{\rd \mu} 
   = 
   \frac{\lambda_r^2}{ 2 \pi \nu} \>.
\end{equation}
This is equivalent to the result from Eq.~\eqref{RD.e:48} that
\begin{equation}\label{RD.e:50}
   \frac{1}{ \lambda_r[\, \mu^2 \,] }
   -
   \frac{1}{ \lambda_r[\, \mu^{\prime 2} \,] } 
   =
   \frac{1}{ 8 \pi \, \nu } \, \LnB{ \frac{ \mu^{\prime 4} }{ \mu^4 } } \>.
\end{equation}

%
%
\section{\label{s:ColeHopf}Cole-Hopf transformation of the KPZ equation}

As discussed in Refs.~\cite{r:Lassiq:1995hs,r:Cardy:1999fk}, one can convert the KPZ equation using a Cole-Hopf transformation to an equation which is very similar to the one we have just discussed for the annihilation process $A+A \rightarrow 0$.  The Cole-Hopf transformation is a change of variables from $\phi(x)$ and $\sphi(x)$ to a new set $w(x)$ and $\sw(x)$ of the form:
\begin{equation}\label{CH.e:1}
   \phi(x)
   =
   \frac{2 \nu}{\lambda} \Ln{ w(x) } \>,
   \Quad{and}
   \sphi(x)
   =   
   \frac{\lambda}{2 \nu} \, \sw(x) \, w(x) \>.
\end{equation}
Then since
\begin{equation}\label{CH.e:2} 
   \sphi(x) \,
   \bigl \{ \,
       D_x \, \phi(x) - f_0 - \lambda \, | \bnabla \phi(x) |^2 / 2 \,
   \bigr \}
   \equiv
   \sw(x) \, [\, D_x - m_0^2 \, ]\, w(x) \>,
\end{equation}
where we have set $m_0^2 = f_0 \sqrt{ 2\lambdaCH }$ and $\lambdaCH = \lambda^2 / ( 8 \nu^2 )$,
the MSR version of the generating functional, Eq.~\eqref{KPZ.MSR.e:1}, and action, Eq.~\eqref{KPZ.MSR.e:2}, become
\begin{equation}\label{CH.e:3}
   Z[ \, j,\sj \, ]
   =
   e^{ W[ \, j,\sj \, ] / \calA } 
   =
   \calN\!
   \iint \calD \sw \, \calD w \, 
   e^{ - S[ \, w,\sw;j,\sj \, ] / \calA } \>,
\end{equation}
where
\begin{align}
   S[ \, w,\sw;j,\sj \, ]
   &=
   \int \rd x \>
   \bigl \{ \,
      \sw(x) \, [\, D_x - m_0^2 \,] \, w(x)
      \label{CH.e:4} \\
      & \hspace{2em}
      -
      \lambdaCH \,
      [ \, \sw(x) \, w(x) \, ]^2
      -
      \sj(x) \, w(x)
      -
      j(x) \, \sw(x) \,
   \bigr \} \>,
   \notag
\end{align}
with corresponding redefinitions of the currents.  The Jacobian for the Cole-Hopf transformation is unity.  

As in Eq.~\eqref{RD.e:10}, introducing auxiliary fields $\sigma(x)$ and $\ssigma(x)$ by means of the Hubbard-Stratonovich transformation,
\begin{equation}\label{CH.e:5}
   S_{\text{HS}}[\, w,\sw \!\!,\sigma,\ssigma \,]
   =
   \int \!\rd x \,
   \bigl \{ \, 
      \bigl [\, \ssigma(x) - \lambdaCH \, w^{\star 2}(x) \,\bigr ] \,
      \bigl [\, \sigma(x) - \lambdaCH \, w^2(x) \,\bigr ] \,
   \bigr \} / \lambdaCH  \>,
\end{equation}
which we add to \eqref{CH.e:4} to obtain an action that becomes quadratic in $w$ and $\sw$\!. 
Adding sources for the auxiliary fields and introducing a two-component notation,
\begin{subequations}\label{CH.e:6}
\begin{align}
   W(x)
   &=
   \begin{pmatrix}
      w(x) \\ \sw(x)
   \end{pmatrix} \>,
   \qquad
   &
   J(x)
   &=
   \begin{pmatrix}
      j(x) \\ \sj(x)
   \end{pmatrix} \>,
   \label{CH.e:6-a} \\
   X(x)
   &=
   \begin{pmatrix}
      \sigma(x) \\ \ssigma(x)
   \end{pmatrix} \>,
   \qquad
   &
   K(x)
   &=
   \begin{pmatrix}
      s(x) \\ s^{\star}(x)
   \end{pmatrix} \>,
   \label{CH.e:6-b}
\end{align}
\end{subequations}
The path integral and action, including auxiliary fields and currents, can then be written as
\begin{equation}\label{CH.e:7}
   Z[\, J, K \, ]
   =
   e^{W[\, J, K \, ]}
   =
   \iint \! \calD W \, \calD X \,
   e^{ - S[\, W, X; J, K \, ] / \calA } \>,
\end{equation}
where
\begin{align}
   S[\, W, X; J, K \, ] 
   &
   =
   \frac{1}{2} 
   \iint \! \rd x \,\rd x' \,
   \sW(x) \, G^{-1}[\, X \, ](x,x') \, W(x)
   \label{CH.e:8} \\
   & \hspace{3em}
   +
   \int \!\rd x \,
   \Bigl \{ \, 
      \frac{ \sX(x) \, X(x) }{ 2 \lambdaCH }
      -
      \sJ(x) \, \Phi(x)
      -
      \sK(x) \, X(x) \,
   \Bigr \} \>,
   \notag
\end{align}
where
\begin{equation}\label{CH.e:9}
   G^{-1}[\, X \, ](x,x')
   =
   \delta(x-x') \,
   \begin{pmatrix}
      \spD_x & - 2 \, \sigma(x) \\
      - 2 \, \ssigma(x) &  \sD_x
   \end{pmatrix} \>,
\end{equation}
and where $\spD_x = \partial_t - \nu \nabla^2$ and  $\sD_x = - \partial_t - \nu \nabla^2$.
We see apart from a sign change in the coupling constant and the absence of a term linear in $\sigma^2$, the Cole-Hopf form of the KPZ actions is quite similar to that for $A+A \rightarrow  0$.  Performing the integration over the fields $W$, we obtain
\begin{equation}\label{CH.e:10}
   Z[\, J,K \,]
   =
   e^{ W[\, J,K \,] }
   =
   \iint \calD X \,
   e^{ - S_{\text{eff}}[\, X;J,K \,] / \calA }
\end{equation}
where
\begin{align}
   S_{\text{eff}}[\, X;J,K \,]
   &=
   -
   \frac{1}{2} \iint \! \rd x \, \rd x' \,
   \sJ(x) \, G[X](x,x') \, J(x)
   \label{CH.e:11} \\
   & \hspace{4em}
   +
   \int \! \rd x \,
   \Bigl \{ \,
      \frac{ \sX(x) \, X(x) }{ 2 \lambdaCH }
      -
      K^{\star}(x) \, X(x)
      +
      \frac{\calA}{2} \, \Tr{ \Ln{ G^{-1}[X](x,x) } } \,
   \Bigr \} \>.
   \notag
\end{align}
Following the same procedure as for the annihilation process, we integrate over the auxiliary fields using the method of steepest descent and keep only the saddle point contribution.  Legendre transforming to the Effective Action we then obtain in leading order in the auxiliary field loop expansion, the effective action:
\begin{align}
   \Gamma[\, W, X \,] \!
   &=
   \int\! \rd x \,
   \bigl \{
      \sJ(x) \, W(x)
      +
      \sK(x) \, X(x) 
   \bigr \}
   -
   W[\, J,K \,]
   \label{CH.e:14} \\
   & 
   =
   \frac{1}{2} \iint \rd x \, \rd x' \,
   \sW(x) \, G^{-1}[X](x,x') \, W(x)
   \notag \\
   & \hspace{4em}
   +
   \int \,\rd x \,
   \Bigl \{ \, 
      \frac{ \sX(x) \, X(x) }{ 2 \lambdaCH }
      +
      \frac{\calA}{2} \, \Tr{ \Ln{ G^{-1}[\,X\,](x,x) } } \,
   \Bigr \}
   +
   \dotsb \>.
   \notag
\end{align}
which is the generating functional of the 1-PI graphs.  Assuming constant fields, the effective potential is then given by
\begin{equation}\label{CH.e:15}
   V_{\text{eff}}[\, W,X \,]
   =
   \frac{\Gamma[\, W, X \,] }{ \Omega }
   =
   -
   \bigl [ \,
      \ssigma \, w^2 + \sigma \, w^{\star 2} \,
   \bigr ]
   +
   \frac{ \ssigma \, \sigma }{\lambdaCH }
   +
   \frac{\calA}{2} \, \Tr{ \Ln{ G^{-1}[\,X\,](x,x) } } \>,
\end{equation}
where now
\begin{align}
   \Tr{ \Ln{ G^{-1}[\,X\,](x,x) } }
   &=
   \int \! \frac{\rd^d k}{(2\pi)^d} \int \! \frac{\rd z}{2\pi i} \,
   \Det{ \tilde{G}^{-1}[\, \sigma,\ssigma \,](\bk,z) }
   \label{CH.e:16}  \\
   &
   =
   \int \!\! \frac{\rd^d k}{(2\pi)^d} \,
   \bigl \{ \, | \, \omega_{k}[\, \sigma,\ssigma \,] \, | + C_{\infty} \, \bigr \} \>,
   \notag
\end{align}
where $\omega_{k}^2[\, \sigma,\ssigma \,] = \nu^2 k^4 - 4 \, \ssigma \sigma$.  We will see below that for a stable Cole-Hopf solution, if we choose $\sigma > 0$, then $\ssigma < 0$.
Absorbing $C_{\infty}$ into the overall effective potential normalization, we find
\begin{equation}\label{CH.e:18}
   V_{\text{eff}}[\, w,\sw,\sigma,\ssigma \,] 
   =
   -
   \bigl [ \,
      \ssigma \, w^2 + \sigma \, w^{\star 2} \,
   \bigr ]
   +
   \frac{ \ssigma \, \sigma }{ \lambdaCH }
   +
   \frac{\nu \calA}{2} 
   \int \!\! \frac{\rd^d k}{(2\pi)^d} \,
   \sqrt{ k^4 - 4 \, \frac{\ssigma \, \sigma }{ \nu^2 } } \>.
\end{equation}
The gap equations are obtained by minimizing the effective potential with respect to $\sigma$ and $\ssigma$,
\begin{subequations}\label{CH.e:19}
\begin{align}
   \frac{ \partial V_{\text{eff}} }
        { \partial \ssigma }
   &=
   -
   w^2
   +
   \frac{\sigma}{\lambdaCH}
   -
   \frac{ \sigma \calA }{\nu}
   \int \!\! \frac{\rd^d k}{(2\pi)^d} \,
   \frac{1}{ [ \, k^4 - 4 \, \ssigma \, \sigma / \nu^2 \,]^{1/2} }
   =
   0 \>,
   \label{CH.e:19-a} \\
   \frac{ \partial V_{\text{eff}} }
        { \partial \sigma }
   &=
   -
   w^{\star 2}
   +
   \frac{\ssigma}{\lambdaCH}
   -
   \frac{ \ssigma \calA }{\nu}
   \int \!\! \frac{\rd^d k}{(2\pi)^d} \,
   \frac{1}{ [ \, k^4 - 4 \, \ssigma \, \sigma / \nu^2 \,]^{1/2} }
   =
   0 \>.   
   \label{CH.e:19-b}
\end{align}
\end{subequations}
Renormalization of Eqs.~\eqref{CH.e:18} and \eqref{CH.e:19} are carried out by our technique of dimensional regularization, as in Section~\ref{RD.ss:Renormalization}.  We start with defining the renormalized Cole-Hopf transformed coupling constant as the second derivative of the effective potential,
\begin{equation}\label{CH.e:20}
   \frac{1}{ \lambdaCH_{r}[\, m^2,d \,] }
   =
   \frac{ \partial^2 V_{\text{eff}} }
        { \partial \ssigma \, \partial \sigma }
   =
   \frac{1}{\lambdaCH}
   -
   \frac{\calA}{\nu} \,
   \Sigma[\, m^2,d \,] \>,
\end{equation}
where $\Sigma[\, m^2,d \,] = \Sigma_1[\, m^2,d \,] - \Sigma_2[\, m^2,d \,]$, which are defined in Eqs.~\eqref{RD.e:28}.  For the Cole-Hopf case, $m^4$ is given by
\begin{equation}\label{CH.e:21}
   m^4
   =
   -
   4 \, \ssigma \sigma / \nu^2 > 0 \>.
\end{equation}
Again separating out the mass and dimension factors, we have $\Sigma[\, m^2,d \,] = m^{d-2} \, H[\, d \,] / ( 2 - d )$, with $H[\, d \, ]$ given in Eq.~\eqref{RD.e:31}.
Again introducing the renormalized coupling constant at scale $\mu$, 
\begin{equation}\label{CH.e:22}
   \frac{1}{ \lambdaCH_{r}[\, \mu^2,d \,] }
   =
   \frac{1}{\lambdaCH}
   -
   \frac{\calA \, \mu^{d-2}}{\nu (2 - d) } \, H[\, d \, ] \>,
\end{equation}
from which we find the renormalization group equation,
\begin{equation}\label{CH.e:23}
   \mu \, \frac{ \rd \lambdaCH_{r} }{ \rd \mu }
   =
   -
   \frac{ \calA \, ( \, \lambdaCH_{r} \, )^2 }{ \nu } \, H[\, d \,] \>.
\end{equation}
Again introducing  the dimensionless renormalized coupling constant 
\begin{equation}\label{CH.e:24}
   g^{\text{CH}}_r 
   =  
   \frac{ \calA \, \lambdaCH_r }{ \nu } \, \mu^{d-2},
\end{equation}
the equation for $\beta^{\text{CH}}_g$ is 
\begin{equation}\label{CH.e:25}
   \beta_g 
   = 
   \mu \, 
   \frac{\partial g^{\text{CH}}_r}{\partial \mu} 
   = 
   (d-2) \, g^{\text{CH}}_r - ( \, g^{\text{CH}}_r \, )^2 \, H[\, d \,] \>.
\end{equation}
So for $d>2$, there is an unstable UV fixed point at
\begin{equation}\label{CH.e:26}
   g^{\text{CH}}_r
   =
   g^\star 
   \equiv 
   \frac{ d-2 }{ H[\, d \,] } \>.
\end{equation}
This leads to the roughening transition at $d=2$ as discussed in \cite{r:Cardy:1999fk}.  Note that our answer for $g^\star$ differs from the exact answer by the ratio $B[d]/H[d]$, which is plotted in    Fig.~\ref{RD.f:BHratio}. 

The renormalized effective potential can be obtained by integrating twice the renormalized second derivative of the potential
\begin{equation}\label{CH.e:27}
   \frac{ \partial^2 V_{\text{eff}}[\, \mu,\sigma,\ssigma,d \,] }
        { \partial \ssigma \, \partial \sigma }
   =
   \frac{1}{\lambdaCH_{r}[\, \mu^2,d \, ] }
   +
   \frac{ \calA \, H[\, d \,] }{ \nu \, ( 2 - d ) } \,
   \Bigl \{ \,
      \mu^{d-2}
      -
      \Bigl [ \, - \frac{ 4 \, \ssigma \sigma }{ \nu^2 } \, \Bigr ]^{(d-2)/4} \,
   \Bigr \} \>.
\end{equation}
Keeping in mind the constants of integration, we find here that
\begin{align}
   V_{\text{eff}}[\, \mu,d \,]
   &=
   -
   [ \, \ssigma \, w^2 + \sigma \, w^{\star 2} \, ]
   +
   \frac{\ssigma \sigma}{ \lambdaCH_r[\, \mu^2,d \, ] }
   \label{CH.e:28} \\
   & \hspace{2em}
   +
   \frac{ \calA \, H[\,d\,] }{ \nu ( d - 2 ) } \,
   \Bigl \{ \,
      \ssigma \sigma \, \mu^{d-2} 
      +
      \frac{4 \nu^2}{(d+2)^2} \,
      \Bigl [ \, - \frac{ 4 \, \ssigma \sigma }{ \nu^2 } \, \Bigr ]^{(d+2)/4} \,      
   \Bigr \} \>.
   \notag
\end{align}
The gap equations are then:
\begin{subequations}\label{CH.e:29}
\begin{align}
   \frac{\sigma}{ \lambdaCH_r[\, \mu^2,d \, ] }
   &=
   w^2
   -
   \frac{ \calA \, H[\,d\,] }{ \nu ( d - 2 ) } \,
   \Bigl \{ \,
      \sigma \, \mu^{d-2} 
      -
      \frac{4 \, \sigma}{d+2} \,
      \Bigl [ \, - \frac{ 4 \, \ssigma \sigma }{ \nu^2 } \, \Bigr ]^{(d-2)/4} \,      
   \Bigr \} \>,
   \label{CH.e:29-a} \\
   \frac{\ssigma}{ \lambdaCH_r[\, \mu^2,d \, ] }
   &=
   w^{\star 2}
   -
   \frac{ \calA \, H[\,d\,] }{ \nu ( d - 2 ) } \,
   \Bigl \{ \,
      \ssigma \, \mu^{d-2} 
      -
      \frac{4 \, \ssigma}{d+2} \,
      \Bigl [ \, - \frac{ 4 \, \ssigma \sigma }{ \nu^2 } \, \Bigr ]^{(d-2)/4} \,      
   \Bigr \} \>.
   \label{CH.e:29-b}
\end{align}
\end{subequations}

%
%
\subsection{\label{CH.ss:d1}One dimension} 

From Eq.~\eqref{CH.e:22}, for $d=1$, we choose to renormalize at $\mu = \infty$, in which case the bare coupling constant becomes equal to the renormalized one: $\lambdaCH_{r}[\, \infty,1 \,] = \lambdaCH$.  So then the effective potential \eqref{CH.e:28} becomes
\begin{equation}\label{CH.e:30}
   V_{\text{eff}}[\, \infty,1 \,]
   =
   -
   [ \, \ssigma \, w^2 + \sigma \, w^{\star 2} \, ]
   +
   \frac{\ssigma \sigma}{ \lambdaCH }
   -
   \frac{ 4 \nu \, \calA \, H[\, 1 \,] }{ 6 } \,
   \Bigl [ \, - \frac{ 4 \, \ssigma \sigma }{ \nu^2 } \, \Bigr ]^{3/4} \>,      
\end{equation}
and the gap equations are
\begin{subequations}\label{CH.e:31}
\begin{align}
   \frac{\sigma}{ \lambdaCH }
   &=
   w^2
   \,\,
   -
   \frac{ 4 \, \sigma\,\, \, \calA \, H[\, 1 \,] }{ 3 \nu } \,
   \Bigl [ \, - \frac{ 4 \, \ssigma \sigma }{ \nu^2 } \, \Bigr ]^{-1/4} \>,     
   \label{CH.e:31-a} \\
   \frac{\ssigma}{ \lambdaCH }
   &=
   w^{\star 2}
   -
   \frac{ 4 \, \ssigma \, \calA \, H[\, 1 \,] }{ 3 \nu } \,
   \Bigl [ \, - \frac{ 4 \, \ssigma \sigma }{ \nu^2 } \, \Bigr ]^{-1/4} \>.     
   \label{CH.e:31-b}
\end{align}
\end{subequations}
Multiplying \eqref{CH.e:31-a} by $-4 \, \ssigma / \nu^2$ and \eqref{CH.e:31-b} by $-4 \, \sigma / \nu^2$, and adding the two equations gives
\begin{equation}\label{CH.e:32}
   \frac{ m^4 }{ \lambdaCH }
   =
   -
   \frac{2}{\nu^2} \,
   [ \, \ssigma \, w^2 + \sigma \, w^{\star 2} \, ]
   -
   \frac{ 4 \, \calA \,H[\, 1 \,] }{ 3 \nu } \, m^3 \>,
\end{equation}
so that
\begin{equation}\label{CH.e:33}
   - [ \, \ssigma \, w^2 + \sigma \, w^{\star 2} \, ]
   =
   \frac{ \nu^2 }{ 2 \lambdaCH } \, m^4 
   +
   \frac{ 4 \nu \, \calA \, H[\, 1 \,] }{ 6 } \, m^3 \>.
\end{equation}
Substitution of this into \eqref{CH.e:30} gives
\begin{equation}\label{CH.e:34}
   V_{\text{eff}}[\, \infty,1 \,]
   =
   \frac{ \nu^2 }{ 4 \lambda } \, m^4 \>.
\end{equation}
So as a function of $m$, the effective potential is quartic in $m$, with a minimum at $m=0$.  

%
%
\subsection{\label{CH.ss:d2}Two dimensions} 

For $d=2$, we take the limit of \eqref{CH.e:28} as $d \rightarrow 2$, and find
\begin{equation}\label{CH.e:40}
   V_{\text{eff}}[\, \mu,d \,]
   =
   -
   [ \, \ssigma \, w^2 + \sigma \, w^{\star 2} \, ]
   +
   \frac{\ssigma \sigma}{ \lambdaCH_r[\, \mu^2,2 \, ] }
   +
   \frac{ \nu \, \calA }{ 8 \pi } \, m^4 \,
   \Bigl \{ \,
      \frac{1}{4} \, \LnB{ \frac{ m^4 }{ \mu^4 } }
      -
      \frac{1}{2} \,
   \Bigr \} \>.
\end{equation}
The renormalized coupling constant at $m^2$ is given by the limit $d \rightarrow 2$ of Eq.~\eqref{CH.e:27},
\begin{equation}\label{CH.e:41} 
   \frac{1}{\lambdaCH_{r}[\, m^2,2 \, ] }
   =
   \frac{1}{\lambdaCH_{r}[\, \mu^2,2 \, ] }
   +
   \frac{ \calA }{ 8 \pi \, \nu } \,
   \LnB{ \frac{ m^4 }{ \mu^4 } } 
\end{equation}
Here we have used $H[\, 2 \,] = 1 / (2 \pi)$.  Differentiation of \eqref{CH.e:41} with respect to $\mu$ gives the RG equation,
\begin{equation}\label{CH.e:42}
   \mu \, \frac{ \partial \lambdaCH_{r} }{ \partial \mu }
   =
   \frac{ \calA \, \bigl [ \, \lambdaCH_r \bigr ]^2 }{ 2 \pi \, \nu} \>,
\end{equation}
which is the same as Eq.~\eqref{RD.e:49} for the $A+A \rightarrow 0$ annihilation case, except for the factor of $\calA$ and that this involves $\lambdaCH = \lambda^2 / (8 \nu^2)$, where $\lambda$ and $\nu$ are the coupling constant and diffusion coefficient respectively for the KPZ equation.  

%
%
\subsection{\label{CH.ss:d3}Three dimensions} 

For $d=3$, we renormalize at $\mu = 0$ so from Eq.~\eqref{CH.e:22}, we again have that 
the bare coupling constant becomes equal to the renormalized one: $\lambdaCH_{r}[\, 0,3 \,] = \lambdaCH$.  Then the effective potential \eqref{CH.e:28} becomes
\begin{equation}\label{CH.e:35}
   V_{\text{eff}}[\, 0,3 \,]
   =
   -
   [ \, \ssigma \, w^2 + \sigma \, w^{\star 2} \, ]
   +
   \frac{\ssigma \sigma}{ \lambdaCH }
   +
   \frac{ 4 \nu \, \calA \, H[\, 3 \,] }{ 25 } \,
   \Bigl [ \, - \frac{ 4 \, \ssigma \sigma }{ \nu^2 } \, \Bigr ]^{5/4} \>,      
\end{equation}
and the gap equations are
\begin{subequations}\label{CH.e:36}
\begin{align}
   \frac{\sigma}{ \lambdaCH }
   &=
   w^2
   \,\,
   +
   \frac{ 4 \, \sigma\,\, \, \calA \, H[\, 3 \,] }{ 5 \nu } \,
   \Bigl [ \, - \frac{ 4 \, \ssigma \sigma }{ \nu^2 } \, \Bigr ]^{1/4} \>,     
   \label{CH.e:36-a} \\
   \frac{\ssigma}{ \lambdaCH }
   &=
   w^{\star 2}
   +
   \frac{ 4 \, \ssigma \, \calA \, H[\, 3 \,] }{ 5 \nu } \,
   \Bigl [ \, - \frac{ 4 \, \ssigma \sigma }{ \nu^2 } \, \Bigr ]^{1/4} \>.     
   \label{CH.e:36-b}
\end{align}
\end{subequations}
Again multiplying \eqref{CH.e:36-a} by $-4 \, \ssigma / \nu^2$ and \eqref{CH.e:36-b} by $-4 \, \sigma / \nu^2$, and adding the two equations gives
\begin{equation}\label{CH.e:37}
   - [ \, \ssigma \, w^2 + \sigma \, w^{\star 2} \, ]
   =
   \frac{ \nu^2 }{ 2 \lambdaCH } \, m^4 
   -
   \frac{ 4 \nu \, \calA \, H[\, 3 \,] }{ 10 } \, m^3 \>.
\end{equation}
Substitution of this into \eqref{CH.e:35} gives
\begin{equation}\label{CH.e:38}
   V_{\text{eff}}[\, 0,3 \,]
   =
   \frac{ \nu^2 }{ 4 \lambdaCH } \, m^4
   -
   \frac{6}{25} \, \nu \, \calA \, H[\, 3 \,] \, m^5
\end{equation}
which has a maximum at
\begin{equation}\label{CH.e:39}
   m
   =
   m_0
   \equiv
   \frac{5}{6} \, \frac{ \nu }{ \lambdaCH \calA \, H[\, 3 \, ] } \>.
\end{equation}
Here $H(3) = 0.0537$. The effective potential becomes negative when $m > 5/4 ~ m_0$. 

%
%
\section{\label{GL.s:GinsburyLandau}The Ginzburg-Landau model} 

The Ginzburg-Landau model is the prototypic relaxation model of an Ising ferromagnet.
To make contact with perturbative renormalization group treatment found in Cardy's lecture notes \cite{r:Cardy:1999fk}, we will work in reduced units, so that $\kB \, T_c=1$.
Here we use $\nu$, rather than $D$, for the diffusion coefficient to correspond to our notation in the rest of this paper.  If $\phi(x)$ is the amplitude of the spins at $x = (\bx,t)$, then 
the Ginzburg-Landau model is described by an equilibrium Hamiltonian
\begin{equation}\label{GL.e:1}
   H[ \, \phi \,]
   =
   \int \rd^d x \,
   \Bigl \{ \,
      \frac{1}{2} \, 
      \bigl [ \,
         | \, \bnabla \phi \, |^2
         +
         f_0 \, \phi^2 \,
      \bigr ]
      +
      \frac{u}{4} \, \phi^4 \,
   \Bigr \} \>,
\end{equation}
where $f_0 = C \, ( \, T - T_c \, )$.
The Langevin equation which relaxes to the equilibrium distribution is then
\begin{equation}\label{GL.e:2}
   \partial_t \phi(x)
   =
   - 
   \nu \, \frac{ \delta H[ \, \phi \,] }{ \delta \phi(x) }
   -
   \eta(x)
   =
   \nu \,
   \bigl [ \,
      \nabla^2 \, \phi(x)
      - 
      f_0 \, \phi(x)
      -
      u \, \phi^3(x) \,
   \bigr ]
   -
   \eta(x) \>.
\end{equation}
Setting $f = \nu \, f_0$ and $\lambda = \nu \, u$, Eq.~\eqref{GL.e:2} can be written as
\begin{equation}\label{GL.e:3}
   D_x \phi(x)
   +
   f \, \phi(x)
   +
   \lambda \, \phi^3(x)
   =
   \eta(x) \>,
\end{equation}
where $D_x = \partial_x - \nu \, \nabla^2$.  
To satisfy the Einstein relation, one requires
\begin{equation}\label{GL.e:4}
   \Expect{ \eta(x) \, \eta(x') }
   =
   2 \nu \, \delta(x-x') \>,
\end{equation}
which means that $\eta(x)$ is a white noise source with amplitude $\calA = 2 \nu$, and distribution functional,
\begin{equation}\label{GL.e:5}
   P[\, \eta \, ]
   =
   \calN \,
   \ExpB{ - \frac{1}{2 \calA} \int \!\! \rd x \, \eta^2(x) } \>.
\end{equation}
The generating functional for this action can be obtain using the formalism of Section~\ref{ss:PI-MSR}, where for the Ginzburg-Landau model, $F[\, \phi \,] = f \, \phi(x) + \lambda \, \phi^4(x)$.  So scaling the star field $\sphi(x)$ and star current $\sj(x)$ by the amplitude $\calA$ of the noise, the MSR form of the generating functional, Eqs.~\eqref{KPZ.MSR.e:1} and \eqref{KPZ.MSR.e:2} become for the Ginzburg-Landau case,
\begin{equation}\label{GL.e:6}
   Z[ \, j,\sj \, ]
   =
   e^{ W[ \, j,\sj \, ] / \calA } 
   =
   \calN\!
   \iint \calD \sphi \, \calD \phi \, 
   e^{ - S[ \, \phi,\sphi;j,\sj \, ] / \calA } \>,
\end{equation}
where
\begin{align}
   S[ \, \phi,\sphi;j,\sj \, ]
   &
   =
   \int \rd x \>
   \bigl \{ \,
      \sphi(x) \,
      \bigl [ \, D_x \phi(x) - f - \lambda \, \phi^2(x) \, \bigr ] \,
      \phi(x)
      \label{GL.e:7} \\
      & \hspace{3em}
      -
      [ \, \sphi(x) \, ]^2 / 2
      -
      \sj(x) \, \phi(x)
      -
      j(x) \, \sphi(x) \,
   \bigr \} \>.
   \notag
\end{align}
One can easily show that for zero currents, this action leads to the Langevin Eq.~\eqref{GL.e:3}.
Auxiliary fields $\sigma(x)$ and $\ssigma(x)$ are introduced by inserting the identity,
\begin{align}
   1
   &=
   \int \calD \sigma \,
   \delta
   \bigl [ \, \sigma  - f - \lambda \, \phi^2(x) \, \bigr ]
   \label{GL.e:8} \\
   &=
   \calN\!
   \iint \calD \sigma \, \calD \ssigma \,
   \ExpB{
      \int \rd x \,
      \frac{ \ssigma(x) }{ \lambda \calA } \,
      \Bigl [ \, 
         \sigma(x) - f - \lambda \, \phi^2(x) \, ] \, 
      \Bigr ]
        } \>,
   \notag
\end{align}
into the path integral \eqref{GL.e:6}.  The integration here over $\ssigma(x)$ is along the imaginary axis.  Using a two-component notion defined by
\begin{equation}\label{GL.e:9}
   \Phi(x)
   =
   \begin{pmatrix}
      \phi(x) \\ \sphi(x)
   \end{pmatrix} \>,
   \Qquad{and}
   X(x)
   =
   \begin{pmatrix}
      \sigma(x) \\ \ssigma(x)
   \end{pmatrix} \>,
\end{equation}
we can write a path integral of the form,
\begin{equation}\label{GL.e:10}
   Z[ \, J,K \, ]
   =
   e^{ W[ \, J,K \, ] / \calA } 
   =
   \calN\!
   \iint \calD \Phi \, \calD X \, 
   e^{ - S[ \, \Phi,X ; J,K \, ] / \calA } \>,
\end{equation}
where
\begin{align}
   S[ \, \Phi,X ; J,K \, ]
   &
   =
   \frac{1}{2}
   \iint \! \rd x \, \rd x' \,
   \sPhi(x) \, G^{-1}[\, X \,](x,x') \, \Phi(x')
   \label{GL.e:11} \\
   & \hspace{3em}
   - 
   \int\! \rd x \,
   \Bigl [ \,
      \frac{ \sX(x) \, X(x) }{ 2 \lambda }
      +
      \sJ(x) \, \Phi(x)
      +
      \sK(x) \, X(x) \,
   \Bigr ] \,,
   \notag
\end{align}
and the inverse Green function $G^{-1}[\, X \,](x,x')$ is given by
\begin{equation}\label{GL.e:12}
   G^{-1}[\, X \,](x,x') 
   =
   \delta(x-x')
   \begin{pmatrix}
      \spD_x + \sigma(x) \,, & - 1 \\
      2 \, \ssigma(x) \,, & \sD_x + \sigma(x)
   \end{pmatrix} \>,
\end{equation}
and the currents by
\begin{equation}\label{GL.e:13}
   J(x)
   =
   \begin{pmatrix}
      j(x) \\ \sj(x)
   \end{pmatrix} \>,
   \Qquad{and}
   K(x)
   =
   \begin{pmatrix}
      s(x) + f / \lambda \\
      s^{\star}(x)
   \end{pmatrix} \>.
\end{equation}
As in Section \ref{ss:PI-MSR}, performing the path integration over the $\Phi(x)$ fields and integrating the remaining auxiliary fields $X(x)$ by the method of steepest descent, we find to leading order,
\begin{align}
   W[\, J,K \,]
   &=
   \frac{1}{2} 
   \iint\! \rd x \, \rd x' \,
   \sJ(x) \, G[\, X_0 \,](x,x') \, J(x')
   \label{GL.e:14} \\
   & \hspace{3em}
   +
   \int\! \rd x \,
   \Bigl \{ \,
      \frac{ \sX_0(x) \, X_0(x) }{ 2 \lambda }
      +
      \sK(x) \, X_0(x)
      -
      \frac{\calA}{2} \,
      \Tr{ \Ln{ G^{-1}[\, X_0 \,](x,x) } } \,
   \Bigr \}
   +
   \dotsb
   \notag
\end{align}
where $X_0$ is evaluated at the saddle point.  Legendre transforming \eqref{GL.e:14}, we obtain the LOAF result for the effective action,
\begin{align}
   \Gamma[\, \Phi,X \,]
   &=
   \!\int\! \rd x 
   \bigl \{ \,
      \sJ(x) \Phi(x) + \sK_0(x) X(x) \,
   \bigr \}
   -
   W[\, J,K \,]
   \label{GL.e:15} \\
   &=
   \frac{1}{2}
   \iint\! \rd x \, \rd x' \,
   \sPhi(x) \, G^{-1}[\, X \,](x,x') \, \Phi(x')
   \notag \\
   & \hspace{3em}
   +
   \int\! \rd x \,
   \Bigl \{ \,
      -
      \frac{ \ssigma(x) \, [ \, \sigma(x) - f \, ] }{ \lambda }
      +
      \frac{\calA}{2} \,
      \Tr{ \Ln{ G^{-1}[\, X \,](x,x) } } \,
   \Bigr \}
   +
   \dotsb
   \notag   
\end{align}
Restricting ourselves to constant fields we obtain for the effective potential 
\begin{align}
   V_{\text{eff}}[\, \Phi, X \,]
   =
   \frac{\Gamma[\, \Phi,X \,] }{ \Omega }
   &=
   \sigma \, \sphi \, \phi
   -
   \phi^{\star 2} / 2
   +
   \ssigma \, \phi^2
   -
   \ssigma \, (\, \sigma - f \,) / \lambda
   \label{GL.e:16} \\
   & \hspace{3em}
   +
   \frac{\calA}{2} \,
   \Tr{ \Ln{ G^{-1}[\, X \,](x,x) } } \>.
   \notag
\end{align}
Expanding $G^{-1}[\, X \,](x,x)$ in a Fourier-Laplace series as in Eq.~\eqref{KPZ.OM.e:23.2}, we have
\begin{equation}\label{GL.e:17}
   \tilde{G}^{-1}[\, \sigma,\ssigma \,](\bk,z)
   =
   \begin{pmatrix}
      \nu k^2 + z + \sigma \,, & -1 \\
      2 \, \ssigma \,, & \nu k^2 - z + \sigma 
   \end{pmatrix} \>,
\end{equation}
so then
\begin{equation}\label{GL.e:18}
   \Det{ \tilde{G}^{-1}[\, \sigma,\ssigma \,](\bk,z) }
   =
   \omega_k^2[\sigma,\ssigma] - z^2 \>,
\end{equation}
where
\begin{equation}\label{GL.e:19}
   \omega_k^2[\sigma,\ssigma]
   =
   ( \, \nu k^2 + \sigma \, )^2 + 2 \, \ssigma \>.
\end{equation}
from which we find
\begin{align}
   \Tr{ \Ln{ G^{-1}[\sigma,\ssigma](x,x) } }
   &
   =
   \int \! \frac{\rd^d k}{(2\pi)^d} \int \! \frac{\rd z}{2\pi i} \,
   \Det{ \tilde{G}^{-1}[\, \sigma,\ssigma \,](\bk,z) }
   \label{GL.e:20} \\
   &
   =
   \int \!\! \frac{\rd^d k}{(2\pi)^d} \,
   \bigl \{ \, | \, \omega_{k}[\, \sigma,\ssigma \,] \, | + C_{\infty} \, \bigr \} \>,
   \notag   
\end{align}
where again $C_{\infty}$ is absorbed into the overall effective potential normalization.  Inserting this result into \eqref{GL.e:16} gives
\begin{align}
   V_{\text{eff}}[\, \phi,\sphi,\sigma,\ssigma \,]
   &
   =
   \sigma \, \sphi \, \phi
   -
   \phi^{\star 2} / 2
   +
   \ssigma \, \phi^2
   -
   \ssigma \, (\, \sigma - f \,) / \lambda
   \label{GL.e:21} \\
   & \hspace{3em}
   +
   \frac{\calA}{2} \,
   \int \!\! \frac{\rd^d k}{(2\pi)^d} \,
   \sqrt{ (\, \nu k^2 + \sigma \,)^2 + 2 \, \ssigma } \>.
   \notag
\end{align}
The gap equations are
\begin{subequations}\label{GL.e:22}
\begin{align}
   \frac{\partial V_{\text{eff}}}{\partial \ssigma}
   &=
   \phi^2
   - 
   (\, \sigma - f \,) / \lambda
   +
   \frac{\calA}{2} \,
   \int \!\! \frac{\rd^d k}{(2\pi)^d} \,
   \frac{1}{ \sqrt{ ( \, \nu k^2 + \sigma \,)^2 + 2 \, \ssigma } }
   =
   0 \>,
   \label{GL.e:22-a} \\
   \frac{\partial V_{\text{eff}}}{\partial \sigma}
   &=
   \sphi \, \phi
   -
   \ssigma / \lambda
   +   
   \frac{\calA}{2} \,
   \int \!\! \frac{\rd^d k}{(2\pi)^d} \,
   \frac{ \nu k^2 + \sigma }
        { \sqrt{ ( \, \nu k^2 + \sigma \,)^2 + 2 \, \ssigma } }
   =
   0 \>.
   \label{GL.e:22-b}
\end{align}
\end{subequations}
The renormalized coupling constant is
\begin{equation}\label{GL.e:23}
   \frac{1}{\lambda_r}
   \equiv
   -
   \frac{\partial^2 V_{\text{eff}}}{\partial \sigma \, \partial \ssigma}
   =
   \frac{1}{\lambda}
   +
   \frac{\calA}{2} \,
   \int \!\! \frac{\rd^d k}{(2\pi)^d} \,
   \frac{ \nu k^2 + \sigma }
        { [ \, (\, \nu k^2 + \sigma \,)^2 + 2 \, \ssigma \,]^{3/2} } \>.
\end{equation}
Defining $m_1^2 = \sigma / \nu$ and $m_2^4 = 2 \ssigma / \nu^2$, Eq.~\eqref{GL.e:23} can be written as
\begin{equation}\label{GL.e:24}
   \frac{1}{\lambda_r[\, m_1^2,m_2^2,d \,]}
   =
   \frac{1}{\lambda}
   +
   \frac{\calA}{2 \nu} \, \Sigma[\, m_1^2,m_2^2,d \,] \>,
\end{equation}
where
\begin{equation}\label{GL.e:25}
   \Sigma[\, m_1^2,m_2^2,d \,] 
   =
   \int \!\! \frac{\rd^d k}{(2\pi)^d} \,
   \frac{ k^2 + m_1^2 }
        { [ \, (\, k^2 + m_1^2 \,)^2 + m_2^4 \,]^{3/2} } \>.   
\end{equation}
We see here that the critical dimension is for $d=4$. 
Expanding the integral in Eq.~\eqref{GL.e:25} in a power series in $m_2^4$ around zero, we notice only the first term ($m_2^4=0$) is divergent for $d=4$, which suggests that we can define a renormalized coupling constant at $m_2^4 =0$ and $m_1^2 = \mu^2$ via
\begin{equation}\label{GL.e:26}
   \frac{1}{\lambda_r[\, \mu^2,0,d \,]}
   =
   \frac{1}{\lambda}
   +
   \frac{\calA}{2 \nu} \, \Sigma[\, \mu^2,0,d \,] \>,
\end{equation}
where
\begin{align}
   \Sigma[\, \mu^2,0,d \,]
   &=
   \int \!\! \frac{\rd^d k}{(2\pi)^d} \,
   \frac{ 1 }{ (\, k^2 + \mu^2 \,)^2 }
   = 
   \frac{\Omega_d}{(2\pi)^d} 
   \int_{0}^{\infty} \!\! \frac{ k^{d-1} \, \rd k }{ (\, k^2 + \mu^2 \,)^2 }
   \label{GL.e:27} \\
   &=
   \frac{\Omega_d}{(2\pi)^d} \,\frac{ \mu^{d-4}  }{ 4 - d } \,
   \Gamma[\, d/2 \,] \, \Gamma[\, 3 - d/2 \,] \>.
\end{align}
Inserting this into \eqref{GL.e:26} gives
\begin{equation}\label{GL.e:28}
   \frac{1}{\lambda_r[\, \mu^2,0,d \,]}
   =
   \frac{1}{\lambda}
   +
   \frac{\calA}{2 \nu} \, 
   \frac{ \mu^{d-4} }{ 4 - d } \, J[\, d \,] \>,   
\end{equation}
where
\begin{equation}\label{GL.e:29}
   J[\, d \,]
   =
   \frac{\Omega_d}{(2\pi)^d} \,
   \Gamma[\, d/2 \,] \, \Gamma[\, 3 - d/2 \,] \>,
\end{equation}
and introducing the dimensionless bare coupling constant
\begin{equation}\label{GL.e:30}
   g_r[\, \mu^2 \, ]
   =
   \frac{\calA}{2 \nu} \,
   \mu^{d-4} \>,
\end{equation}
then the dimensionless renormalized coupling constant is
\begin{equation}\label{GL.e:31}
   g_r[\, \mu^2 \,] 
   = 
   \frac{ g_0[\, \mu^2 \,] }
        { 1 + g_0[\, \mu^2 \,]  \, J[\, d \,] / (4-d) } \>,
\end{equation}
and the $\beta$ function for $g_r[\mu^2]$ is given by
\begin{equation}\label{GL.e:32}
   \beta[\, g_r \,] 
   = 
   \mu \, \frac{\partial g_r[\, \mu^2 \,]}{\partial \mu} 
   = 
   [d-4] \, g_r[\, \mu \,] 
   + 
   \bigl [\, 
      g_r[\,\mu^2 \,] \,
   \bigr ]^2 \, J[\, d \,] \>.
\end{equation}
So there is a fixed point for $d \leq 4$ at
\begin{equation}
   g^\star =  \frac{\epsilon}{J[\, d \,]},
\end{equation}
where $\epsilon = 4-d$.  This is quite similar to the result of a perturbative analysis of the problem with the distinction that $J[\, d \,]$ is replaced by a related $d$ dependent function $K[\, d \,]$ (See Ref.~\cite{r:Cardy:1999fk}). 

As in the previous examples we can now regulate the second derivative of $V_{\text{eff}}$ using the definition of $\lambda_r[\, \mu^2,0,d \,]$, so that
\begin{align}
   \frac{1}{ \lambda_r[\, m_1^2,m_2^2,d \,] }
   &\equiv
   -
   \frac{ \partial^2 V_{\text{eff}} }
        { \partial \sigma \, \partial \ssigma }
   \label{GL.e:33} \\
   &
   =
   \frac{1}{ \lambda_r[\, \mu^2,0,d \,] }
   +
   \frac{\calA}{2 \nu} \!
   \int \!\! \frac{\rd^d k}{(2\pi)^d} \,
   \biggl \{
      \frac{ k^2 + m_1^2 }
           { [ \, (\, k^2 + m_1^2 \,)^2 + m_2^4 \,]^{3/2} }      
      -
      \frac{ 1 }{ (\, k^2 + \mu^2 \,)^2 } 
   \biggr \} \>.
   \notag
\end{align}
Eq.~\eqref{GL.e:33} can be evaluated analytically as a power series in $m_2^4$ for example and then one can reconstruct the full effective potential by integrating this result with respect to $\sigma$ and $\ssigma$ and adding the appropriate classical terms.  Evaluating the resulting  effective potential at the solution of the gap equation then gives the effective potential in terms of $\phi, \sphi$.  Finally, solving for the Lagrange multiplier field $\sphi[\, \phi \,]$ and substituting that in the potential gives the Onsager-Machlup form of the potential discussed earlier.

%
%
\section{\label{SchwingerDyson}Schwinger-Dyson equations}

To go beyond our LOAF approximation, one can systematically calculate the 1-PI action order-by-order in $\epsilon$, as discussed in detail in Ref.~\cite{{r:Bender:1977bh}}.  However for time-dependent problems such an expansion becomes secular, as shown in Ref.~\cite{r:MCD01}.  To solve the secularity problem requires a further resummation which can be performed using the exact Schwinger-Dyson (SD) equations.  This can be equivalently formulated using the second Legendre transform of the generating functional which yields the generating functional of the 2-PI vertices.  In this section we sketch this approach.  Using auxiliary fields, the Lagrangian is then made trilinear in all interactions.  Then the original and auxiliary fields are incorporated into one vector field $\phi_{\alpha}(x) = \Set{ \phi(x), \sphi(x), \sigma(x), \chi(x) }$, with $\phi^{\alpha}(x) = \Set{ \sphi(x), \phi(x), \sigma(x), \chi(x) }$, and the SD equations truncated at the cubic vertex level, what was called a ``bare vertex approximation'' (BVA) in Ref.~\cite{r:MCD01}.  
  
Adding source terms, the action in the generating functional for the correlation functions becomes symbolically
\begin{align}
   S[ \Phi, J ]
   &=
   -
   \frac{1}{2} \,
   \iint \rd x \, \rd x' \,
      \phi_{\alpha}(x) \, G^{-1}_0{}^{\alpha}{}_{\beta}(x,x') \, \phi^{\beta}(x')
   \label{SD.e:1} \\ 
   & \hspace{3em}
   \int \rd x \, \frac{1}{6} \, \gamma_{\alpha\beta\gamma}(\theta) \,
   \phi^{\alpha}(x) \phi^{\beta}(x) \phi^{\gamma}(x)
   +
   \int \rd x \, j_{\alpha}(x) \, \phi^{\alpha}(x) \>, 
   \notag
\end{align}
where $\gamma_{\alpha\beta\gamma}(\theta)$ is a  matrix which describes all the trilinear couplings among the fields and their conjugates. 
The SD equations are generated by considering the identity, 
\begin{equation}\label{SD.e:2}
   \int \rD \Phi \, \frac{ \delta \, e^{i S[\, \phi;j \, ] / \hbar} }{ \delta \phi_{\alpha}(x) } 
   = 
   0 \>,
\end{equation}
which gives the exact inverse Green function as
\begin{equation}\label{SD.e:3}
   G^{-1}{}^{\alpha\beta}[\phi](x,x')
   =
   G^{-1}_1{}^{\alpha\beta}[\phi](x,x')
   -
   \Sigma^{\alpha\beta}[\phi](x,x') \>,
\end{equation}
where
\begin{subequations}\label{SD.e:4}
\begin{align}
   G^{-1}_1{}^{\alpha\beta}(x,x')
   &=
   \bigl \{ \,
      G^{-1}_0{}^{\alpha\beta}(x)
      +
      \gamma^{\alpha\beta\gamma} \, \phi_{\gamma}(x) \,
   \bigr \} \, \delta(x,x') \>,
   \label{SD.e:4a} \\
   \Sigma^{\alpha\beta}[\phi]
   &=
   \frac{\hbar}{2 i} \,
   \gamma^{\alpha\alpha'\beta'} \,
   G_{\alpha'\alpha''}[\phi] \,
   G_{\beta'\beta''}[\phi] \,    
   \Gamma^{\alpha''\beta''\beta}[\phi] \>,
   \label{SD.e:4b}
\end{align}
\end{subequations}
Here $\Gamma^{\alpha''\beta''\beta}[\phi]$ is the exact one particle irreducible vertex function. 
The BVA consists of truncating the SD infinite hierarchy of equations by replacing the exact vertex $\Gamma^{\alpha''\beta''\beta}[\phi]$ in Eq.~\eqref{SD.e:4b} by the bare one,
\begin{equation}\label{SD.e:5}
   \Gamma^{\alpha''\beta''\beta}[\phi]
   \rightarrow
   \Gamma^{\alpha\beta\gamma}_{\text{BVA}}(x,x',x'') 
   = 
   \gamma^{\alpha\beta\gamma} \, \delta(x,x') \>.
\end{equation}
The BVA is a conserving approximation and the equations can be obtained from the 2-PI generating functional by functional differentiation.  This action is given by
\begin{equation}\label{SD.e:6}
   S[\Phi,G]
   =
   S_{\text{class}}[\Phi]
   +
   \frac{i}{2} \, \Tr{ \Ln{ G_1^{-1} } }
   +
   \frac{i}{2} \, \Tr{ \Ln{  G_1^{-1} \, G - 1 } }
   +
   \Gamma_2[G] \>,
\end{equation}
where for the BVA, $\Gamma_2[G] = - \Tr { \gamma \, G \, G \, G \, \gamma} / 12$.  This approximation to the 2-PI generating functional was first discussed in Ref.~\cite{r:MCD01} for the case of $N$ fields with O(N) symmetry, and then later related to the 2-PI-1/N model.  This approximation was then used to do dynamical simulations for both quantum field theories and Bose gases \cite{r:CDM02ii,PhysRevD.67.056003,r:gasenzer:2005}.

%
%
\section{\label{s:Conclude}Conclusions}

In this paper, we have shown how to apply the auxiliary field loop expansion method to obtain, in leading order, the effective action, effective potential, and the renormalization group flows for several examples of stochastic partial differential equations in arbitrary dimensions.  We have discussed both the Onsager-Machlup formulation and the Janssen-de Dominicis  versions of the path integral for reaction diffusion equations in the presence of external noise. We have shown how to obtain the effective potential of the Onsager-Machlup variety from the Janssen-de Dominicis variety in general, and worked this out explicitly for the case of the KPZ effective potential in the LOAF approximation. This required determining the value of the conjugate field in the MSR action by minimizing the JD effective potential with respect to the conjugate field. We believe this is the first discussion of this procedure in the literature. 

Using this formalism, we  re-examined earlier studies of possible fluctuation induced symmetry breaking in the KPZ equation using our method.  Our results contradicted earlier studies using a loop expansion in the fluctuation strength $\calA$ as well as a gaussian self-consistent approximation.  These previous approaches either violated Ward identities (Hartree approximation) or were not applicable at large values of $\calA$.  We found no evidence for fluctuation induced symmetry breaking. This is in agreement with a recent renormalization group improvement study of the one loop result by Bork and Ogarkov \cite{r:Bork:2014yg}.

  We then discussed how to obtain the effective potential of the MSR type when there is internal noise arising from the probabilistic nature of the underlying chemical reactions which are described by a master equation.  We used this approach to obtain the effective potential for the reaction-diffusion annihilation process $A+ A \rightarrow 0$ which has been much studied in the literature using perturbative diagrammatic methods.  The renormalization group flows for this problem have been exactly determined by summing an infinite series of Feynman graphs.
Our non-perturbative evaluation of the effective potential led to a renormalization group flow that qualitatively agreed with the exact answer in all dimensions.
We then considered a Cole-Hopf transformed approach to understanding the KPZ equation which bears many similarities to the annihilation problem.  Using the LOAF approximation, we then were able to obtain the renormalized effective potential and renormalization group flow that again qualitatively agreed with known exact results.
Finally, we obtained the  MSR effective potential for the Ginzburg-Landau model and the associated renormalization group flows.  For this problem our results were quite similar to those found using renormalization group improved perturbation theory.
 
The formalism presented here can be used to study the real time dynamics of noise averaged evolutions of stochastic fields as well as their correlation functions.
In the case of dilute BECs the LOAF approximation to the dynamics was able to predict dynamical phase transitions such as phase separation in multicomponent dilute Bose gases.
We intend to pursue the dynamical questions in the future to compare the results of the LOAF approximation as well as the BVA  with numerical simulation of the Langevin equations. In this paper we focused on the leading order LOAF approximation, which is expected to give qualitative results for Phase diagrams. The auxiliary field loop expansion provides a systematic way of calculating corrections to this approximation to the phase diagram when  that is needed.

%
%
\section*{Acknowledgments}
We would like to thank Juan Perez-Mercader for thoughtful discussions and suggesting this research.  We would also like to thank Gourab Ghoshal for valuable discussions.  
We would like to thank the Santa Fe Institute for hospitality where some of this work was done.

%
%
\appendix
\section{\label{append-A}Gaussian integrals and path integration}

In this work we made extensive use of Gaussian Path Integrals which are a continuum generalization of Gaussian integrals for Matrices. This is a topic described in many text books such as those by Kamenev  \cite{r:Kamenev:2011fk} and Peskin and Schroeder \cite{ref:PeskinSchroeder} and we refer the reader to these texts for elaboration and derivation of these results.

The basic result from integrating Gaussians over the real line is
\begin{equation}
\int \! \frac{\rd x }{\sqrt {2 \pi} }  ~ e^{- ax^2/2 + J x } = \frac{ e^{J a J/2}}{\sqrt{a}} 
\end{equation} 
This generalizes to symmetric complex matrices $A$, whose eigenvalues have non-negative real parts,  and vectors $J_i$ as follows
\begin{align}
   F[J] 
   &= 
   \prod_{i=1}^N 
   \biggl \{ \,
      \int \! \frac{\rd x_i }{\sqrt {2 \pi} } \,
   \biggr \} \,
   \ExpB{ - \frac{1}{2}\sum_{ij}^N  x_i A_{ij} x_j 
   +   
   \sum_{j}^N x_j J_j } 
   \\
   &= 
   \frac{1}{\sqrt {\det A}} 
   \ExpB{ \frac{1}{2}\sum_{ij}^N  J_i \, A^{-1}_{ij} \, J_j } \>.
   \notag
\end{align}
Generalizing to complex numbers $z$ and $J$, one has 
\begin{align}
   F[J,J^{\ast}] 
   &= 
   \prod_{i=1}^N 
   \biggl \{ \,
      \int \! \frac{\rd z_i \, \rd z^{\ast}_i}{2 \pi } \,
   \biggr \} \,
   \ExpB{ 
      - \sum_{ij}^N  z^{\ast}_i A_{ij} z_j 
      + 
      \sum_{j}^N 
      \bigl [ \,
         z^{\ast}_j J_j + z_j J^{\ast}_j \,
      \bigr ] }
   \\
   &= 
   \frac{1}{\det A} 
   \ExpB{ \sum_{ij}^N  J^{\ast}_i \, A^{-1}_{ij} \, J_j } \>.
   \notag
\end{align}
The basic definition of a path integral comes from quantum mechanics where one has an infinite number of trajectories $q(t)$ going from $t_i$ to $t_f$ with the ends held
fixed.  One breaks each trajectory in time into N segments of length $\epsilon= (t_f-t_i)/N$  and defines 
\begin{equation}
   \int \calD q 
   =  
   \frac{1}{C( \epsilon) } \prod_{i=1}^{N-1}
   \int_{-\infty}^{\infty}  \frac {\rd q_i}{C( \epsilon )}
\end{equation}
where $C(\epsilon)$ can be determined by continuity arguments as discussed in Peskin and Schroeder \cite{ref:PeskinSchroeder}.  Similarly to evaluate $\calD \, \phi(\vec x, t)$, we imagine that there is a square lattice in the space-time volume $L^d T$ with equal lattice spacing $\epsilon$.  So we use the notation that on the lattice $\phi(\vec x, t) \rightarrow \phi(x_i)$, and up to an irrelevant
overall constant, we define \cite{ref:PeskinSchroeder}:
\begin{equation}
   \calD \, \phi(\vec x, t) 
   \equiv  
   \calD \, \phi(x)  
   =  
   \prod_i  \rd \phi(x_i) \>.
 \end{equation}
The lattice version of the Gaussian path integral is often performed by introducing a discrete Fourier series for $\phi(x_i)$, and performing the integrals in momentum space. 
For fields we then get the following results for Gaussian Integrals over  real and complex fields with the identification that $x_i \rightarrow \phi(x)$.
For real fields,
\begin{align}
   F[J] 
   &=
   \int  \calD \phi(x) \,
   \ExpB{ 
      -
      \frac{1}{2} \int \! \rd x \, \rd y \, \phi(x) \, A(x,y) \, \phi(y)  
      +   
      \int \! \rd x \, \phi(x) \, J(x) }
   \\
   &= 
   \frac{1} {\sqrt {\det A}}\,
   \ExpB{ 
      \frac{1}{2} \int \! \rd x \, \rd y \, J(x) \, A^{-1}(x,y) \, J(y) }
   \notag
\end{align}
and for complex fields,
\begin{align}
   F[J, J^\ast ] 
   &=
   \int \! \calD \phi(x) \calD \phi^\dag(x)
   \\
   & \hspace{3em} 
   \times
   \ExpB{
      - 
      \int \! \rd x \, \rd y \,
      \phi^\dag(x) \, A (x,y) \, \phi(y)    
      +  
      \int \! \rd x \,
      \bigl [ \,
         \phi^\dag(x) \, J(x) + J^\dag(x) \phi(x) 
      \bigr ] }
   \notag \\
   &= 
   \frac{1}{\det A} \,
   \ExpB{ \int \rd x \, \rd y \, J^\dag (x) \, A^{-1}(x,y) \, J(y) } \>.
   \notag
\end{align}

%
%
\section{\label{append-B}Identities using functional Dirac delta functions} 

In introducing composite fields it is often useful to introduce them in a way that does not change the path integral.  Starting from the identity 
\begin{equation}\label{DDF.e:1}
   1 = \int_{-\infty}^{+\infty} \!\!\!\! \rd \sigma \, \delta( \sigma-F ) \>,
\end{equation}
one then introduces the Fourier representation of the delta function
\begin{equation}\label{DDF.e:2}
   1
   = 
   \int _{-\infty}^{+\infty} \frac{dk}{2 \pi}
   \int_{-\infty}^{+\infty} \!\!\!\! \rd \sigma \,  e^{i k(\sigma-F)} \>.
\end{equation}
For notational convenience, one often lets $k = i \chi $, and writes the integral over $\chi$ along the imaginary axis so that 
\begin{equation}\label{DDF.e:3}
   1
   =
   \int_{-i \infty}^ {i \infty } \frac{\rd \chi}{2 \pi i}  
   \int_{-\infty}^{+\infty} \!\!\!\! \rd \sigma \,  e^{- \chi (\sigma-F)} \>.
\end{equation}
This translates to the functional identity 
\begin{equation}\label{DDF.e:4}
   1
   =  
   \calN \iint_{\calC} \! \calD  \chi(x) \, \calD \sigma(x) \,
   e^{-\int \! \rd x \, \chi(x) \, [\, \sigma(x) - F(x) \,] } \>,
 \end{equation}
where $\calN$ is a normalization constant, and $F(x)$ and arbitrary function.  The integration region $\calC$ is over complex functions.  We use this identity in this paper in several ways. One way was  to introduce the  auxiliary fields  $\sigma(x) = F[\phi(x)]$ into the path integral for the generating functional.  The constant $\calN$ can be determined from the lattice definitions, but since we are only interested in connected correlation functions which are derived from the log of the generating functional, this constant is inessential to the dynamics and thus will be ignored in this paper.  Similarly, for convenience of organizing a particular expansion one may want to rescale $\chi$ and $\sigma$ in Eq.~\eqref{DDF.e:4} which only changes an overall irrelevant constant.

%
%
\section*{References}
\bibliographystyle{elsarticle-num}
%
%

%
%
\end{document}